\documentclass[alpha-refs]{wiley-article}

\usepackage{color}
\usepackage{ulem}

\usepackage{caption}
\usepackage{graphicx}
\usepackage{siunitx}
\usepackage{subcaption}
\usepackage{rotating}
\usepackage{float}
\usepackage{array}
\usepackage{ragged2e}
\usepackage{chngcntr} 
\usepackage{multirow}
\usepackage{multicol}
\usepackage{adjustbox}

\papertype{Original Article}

\title{Intercomparison of a High-Resolution Regional Climate Model Ensemble for Catchment-Scale Water Cycle Processes under Human Influence}

\author[1]{Jane Leonor Roque}
\author[2]{Francis Da Silva Lopes}
\author[1]{Julian Alberto Giles}
\author[2]{Benjamin D. Gutknecht}
\author[3]{Bernd Schalge}
\author[4]{Yikui Zhang}
\author[5]{Marco Ferro}
\author[1]{Petra Friederichs}
\author[4]{Klaus Goergen}
\author[6]{Stefan Poll}
\author[3]{Arianna Valmassoi}

\affil[1]{Institute of Geosciences, Meteorology Section, University of Bonn, Germany}
\affil[2]{Institute of Geodesy and Geoinformation, University of Bonn, Germany}
\affil[3]{Deutscher Wetterdienst, Offenbach am Main, Germany}
\affil[4]{Institute for Bio- and Geosciences (Agrosphere, IBG-3), Forschungszentrum Jülich, Jülich, Germany}
\affil[5]{Institute for Food and Resource Economics, University of Bonn, Germany}
\affil[6]{CASA Simulation and Data Laboratory Terrestrial Systems, Jülich Supercomputing Centre (JSC), Forschungszentrum Jülich, Jülich, Germany}

\corraddress{Jane Leonor Roque, Meteorology Section, Institute of Geosciences, University of Bonn, Bonn, North Rhine-Westphalia, 53121, Germany}
\corremail{jroquema@uni-bonn.de}

\fundinginfo{Funded by the Deutsche Forschungsgemeinschaft (DFG, German Research Foundation) – SFB 1502/1–2022 - Projektnummer: 450058266. }

\runningauthor{Roque et al.}

\begin{document} %

\maketitle

\begin{abstract}

Understanding regional hydroclimatic variability and its drivers is essential for anticipating the impacts of climate change on water resources and sustainability. Yet, considerable uncertainty remains in the simulation of the coupled land–atmosphere water and energy cycles, largely due to structural model limitations, simplified process representations, and insufficient spatial resolution. Within the framework of the Collaborative Research Center 1502 – DETECT, this study presents a coordinated intercomparison of regional climate model simulations designed for water cycle process analysis over Europe. We analyze the performance of simulations using the ICON and TSMP1 model systems and covering the period 1990–2020, comparing against reference datasets (E-OBS, GPCC, and GLEAM). We focus on 2~m air temperature, precipitation and evapotranspiration over four representative basins — the Ebro, Po, Rhine, and Tisa — within the EURO-CORDEX domain.

Our analysis reveals systematic cold biases across all basins and seasons, with ICON generally outperforming TSMP1. Precipitation biases exhibit substantial spread, particularly in summer, reflecting the persistent challenge of accurately simulating precipitation. ICON tends to underestimate evapotranspiration, while TSMP1 performs better some seasons. Sensitivity experiments further indicate that the inclusion of irrigation improves simulation performance in the Po basin, which is intensively irrigated, and that higher-resolution sea surface temperature forcing data improves the overall precipitation representation. This baseline evaluation provides a first assessment of DETECT’s multi-model ensemble and highlights key structural differences influencing model skill across hydroclimatic regimes.

\end{abstract}

\section{INTRODUCTION} 

Anthropogenic changes of land use and land cover affect biophysical processes in the energy and water cycle \citep{Pielke_2011,Mahmood}, from the local to regional scale and remotely. Accurate climate simulations are needed to provide decision makers with reliable information to develop effective climate change adaptation and mitigation strategies \citep{Gutowski2020}. In this context, regional drying and wetting trends emerged in recent years and pose a challenge in terms of process representation and understanding. While the presence of large-scale drying and wetting patterns has been described extensively \citep{ipcc_ar6_wg1_2021}, both satellite observations and in-situ hydroclimatic trends cannot be explained by simple paradigms such as the “dry-gets-drier, wet-gets-wetter” or “global aridification” hypotheses, due to their complex time-varying spatial heterogeneous footprint \citep{Zaitchik2023}.
Further comparative studies of global and regional climate simulations have revealed that even the sign of the terrestrial water storage trends differs between models and observations \citep{greve_global_2014,gudmundsson_globally_2021,Jensen2024, hirschi_potential_2025}, highlighting the need for an improved representation of the coupled water-energy cycles representation across different Earth System compartments.

The underlying modeling uncertainty can be attributed to a number of factors, e.g., model structural deficits, unrepresented processes -such as groundwater dynamics-, or to the low resolution of the current modeling systems. In fact, global models typically operate at grid resolutions of 50-100~km while many regional climate models run on resolutions of around 10-25~km, relying on the parameterization of many water cycle related processes. Deep and shallow convection at these resolutions must be parametrized and many key land-surface and hydrologic processes are only represented in a simplified manner \citep{schar_kilometer-scale_2020,lucas-picher_convection-permitting_2021}. In recent decades, the development and evaluation of regional climate models have revealed substantial improvements in the simulation of regional climate patterns, but also persistent biases and spread among models, particularly for precipitation and land-surface fluxes \citep{christensen2007,Jacob2020,Prein2016}.

The models differ in their physical parameterizations, soil and vegetation schemes, treatment of lateral boundary conditions, etc., which strongly affect their representation of surface fluxes and the terrestrial water balance \citep{kotlarski_regional_2014,Poshyvailo2024,Silwimba2025}. Furthermore, the general lack of explicit coupling to subsurface or groundwater components limits their ability to capture the full spectrum of land-atmosphere interactions. In addition, most model systems do not account for lateral subsurface flow, i.e., 3D hydrodynamics, but feature only 1D gravitational subsurface flow with open lower boundary conditions. Therefore, subsurface-land-atmosphere feedbacks do not account for groundwater dynamics, which has an impact on land-atmosphere coupling and on soil moisture-temperature or -precipitation feedback loops \citep{keune_studying_2016}. Also, human interventions, such as irrigation or groundwater abstraction, which have a substantial impact on the terrestrial water cycle \citep{siebert_groundwater_2010, wada2010}, strongly affect the surface energy partitioning and hydrometeorology but are either over-simplified or absent in models \citep{hirsch_biogeophysical_2018,Seneviratne_2018,avalm2023}. Such limitations may lead models to produce contrasting water balance representations and underestimate, or even obscure, the anthropogenic impact on the hydroclimatic variability \citep{winkler2023, casirati2025}.  

In recent years, party due to the increase in computational resources, there is a tendency towards the use of convection-permitting regional climate models at grid resolutions below 4~km, which overcome some of the aforementioned limitations \citep{prein_review_2015,schar_kilometer-scale_2020}. At these resolutions, atmospheric models show improved representations of key processes (e.g., atmospheric convection, mesoscale circulations, small-scale weather events), leading to a more accurate representation of precipitation intensity, spatial distribution, and timing \citep[e.g.,][]{ban_first_2021,Fosser}. \citet{barlage_importance_2021} further showed that the representation of terrestrial hydrologic processes is scale-dependent and that an enhanced treatment of groundwater–land interactions can influence hydrometeorology at kilometer scales in the WRF model. However, the representation of hydrologic processes in convection-permitting models often remains simplified and human interventions are usually not represented. 

To address these limitations, modeling efforts are increasingly moving towards fully coupled regional Earth or climate system models. These models explicitly link the subsurface, land surface and atmosphere. The Terrestrial System Modelling Platform (TSMP) is such a model \citep{Shrestha2014,Gasper2014}. In its first version, TSMP1, the COSMO atmospheric model is coupled to the Community Land Model and the ParFlow integrated hydrologic model via the OASIS3-MCT coupler \citep{Shrestha2014}. TSMP1 thus enables a physically consistent representation of water and energy fluxes from the groundwater to the atmosphere, advancing the simulation of regional hydroclimate dynamics.

The Collaborative Research Center 1502 – DETECT – \citep{DETECT_citation}, funded by the German Research Foundation, builds upon these recent advances in regional climate modeling. In light of the intensification of the hydrologic cycle under climate change—with major implications for water resources, sustainability, and water security \citep{huntington_evidence_2006,wada_sustainability_2014}—there are still significant gaps in our understanding of how feedbacks within the coupled water–energy system respond to and amplify changes in the system state. These uncertainties are reflected in the often-contradictory patterns of observed hydrologic change \citep{Jensen2024}. The central hypothesis driving the CRC 1502 DETECT is that, in addition to greenhouse gas forcing and natural variability, human water use as well as land-use and land-cover change have induced persistent modifications of the coupled land–atmosphere water and energy cycles, leading to multiple local and non-local hydroclimatic effects and contributing to observed regional- to continental-scale water storage trends \citep{Ferro_2025}.

To answer the central hypothesis, coordinated regional simulation experiments with a state-of-the-art multi-physics multi-model ensemble over Europe were designed and performed within DETECT. Similar to studies such as \citet{kotlarski_regional_2014}, a model intercomparison and evaluation to observational reference data is a pivotal step to assess simulation performances, e.g., with respect to terrestrial water cycle processes and how the models represent key aspects of land-atmosphere interactions. 

This study analyzes the regional climate model ensemble for water cycle and water resource research from the first funding phase of DETECT. We rely on the framework introduced by \citet{Seneviratne_2018} to compare ten simulations from different models for a 30 year time span. We use 2-meter air temperature, precipitation, and evapotranspiration over land as key indicators of the energy and water cycle, and we assess their performance through an inter-model comparison and against external reference datasets. We focus on four European basins that are representative of the continent's different bioregions, i.e., where climatic conditions can vary considerably: the Rhine, the Tisa, the Ebro, and the Po river basins. Throughout our analysis, we address the following research questions to guide our model intercomparison: (1) What are the key similarities and differences in the representation of the water cycle across model systems? (2) What do these reveal about model behavior at the basin scale? (3) What are the capabilities of the simulations to represent the human impact? In addition, the study introduces some of the datasets produced during the first funding phase of DETECT.

\section{METHODOLOGY} 
\subsection{Simulation data} \label{sect:model_descriptions}
Ten simulations are available for comparison, all covering the EURO-CORDEX model domain (see Figure \ref{fig:MapEuroCordex}). 
Table \ref{tab:simulation_overview} summarizes some key characteristics of the model simulations, and each experiment acronym refers to the model name, horizontal grid spacing and experiment type. 
\begin{sidewaystable}
    \caption{Model characteristics. Please note: If no information is provided in the SST column, then the SST is prescribed jointly with the boundary condition data and stems form the same dataset; likewise for the irrigation methods column, no information means the model is run without irrigation.}\label{tab:simulation_overview}
    \begin{adjustbox}{scale=0.90,center}
    \begin{threeparttable}
        \begin{tabular}{c c m{1.4cm} c m{1.4cm} m{1cm} m{1.3cm} m{1.4cm} c c c}
        \headrow
        Acronym & Model & Boundary conditions & Full Period & Horizontal resolution & No. of vertical levels & Radiation scheme & Land surface scheme & Land use external data & Irrigation method & SST\\
        TSMP1\_12\_ctrl & TSMP1 & ERA5 & 1979--2022 & 12~km & 50 & RRTM & CLM 3.5 & GLC-2000 & - & - \\
        TSMP1\_12\_irri\_w & TSMP1 & ERA5 & 1984--2022 & 12~km & 50 & RRTM & CLM 3.5 & GLC-2000 & HWU1\textsuperscript{1} & - \\
        TSMP1\_12\_irri\_s & TSMP1 & ERA5 & 1990--2022 & 12~km & 50 & RRTM & CLM 3.5 & GLC-2000 & HWU2\textsuperscript{2} & - \\
        ICON\_12\_ctrl & ICON & ERA5 & 2002--2020 & 12~km & 60 & ecRad & TERRA & GlobCover2009 & - & -\\
        ICON\_12\_sst\_era5 & ICON & ERA5 & 2002--2011 & 12~km & 60 & ecRad & TERRA & GlobCover2009 & - & ERA5$^{SST}$\textsuperscript{3}\\
        ICON\_12\_sst\_mur & ICON & ERA5 & 2002--2011 & 12~km & 60 & ecRad & TERRA & GlobCover2009 & - & MURSST\textsuperscript{4}\\
        ICON\_12\_sst\_ostia & ICON & ERA5 & 2002--2011 & 12~km & 60 & ecRad & TERRA & GlobCover2009 & - & OSTIA\textsuperscript{5}\\
        ICON\_3\_ctrl & ICON & ICON-DREAM & 2010--2022 & 3~km & 75 & ecRad & TERRA & GlobCover2009 & - & -\\
        ICON\_3\_irr & ICON & ICON-DREAM & 2010--2022 & 3~km & 75 & ecRad & TERRA & GlobCover2009 & RAW\textsuperscript{6} & -\\
        ICON\_3\_rea & ICON & ERA5 & 2018--2021 & 3~km & 75 & ecRad & TERRA & GlobCover2009 & - & -\\
        \hline  
    \end{tabular}
    \begin{tablenotes}
        \item \textsuperscript{1} Human water use 1 \citep{Wada2016}
        \item \textsuperscript{2} Human water use 2 \citep{Siebert2010}
        \item \textsuperscript{3} ERA5$^{SST}$ \citep{Yang2021}
        \item \textsuperscript{4} MURSST \citep{Chin2017} 
        \item \textsuperscript{5} OSTIA \citep{Donlon2012}
        \item \textsuperscript{6} Readily available water
        \end{tablenotes}
    \end{threeparttable}
    \label{sim_charac}
    \end{adjustbox}
\end{sidewaystable}

Seven of the ten experiments use the ICOsahedral Nonhydrostatic model (ICON v2.6.4, except v2.6.6-nwp3c for the ICON\_3\_irr/ctrl), developed by the German Weather Service and the Max Planck Institute for Meteorology \citep{Zängl2015,Heinze2017,Hohenegger2023,Gassmmann2008} and operationally used for numerical weather prediction as well as climate simulations.
The other three experiments use a fully integrated regional climate modelling framework, the TSMP1, designed to explicitly resolve land–atmosphere–subsurface interactions. 
The modelling system couples three component models via the OASIS3-MCT coupler \citep{Valcke2013}: (i) the COnsortium for Small Scale MOdelling (COSMO) atmosphere model v5.01 \citep{Baldauf2011,Doms2011}, which is the predecessor of the ICON model; (ii) the Community Land Model (CLM v3.5) developed at the National Center for Atmospheric Research \citep{Oleson2008}; and (iii) the ParFlow v3.12.0 surface–subsurface flow integrated hydrologic model \citep{kuffour_simulating_2020,Jones2001,Kollet2006,Maxwell2013}. 

As can be seen from Table \ref{tab:simulation_overview}, most of the simulations are forced with the ERA5 reanalysis \citep{Hersbach2020}, a global reanalysis with around 31~km of horizontal resolution and 137 vertical levels, which is widely used as initial and boundary conditions for regional climate model studies. It is the recommended dataset to drive CORDEX evaluation simulations \citep{katragkou_delivering_2024}. The original resolution was remapped to 12~km resolution using routines from the model’s runtime environment \citep{Rockel2022}, and, if needed, using sea surface temperature (SST) as the prescribed lower boundary condition over the EURO-CORDEX domain.

The two convection-permitting ICON experiments (ICON\_3\_irr and ctrl) use instead the newly produced ICON-DREAM reanalysis \citep{valmassoi2024ems} as boundary conditions, since it provides a higher spatial resolution (approximately 13~km, compared to 31~km of ERA5), which is preferable for the target resolution. To ensure consistency, we used the same ICON-nwp version as ICON-DREAM reanalysis.
The 3~km ICON reanalysis (ICON\_3\_rea) uses the ERA5 and its 10 ensemble members, which are at a coarser resolution (62~km), as ICON-DREAM was not yet available at the start of the DETECT simulation experiment.

The first seven simulations, which are convection-parameterized, have a 12~km horizontal grid spacing and 50/60 vertical levels, TSMP1 and ICON respectively \citep[for more information:][]{zhang2024,daSilvaLopes2025}, extending up to 22~km and 23.5~km, respectively. 
The last three convection-permitting simulations have a horizontal resolution of 3~km with 75 vertical levels extending up to 22~km (ICON\_3\_irr/ctrl) and 30~km (ICON\_3\_rea).

All experiments use the single-moment cloud microphysics and precipitation scheme \citep{Doms2011,Seifert2008}, and the prognostic turbulent kinetic energy scheme for vertical mixing \citep{Raschendorfer2001}. 
The Tiedtke bulk mass flux scheme \citep{Tiedtke1989} is used for shallow convection for all experiments, but disabled for deep and mid-level convection \citep{DeLucia} in the 3~km simulations. 

TSMP1, based on COSMO and with coupled CLM and ParFlow constitutes a different model system than ICON. Nevertheless, the atmospheric model COSMO and ICON share some common parametrizations and physics packages and diagnostics, e.g., for cloud cover.
Notable difference in parameterizations between the TSMP1 and ICON-experiments are, e.g., the radiation scheme and the land surface model components. 
TSMP1 uses the RTTM radiation parameterization available to the COSMO predecessor, all ICON experiments use the more advanced ecRad radiation scheme, which brings improvements in process representation \citep{Hogan2018}.
While the ICON-experiments use the TERRA land surface model \citep{Schrodin2001,Schulz2016}, which was also default in the COSMO model, TSMP1 replaces it with the CLM3.5.

Now each set of experiments is presented to further contextualize the respective experiment designs, depending on each simulation's objectives. 

\subsubsection{Terrestrial System Modeling Platform (TSMP) simulations}

The Terrestrial System Modelling Platform (TSMP) is a fully integrated regional climate modelling framework designed to explicitly resolve subsurface-land–atmosphere interactions. In contrast to regional climate models that employ simplified hydrologic schemes, TSMP incorporates a full three-dimensional representation of subsurface hydrodynamics, including groundwater flow. This capability enables a more realistic simulation of soil moisture states, energy fluxes, and associated land–atmosphere feedbacks, thereby improving the representation of heatwave intensity and reducing temperature biases, as demonstrated for both the 2003 European heatwave and multi-decadal climatologies \citep{Keune2018,FurushoPercot2022,Poshyvailo2024}. This bedrock-to-atmosphere coupling allows TSMP to capture surface–subsurface exchange processes and groundwater dynamics with high fidelity, leading to improved soil moisture profiles and evapotranspiration behaviour \citep{Zipper2019,Naz2022}. Previous work by \citep{Hartick2022} has shown that TSMP’s explicit groundwater representation enhances the persistence of drought events through groundwater memory effects.

In this study, we conducted three TSMP1 experiments covering the period 1990--2022: one control simulation without irrigation (CTL) and two with irrigation (HWU-W and HWU-S). The irrigation forcing is based on two monthly datasets spanning 1984–2022: HWU-W from \citep{Wada2016} and HWU-S from \citep{Siebert2010}. The irrigation forcing is implemented in ParFlow through source–sink terms applied to the relevant model layers. \citet{Keune2018} gives an overview on the use of TSMP1 including human water use.

\subsubsection{Sea surface temperature experiments}

The SST product choice as a prescribed lower boundary in limited-area climate simulations remains a critical issue, especially when the model domains partly cover the ocean \citep[cf.][]{Katragkou2015,Cassola2016,Liu2024}. Different SST products, ranging from global model outputs \citep{Donlon2002} to high-resolution satellite-based datasets, can lead to substantial variability in the lower boundary forcing. This is particularly important for water cycle studies, where SST patterns influence the low-level atmospheric dynamics \citep[e.g.,][]{Zhou2019}, ocean latent heat flux, and thus the moisture transport to land \citep{Donlon2002}, with an impact on precipitation \citep{goergen_boundary_2021}. The DETECT SST sensitivity experiments aim to assess the sensitivity of the simulated water budget components to SST forcing over the EURO-CORDEX \citep{Jacob2020} domain ($\sim$27$^\circ$N–72$^\circ$N, $\sim$22$^\circ$W–45$^\circ$E). The setup, parametrizations and experimental design for ICON 12~km SST experiments are described in detail by \cite{daSilvaLopes2025}.

The SST experiments cover a time span from June 2002 to May 2011, each using the same ICON-CLM configuration, and they differ in the imposed SST dataset. Changing only the lower boundary condition (SST) while retaining the lateral boundary forcing can introduce inconsistencies in the regional energy and moisture budgets. Nonetheless, \cite{daSilvaLopes2025} showed that these effects were negligible, with basin-scale water budget offsets (see their Table 2), indicating no significant imbalance across experiments.

The four SST products tested in our regional climate model simulations are: (1) hourly ERA5 skin temperature (ERA5$^{SKT}$), (2) daily ERA5 SST \citep[ERA5$^{SST}$,][]{Yang2021}, (3) daily Multi-scale Ultra-high Resolution SST \citep[MURSST,][]{Chin2017}, and (4) daily gap-free maps of the Operational Sea Surface Temperature and Ice Analysis foundation SST \citep[OSTIA,][]{Donlon2012}. ERA5$^{SKT}$ reflects the radiative equilibrium temperature of an infinitesimally thin surface layer and includes diurnal variability \citep{Donlon2002,Minnett2019,IFS2020}, based on model-data synthesis \citep{Zeng2005,Luo2020}. ERA5$^{SST}$, MURSST and OSTIA, by contrast, provide foundation SST \citep{Minnett2019}, a bulk near-surface temperature unaffected by diurnal warming, derived from observation-based datasets \citep{Titchner2014,Donlon2012,Chin2017,Hirahara2016,Hersbach2020}. MURSST’s higher resolution (0.01$^\circ$) enables finer representation of mesoscale ocean features \citep[e.g.,][]{Lebeaupin2015,Seo2023}. However, its availability only since June 2002 constrained the temporal extent of the simulations. For preprocessing, we use Climate Data Operators \citep[][]{Schulzweida2023}, interpolating SSTs to the ICON-CLM 0.11$^\circ$ grid. Skin temperature over land and ice, as well as the land-sea mask, remain unchanged. ERA5$^{SKT}$ serves as our reference, consistent with its use in atmospheric data assimilation \citep{Akella2017,Zeng2005}, weather prediction \citep{Scanlon2024}, and air-sea interaction studies \citep{Fairall1996}.

\subsubsection{Irrigation experiment with ICON-nwp} 
The main purpose of irrigation is to fulfill crop water requirements, leading to higher soil moisture values. This increase in soil moisture initiates a range of feedbacks and interactions with various components and processes of the Earth system \citep{McDermid2023}. However, the representation of irrigation in weather and climate models remains a significant limitation. Omitting irrigation introduces significant biases even in reanalysis datasets \citep{Tuinenburg2017}, as they miss displaying heterogeneity at sub-grid scale \citep{Mangan2023}. The cooling effect is one of the most recognized irrigation effects, yet the specific extent of this impact remains uncertain locally and globally \citep{avalm2023}. As shown by \citet{Keune2018} irrigation in combination with groundwater abstraction may not only have local effects but can also lead to remote impact due to atmospheric moisture transport.

In this study, we performed one control and one irrigation experiment with the model ICON-nwp in Limited Area Mode (LAM) to quantify the long-term impact of irrigation on surface and atmospheric variables.  

The control experiment covers the time span 2010-01-01/2022-12-31. The irrigation experiment has a parameterization located in the interface between ICON and the land surface scheme. The amount of irrigation is fixed, equal to the readily available water (RAW) \citep{Allen1998}. The irrigation experiment started from a restart on 2010-04-11 00UTC and covers the same period as the control. In each year, the irrigation event starts on May 1, at 5 UTC and runs for 5 hours. It repeats every 12 days until Aug 17, with a total of 10 irrigation events. We considered 2010 the spin-up time for both experiments.

\subsubsection{Sparse Input Reanalysis (2018--2021)}
For the sparse-input reanalysis, the ICON model is run in conjunction with the KENDA \citep{Schraff2016} data assimilation framework. Unlike all other experiments, the reanalysis is not a free forecast but runs in an assimilation loop where every three hours observations are assimilated. The observation types assimilated here include synoptic stations, buoys, aircraft data, pilot observations and radiosondes, but critically do not include any radar or satellite data, except atmospheric motion vectors. This setup is therefore considered a sparse-input reanalysis, which is a  compromise between computational demand and accuracy of the analysis.

The data assimilation system KENDA uses the localized ensemble transform Kalman filter (LETKF) and therefore needs an ensemble in addition to the main deterministic run. In operational systems (which mostly use 3/4DVar or similar variational methods), the ensemble is usually run at a lower resolution than the deterministic run to reduce the computational costs while still adequately sampling uncertainty. In our case, this approach would have meant to run the ensemble in lower resolution, non-convection-permitting mode, leading to different precipitation and cloud statistics. Consequently, it was decided to keep the resolution at 3~km for the ensemble as well.

As mentioned before, ERA5 was chosen as boundary data. Not only is it a reanalysis itself, it also features 10 ensemble members which is ideal, since boundary data is also needed for the ensemble. However, this comes with two constraints. First, this limits the ensemble to just 10 members, which is rather low for such data assimilation methods. Several tests have been conducted and results have shown that indeed adding ensemble members does improve the quality of the analysis further, but we have also noted that there are no adverse effects to using just 10 members, the system works in a stable and predictable way. Secondly, ERA5 data is only available at a rather coarse resolution so the transition to the 3~km model domain is rather extreme. This was tested and only a very minor improvement can be achieved by using operational ICON fields ($\sim$13~km) instead. 

With January 2018 being considered a spin-up phase, currently the reanalysis covers the time February 2018 to December 2021, with an extension until the end of 2023 underway. Unlike other simulations, evapotranspiration is not directly available as an output variable but is instead derived from surface latent heat flux.

\subsection{Reference datasets} \label{sect:ref_datasets}
For comparison of simulated 2 m air temperature, precipitation, and evapotranspiration against external references, we selected several gridded observational and calibrated reanalysis products (Table \ref{tab:reference_data_table}) as follows.

\noindent
\begin{enumerate} [label=(\alph*)]
    \item \textbf{E-OBS} \\
    We used daily mean air temperature data from the E-OBS gridded dataset, developed by the European Climate Assessment \& Dataset (ECA\&D) and the Copernicus Climate Change Service \citep{Haylock2008_eobs,Cornes2018_eobs}. E-OBS provides daily meteorological variables across Europe based on quality-controlled station observations, interpolated onto regular latitude–longitude grids at 0.1$^\circ$ and 0.25$^\circ$ resolution. The dataset spans from 1950 to 2024 (v31.0e) and covers the European land domain from 25$^\circ$N-- 71.5$^\circ$N and 25$^\circ$W--45$^\circ$E. E-OBS mean temperature usually describes the temperature measured at 2 meters above the surface.
    
    E-OBS is based on a 20-member ensemble to account for interpolation uncertainty, with the ensemble mean typically used as a best estimate. The gridded fields are produced using conditional simulation methods that preserve spatial correlations while providing uncertainty information. While the dataset is widely used as a high-resolution reference for climate studies in Europe, its accuracy depends on the underlying station network, and uncertainties are larger in regions with sparse observational coverage. In this study, we used the ensemble-mean daily-mean temperature at 0.25$^\circ$ and computed the monthly means.

    \item \textbf{GPCC} \\
    We compare simulated land precipitation with monthly precipitation sums from the Global Precipitation Climatology Centre (GPCC) 'Full Data Monthly' product version 2022 \citep{Schneider2022_GPCC}.
    Operated by the German Meteorological Service (DWD) under the auspices of the World Meteorological Organization (WMO), GPCC provides globally gridded monthly observations of land-surface precipitation. 
    These are based on quality-controlled data from rain gauges and are optimized for spatial coverage.
    Station density varies with time and region, but it is particularly high across Europe and is generally of the order of 1--2 gauges per 0.25$^\circ$ cell.

    \item \textbf{GLEAM} \\
    We utilized evaporation estimates from the GLEAM (Global Land Evaporation Amsterdam Model) dataset (gleam.eu). GLEAM provides daily evaporation (or evapotranspiration) and its component fluxes (transpiration, soil evaporation, interception loss, open-water evaporation, snow sublimation) over land surfaces, along with auxiliary variables such as potential evaporation, root-zone and surface soil moisture, evaporative stress, and sensible heat flux. 
    
    The latest version (GLEAM 4.2a, \cite{Miralles2025_GLEAM}) offers continuous global coverage from 1980 to 2023 at 0.1$^\circ$ spatial resolution, using a hybrid machine-learning and physical model approach. The model computes potential evaporation based on surface net radiation, temperature, wind speed, vapor pressure deficit, and vegetation properties, and then applies evaporative stress constraints (derived from remote sensing of soil moisture, vegetation optical depth, etc.) to partition into actual evaporation components. As with any modeled dataset, uncertainties arise from forcing errors, remote sensing uncertainties, and limitations in representing spatial heterogeneity, particularly in regions with sparse sensor coverage or complex terrain. Here we use GLEAM monthly Actual Evaporation averaged over the selected basins as a reference.
\end{enumerate}

\begin{table}[bt]
    \caption{Reference data sets}
    \begin{threeparttable}
        \begin{tabular}{llcc}
            \headrow
            \thead{Model} & \thead{Version} & \thead{Variable} & \thead{Gridded native resolution}\\
            E-OBS & v31.0e & Monthly mean of daily mean air temperature & 1/4$^\circ$ \\
            GPCC  & v2022  & Monthly precipitation (full data monthly)  & 1/4$^\circ$ \\
            GLEAM & v4.2a  & Monthly Actual Evaporation & 1/10$^\circ$\\\hline
        \end{tabular}
    \end{threeparttable}
    \label{tab:reference_data_table}
\end{table}

\subsection{Statistical validation}

\begin{figure}[!htb]
    \centering
    \includegraphics[width=0.9\linewidth]{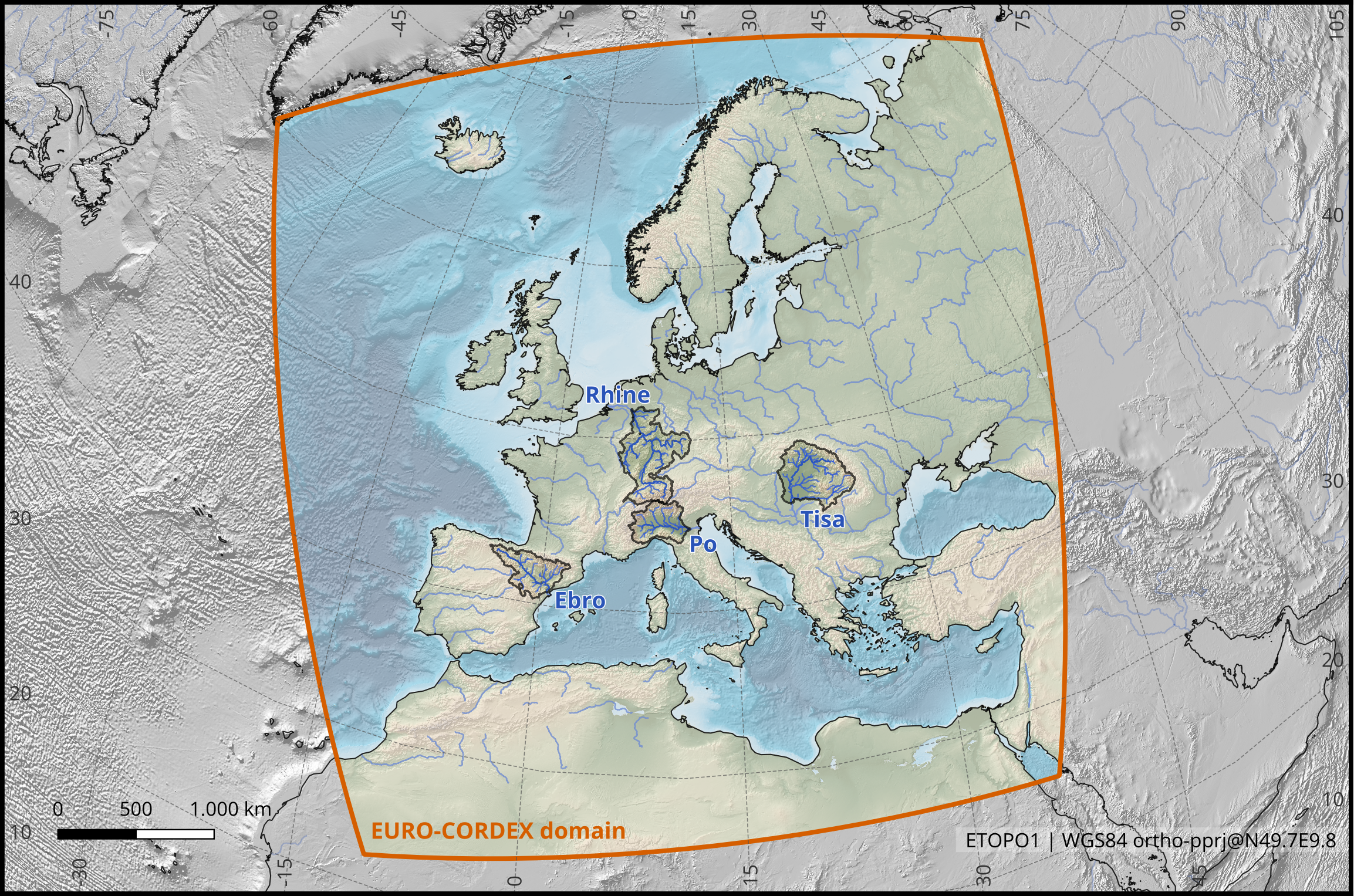}
    \caption{Overview map of our study area in the EURO-CORDEX domain. Highlighted are the four selected river catchments used in this study for area-averaged analysis: Rhine River (Rees station), Rio Ebro (Tortosa station), Fiume Po (Pontelagoscuro station), and Tisa River (Senta station, tributary to Danube River). The named stations are GRDC-registered locations of river gauges and thus define the upstream portion of the drainage basins, respectively \citep{GRDC_hydrosheds}.}
    \label{fig:MapEuroCordex}
\end{figure}
To compare the model performance of the simulations described above, we focused on three key variables: 2 m air temperature (T2m), precipitation (P) and evapotranspiration (ET). The corresponding observational datasets used for validation, as stated in the previous section, are E-OBS (T2m), GPCC (P) and GLEAM (ET). Initially, due to the long temporal coverage, we evaluate the TSMP1 simulations against observations based on the climatological mean of each variable at grid-point scale. Then, given our interest in the water cycle representation, we aggregate data from all simulations and observations at the watershed level. We chose four representative watersheds, two located in mid-Europe and two located in southern Europe: Rhine, Tisa (sub-watershed from the Danube), Ebro and Po (Figure \ref{fig:MapEuroCordex}). Therefore, at a watershed level, we work with timeseries of spatially aggregated monthly means (T2m) and  monthly sums (P and ET) .

The validation consists of five statistical metrics: The bias, standard deviation (STD), root mean square error (RMSE), mean absolute error (MAE) and the Pearson correlation coefficient. We calculate these metrics for an interannual seasonal and yearly evaluation with data separated to two simulation periods. We opt to separate the data to account for overlapping simulation years across experiments. The first period corresponds to 2003--2010 (January 1, 2003 - December 31, 2010). The second period corresponds to 2011--2020 (January 1, 2011 - December 31, 2020).

Due to horizontal grid resolution differences between the simulations and the reference datasets, we remap the modeled data to the grids of the reference datasets for the comparisons on a grid-point scale. T2m is remapped using bilinear interpolation and P and ET are remapped using conservative methods to preserve the total amount of each field \citep{Taylor2024}. For this task we use the xESMF Python package \citep{zhuang_2025_15304267}.

\section{RESULTS AND DISCUSSION} 
\subsection{Climatological biases of TSMP1 CTRL}
\label{tsmp_clim}

\begin{figure}[!htb]
\centering
\begin{subfigure}{0.45\textwidth}
  \caption{Temperature 2 m}
  \includegraphics[width=\linewidth]{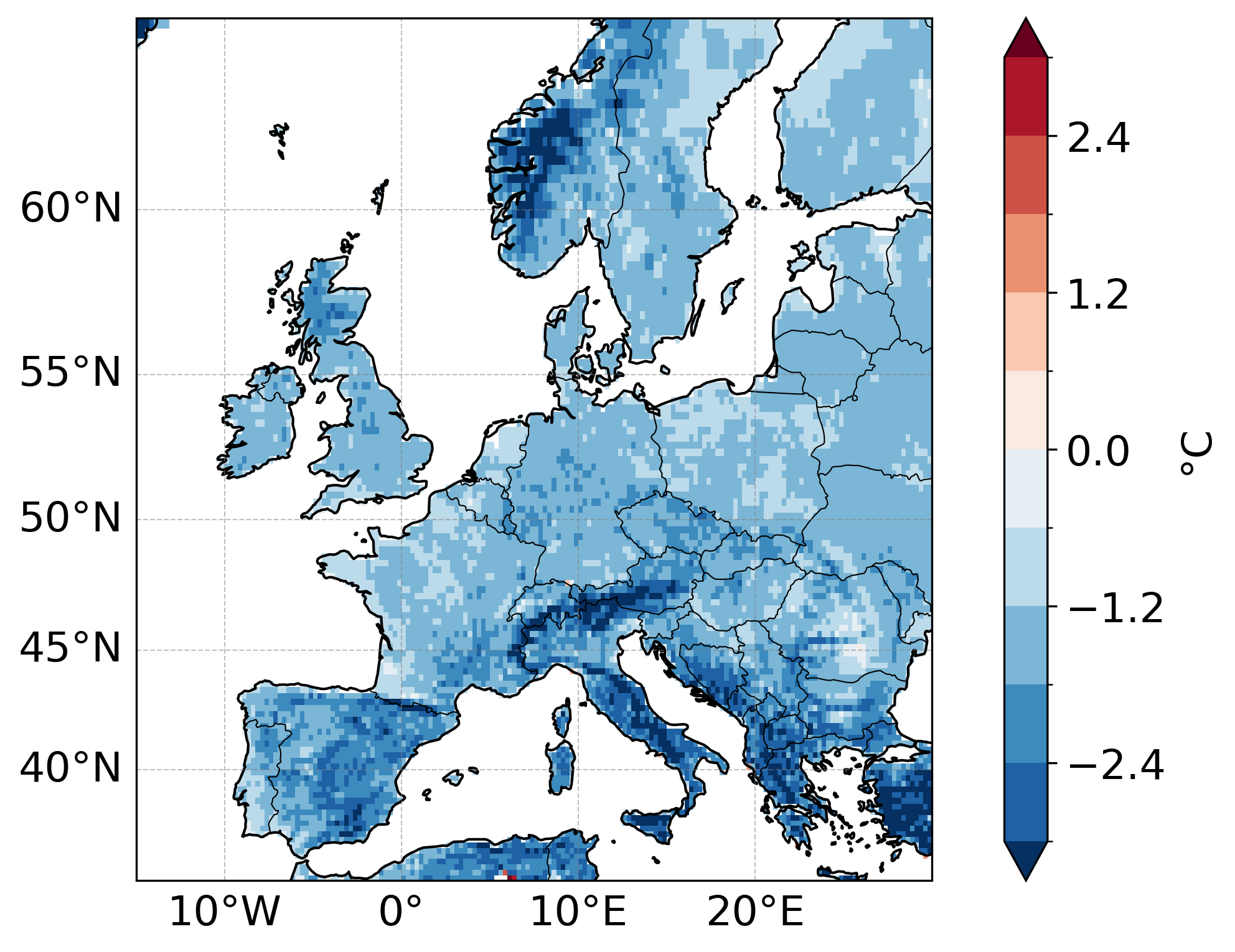}
  \label{clim_T2m}
\end{subfigure}\hspace{1mm} 
\begin{subfigure}{0.45\textwidth}
  \caption{Precipitation}
  \includegraphics[width=\linewidth]{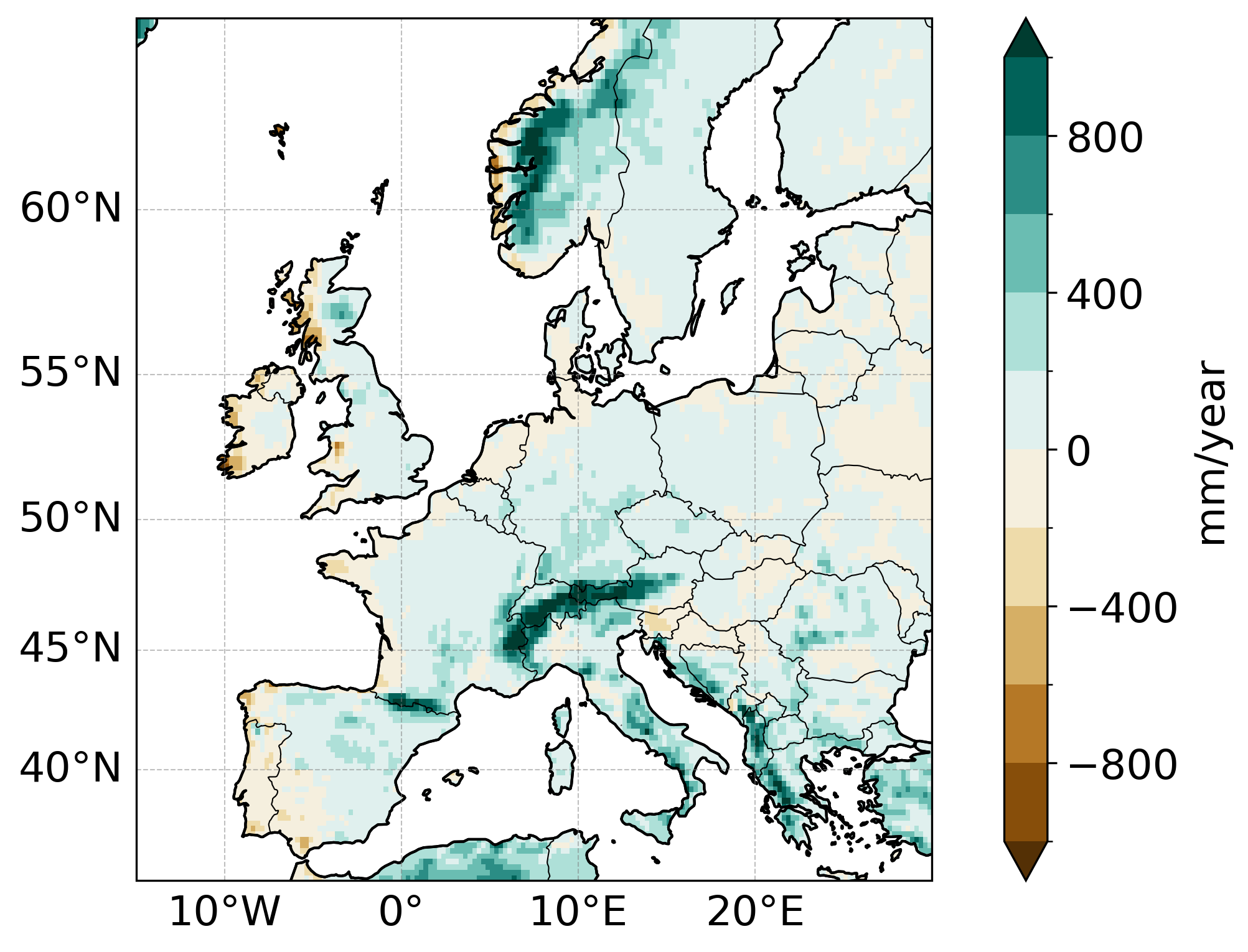}
  \label{clim_PP}
\end{subfigure}
\vspace{-9mm}
\medskip
\begin{subfigure}{0.45\textwidth}
  \vspace{-5mm}
  \caption{Evapotranspiration}
  \includegraphics[width=\linewidth]{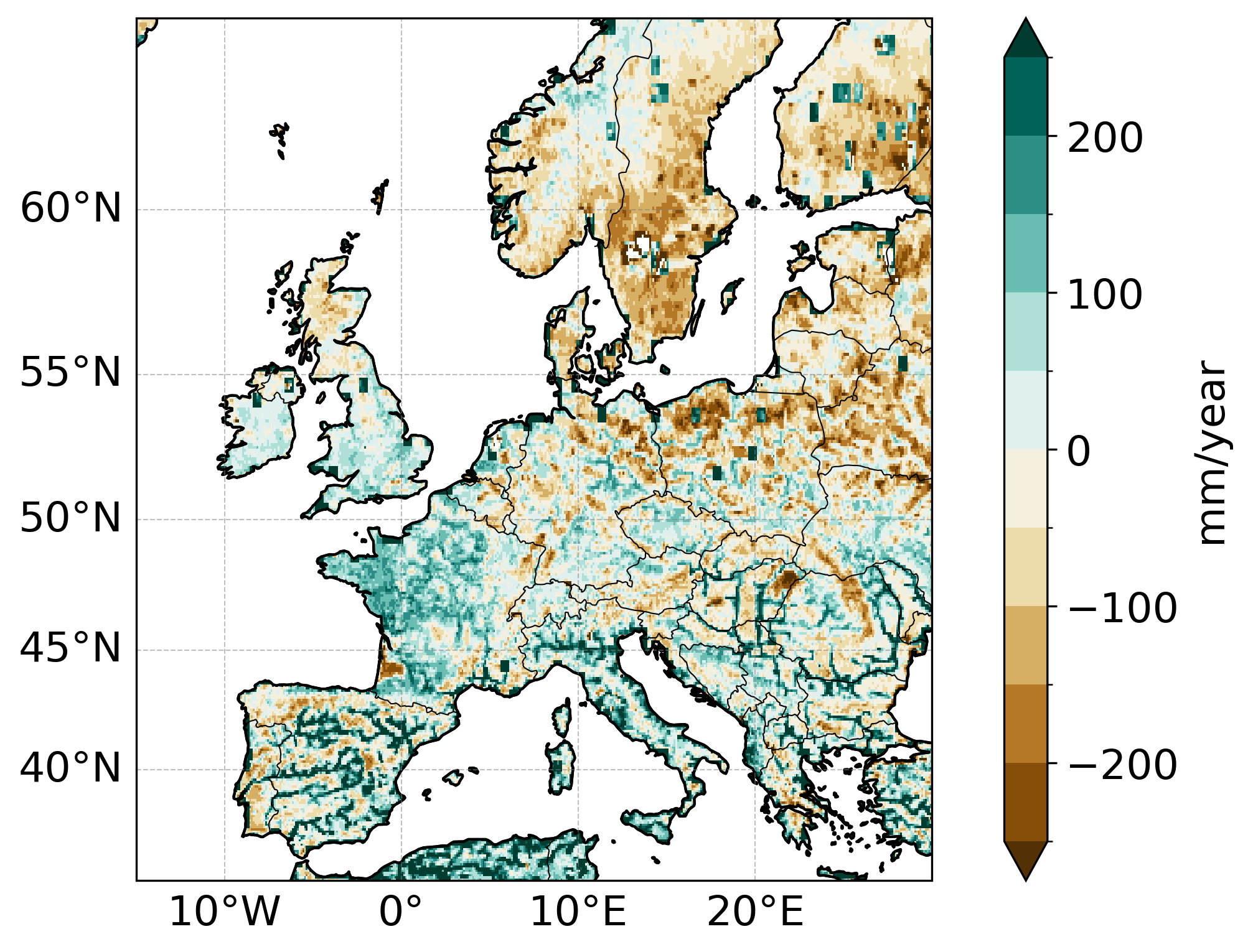}
  \label{clim_ET}
\end{subfigure}\hspace{1mm} 
  \caption{TSMP1 CTRL (TSMP1\_12\_ctrl) 1990-2020 mean biases of T2m, P and ET with respect to E-OBS, GPCC and GLEAM respectively.}
    \label{climatology_biases}
\end{figure}

The TSMP1 CTRL climatological biases (i.e., differences of long-term means, simulation minus reference) of T2m, P and ET for the period 1990--2020 are shown in Figure \ref{climatology_biases}. Regarding T2m, Figure \ref{climatology_biases}a shows that TSMP1 CTRL has a negative T2m bias over all land areas. The bias is more negative (stronger) over areas with complex topography.

Figure \ref{climatology_biases}b shows that a positive precipitation bias that covers almost all of Europe, with some scattered areas having a slight negative bias. Regions of high positive bias are directly associated with elevated and complex topography, like over the Alps, the Pyrenees, Norway and the Balkans. 
Finally, ET, Figure \ref{climatology_biases}c shows positive biases mostly over Italy, the Balkans, France, Spain, Ireland and southern Great Britain. Meanwhile, negative biases are visible mostly over northern and eastern Europe. Interestingly, high positive biases are seen following many rivers, most noticeably in the Iberian peninsula, Italy and the Balkans. This effect is attributed to how the rivers are represented in ParFlow. ParFlow does not use a river routing scheme. Overland flow water, simulated with a 2D kinematic wave equation, is flowing to local depressions and then along river corridors, as convergence zones. At the same time, these convergence zones are characterized by shallow water tables, aquifer-streamflow interaction occurs. Due to ParFlow's 3D subsurface hydrodynamics, lateral subsurface flow along topographic gradients is also directed to the river valleys. In the DETECT setup Parflow uses a hydrofacies distribution where the alluvia in the riverbeds have high permeability and porosity values assigned to mimic river bed properties. In addition, ParFlow as an integrated model treats overland flow and surface hydrology and subsurface 3D hydrodynamics in a continuum approach, which means that saturation and infiltration excess overland flow can be simulated.

Two primary factors may contribute to the cold bias of T2m and the positive bias of ET: First, the ParFlow model in TSMP1\_12\_ctrl tends to simulate a relatively shallow groundwater table across most of the European continent, except in parts of Eastern Europe. This may result from a revised spinup in combination with a revised hydrogeological parametrisation that leads to a higher water table depth in the initial conditions. The shallower water table likely enhances soil moisture availability near the surface, leading to higher root water uptake, capillary rise and eventually ET in regions such as France, the Mediterranean region, and Central Europe, thereby contributing to surface cooling and the cold bias. Second, the cold bias may also be linked to the SST dataset used as the lower boundary condition in the COSMO model. The dataset, appears to introduce cold anomalies not only over land but also over adjacent ocean areas, which may hint at a  potentially large-scale forcing issue. Both factors are currently under investigation and require further validation.

\subsection{Model inter-comparison with Reference datasets}

\subsubsection{Seasonal interannual variability (Period 2003--2010)}

\begin{figure}[htp!]
\centering
\begin{subfigure}{0.45\textwidth}
  \caption{Ebro}
  \includegraphics[width=\linewidth]{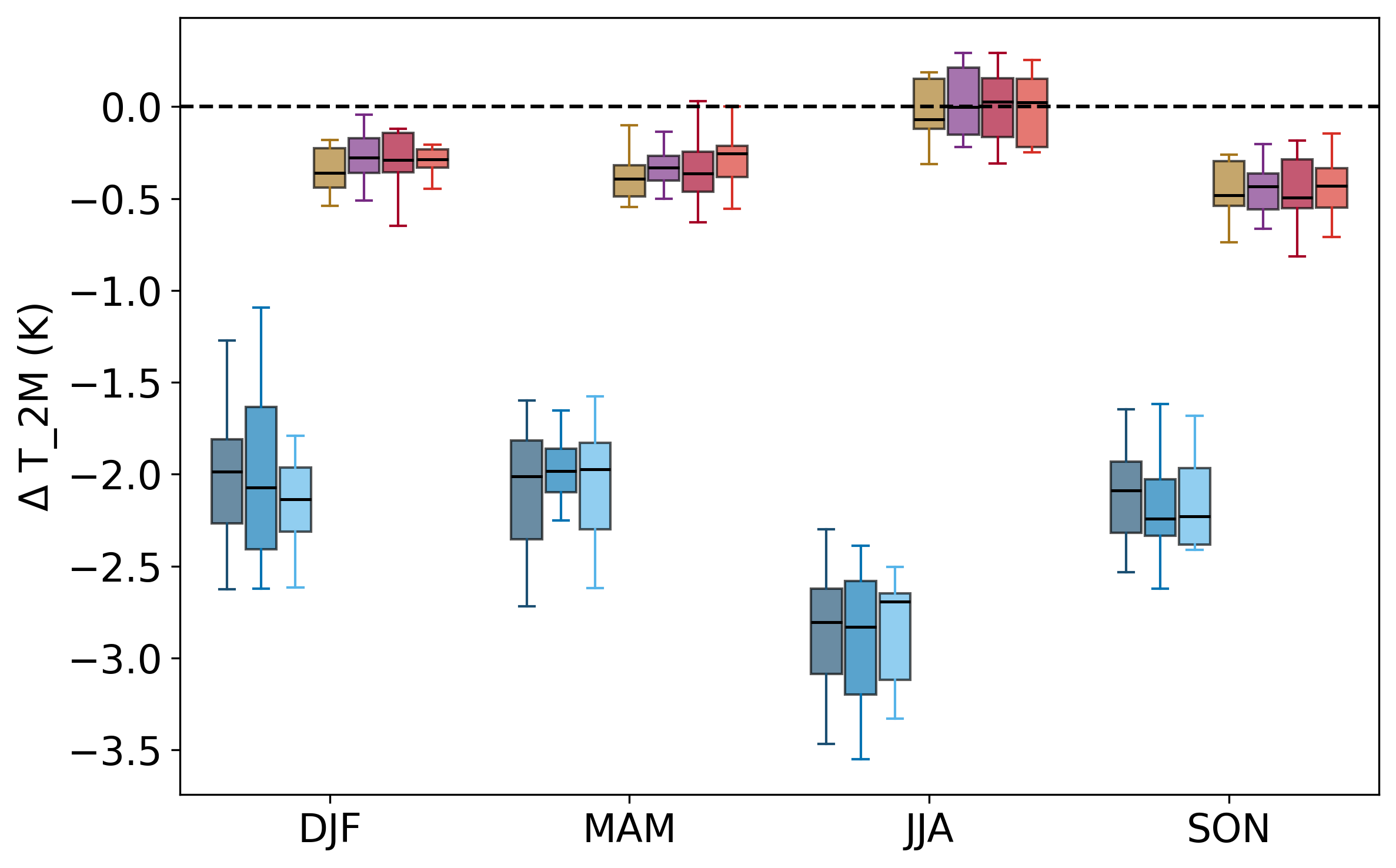}
  \label{fig:2.Ebro_t2m}
\end{subfigure}\hspace{1mm} 
\begin{subfigure}{0.45\textwidth}
  \caption{Po}
  \includegraphics[width=\linewidth]{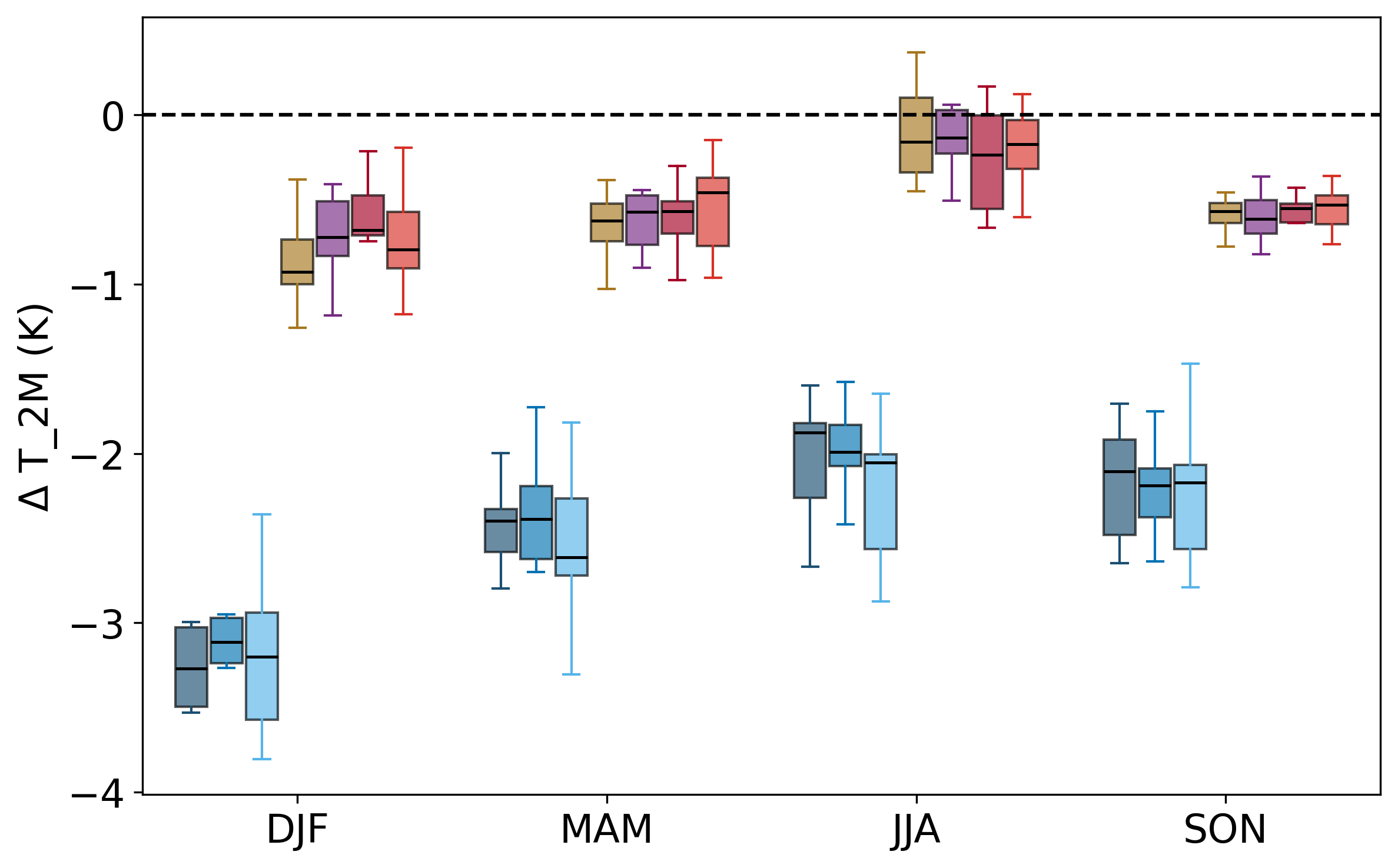}
  \label{fig:3.Po_t2m}
\end{subfigure}
\vspace{-9mm}
\medskip
\begin{subfigure}{0.45\textwidth}
  \vspace{-5mm}
  \caption{Rhine}
  \includegraphics[width=\linewidth]{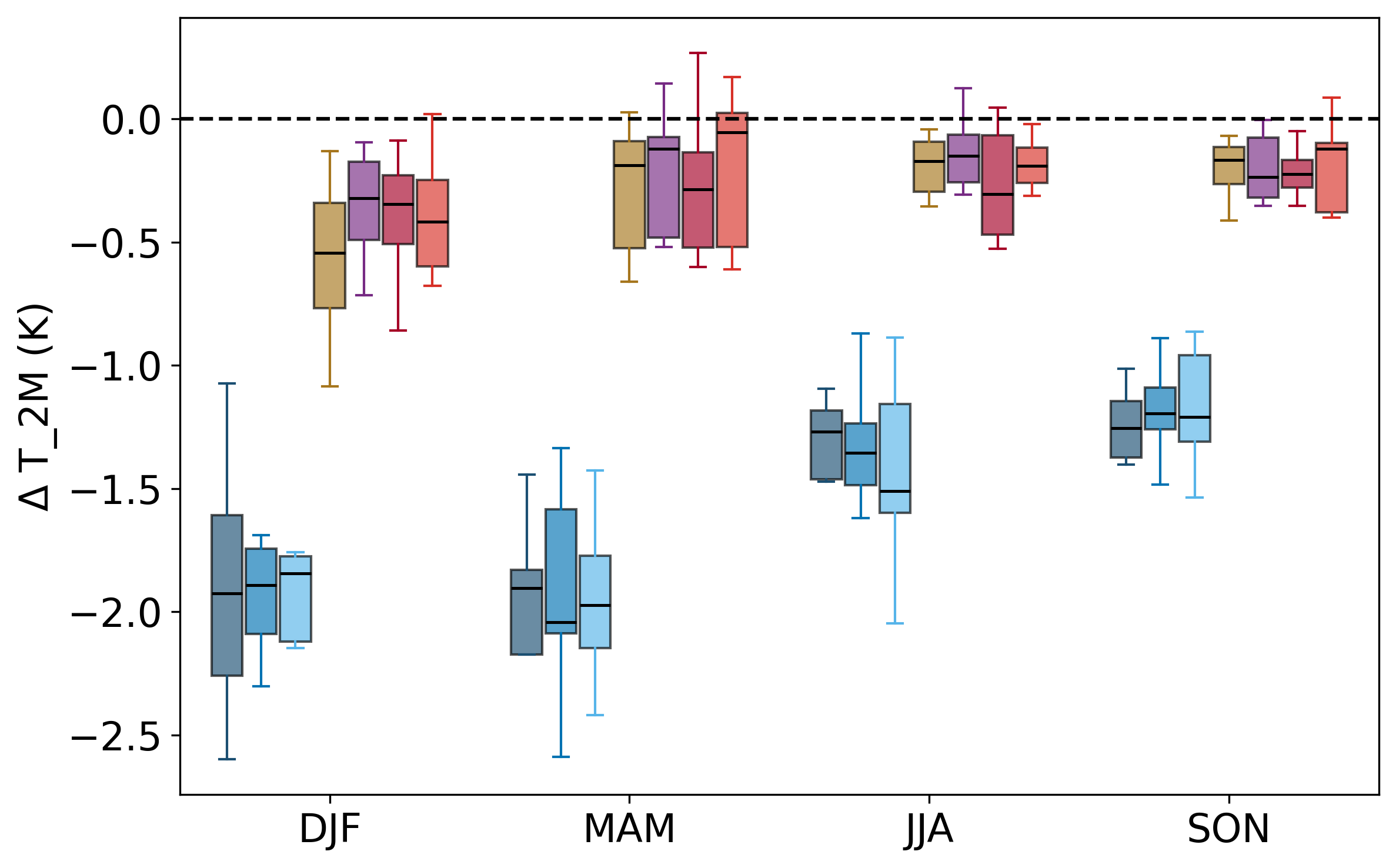}
  \label{fig:4.Rhein_t2m}
\end{subfigure}\hspace{1mm} 
\begin{subfigure}{0.45\textwidth}
  \vspace{-5mm}
  \caption{Tisa}
  \includegraphics[width=\linewidth]{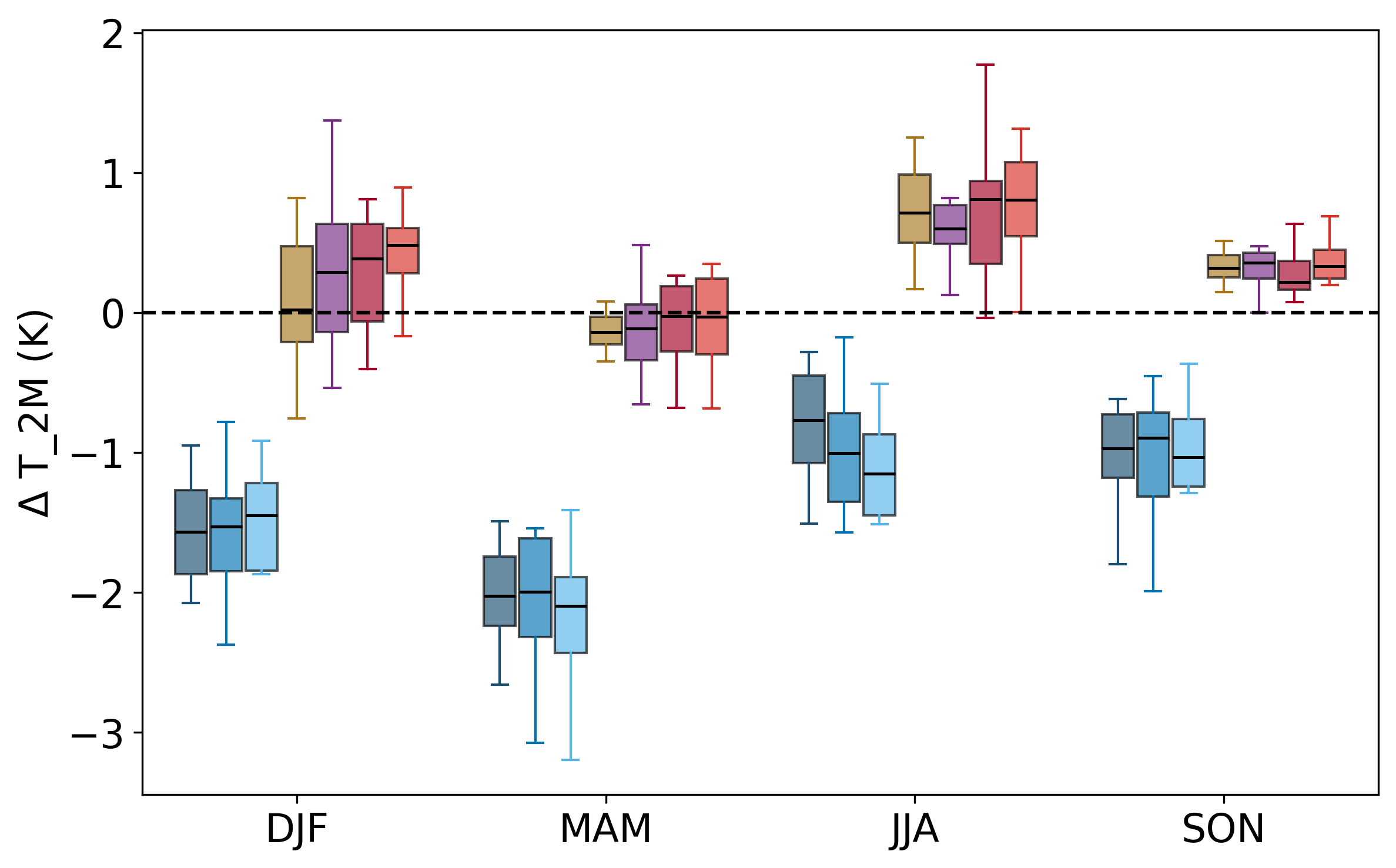}
  \label{fig:5.Tisa_t2m}
\end{subfigure}\hspace{1mm}
\begin{subfigure}{0.45\textwidth}
  \vspace{-21mm}
  \includegraphics[width=\linewidth]{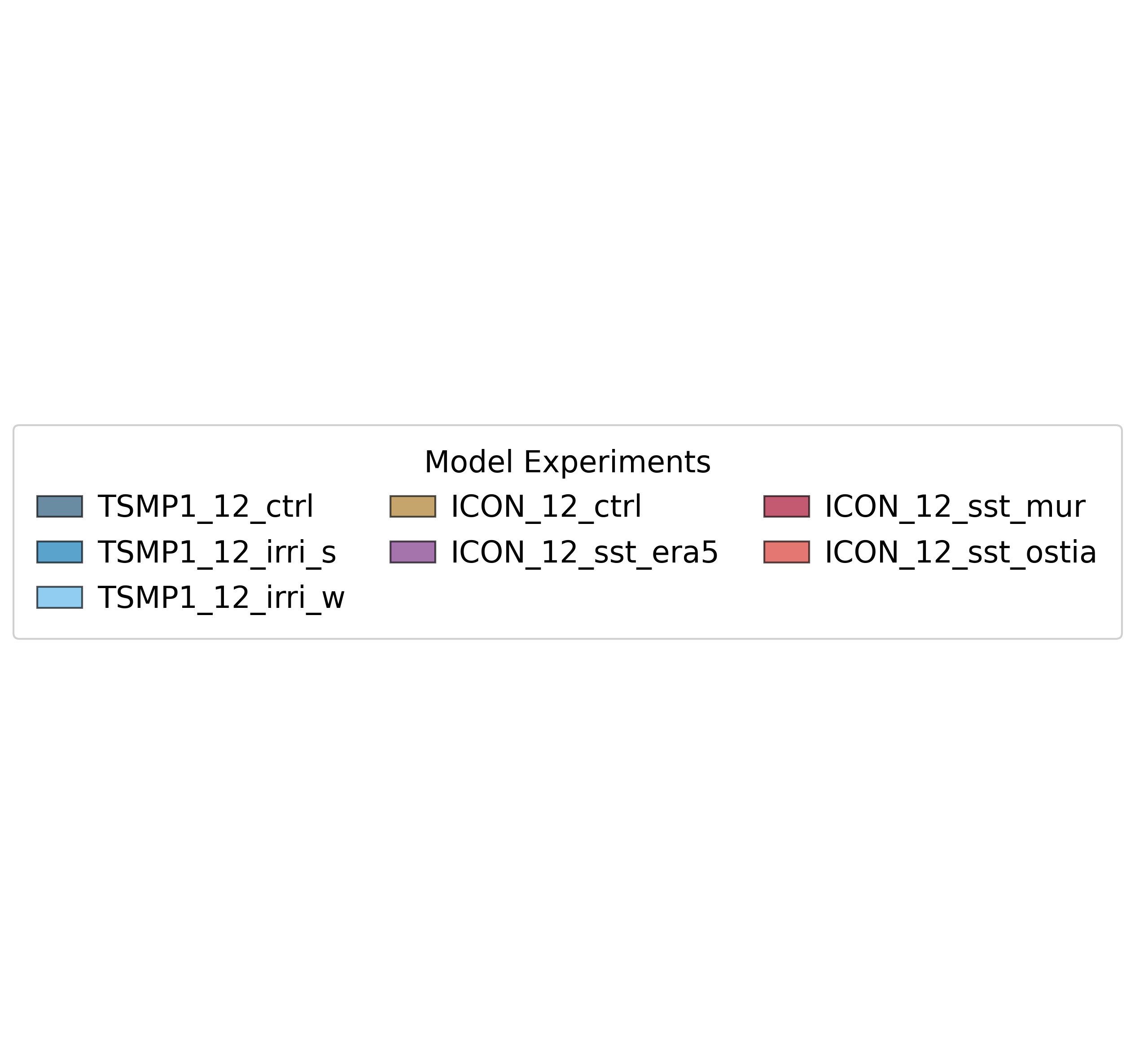}
\end{subfigure}
\vspace{-26mm}
  \caption{Boxplots of seasonal interannual T2m biases for different watersheds for the period 2003--2010.}
    \label{T2M_SEASDIFF_2003}
\end{figure}

The seasonal T2m biases averaged per basin are presented as boxplots showing the interannual data spread across each simulation (Figure~\ref{T2M_SEASDIFF_2003}). The biases for this variable exhibit larger differences between the medians of TSMP1 and ICON experiments. Overall, all experiments systematically underestimate T2m across seasons, with the exception of the Tisa basin, where only ICON shows a warm bias in most seasons. 

In nothern hemispheric winter (DJF) and spring (MAM), ICON median biases in the Ebro, Po, and Rhine basins remain close to -0.5~K, whereas TSMP1 biases reach approximately -2~K and up to -3~K in DJF for the Po basin. The bias spread, similar to other variables, is larger in TSMP1 experiments. In the Tisa basin, ICON maintains a warm bias in most of the experiments, with a maximum of 0.36~K for ICON\_12\_sst\_mur, and for MAM these are almost distributed around zero. Moreover, TSMP1 consistently shows a cold bias exceeding -1.5~K. 

In northern hemispheric summer (JJA), ICON achieves its best performance, with median biases near zero in the Ebro, Po, and Rhine basins. In contrast to this, compared to the preceding seasons, biases in the Tisa basin increase for the ICON model and reduce for the TSMP1, being 0.85~K the highest mean bias for ICON\_12\_sst\_mur and -1.97~K the lowest for TSMP1\_12\_irri\_s. In all cases, TSMP1 maintains the cold bias in all basins. 

In autumn (SON), both models show cold biases in the Ebro, Po, and Rhine basins, while ICON continues to display a warm bias in the Tisa basin. Overall, ICON biases are consistently closer to zero than those of TSMP1, indicating better performance in representing T2m. 

In this regard, our findings are consistent with \citet{Pham2021}, who reported smaller biases for ICON-CLM and COSMO-CLM in winter and autumn and larger biases in spring and summer. The persistent cold bias across all seasons in TSMP1, therefore, cannot be attributed solely to the performance of COSMO. As discussed in section \ref{tsmp_clim}, the systematic cold bias in TSMP1 is largely associated with the groundwater configuration in ParFlow and possible biases in the prescribed SST lower boundary conditions, rather than with the atmospheric component COSMO.

\begin{figure}[htp!]
\centering
\begin{subfigure}{0.45\textwidth}
  \caption{Ebro}
  \includegraphics[width=\linewidth]{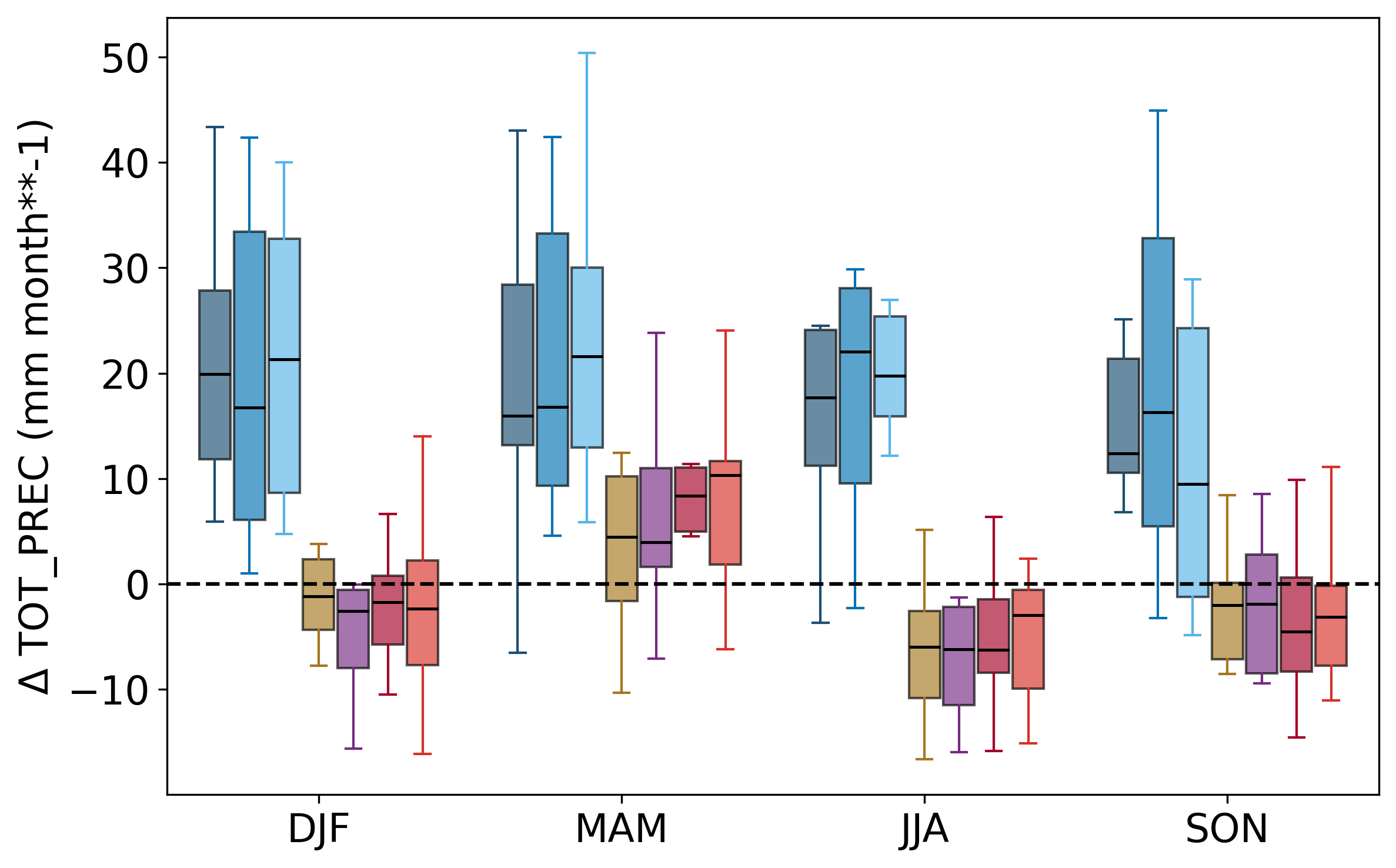}
  \label{fig:2.Ebro_pp}
\end{subfigure}\hspace{1mm} 
\begin{subfigure}{0.45\textwidth}
  \caption{Po}
  \includegraphics[width=\linewidth]{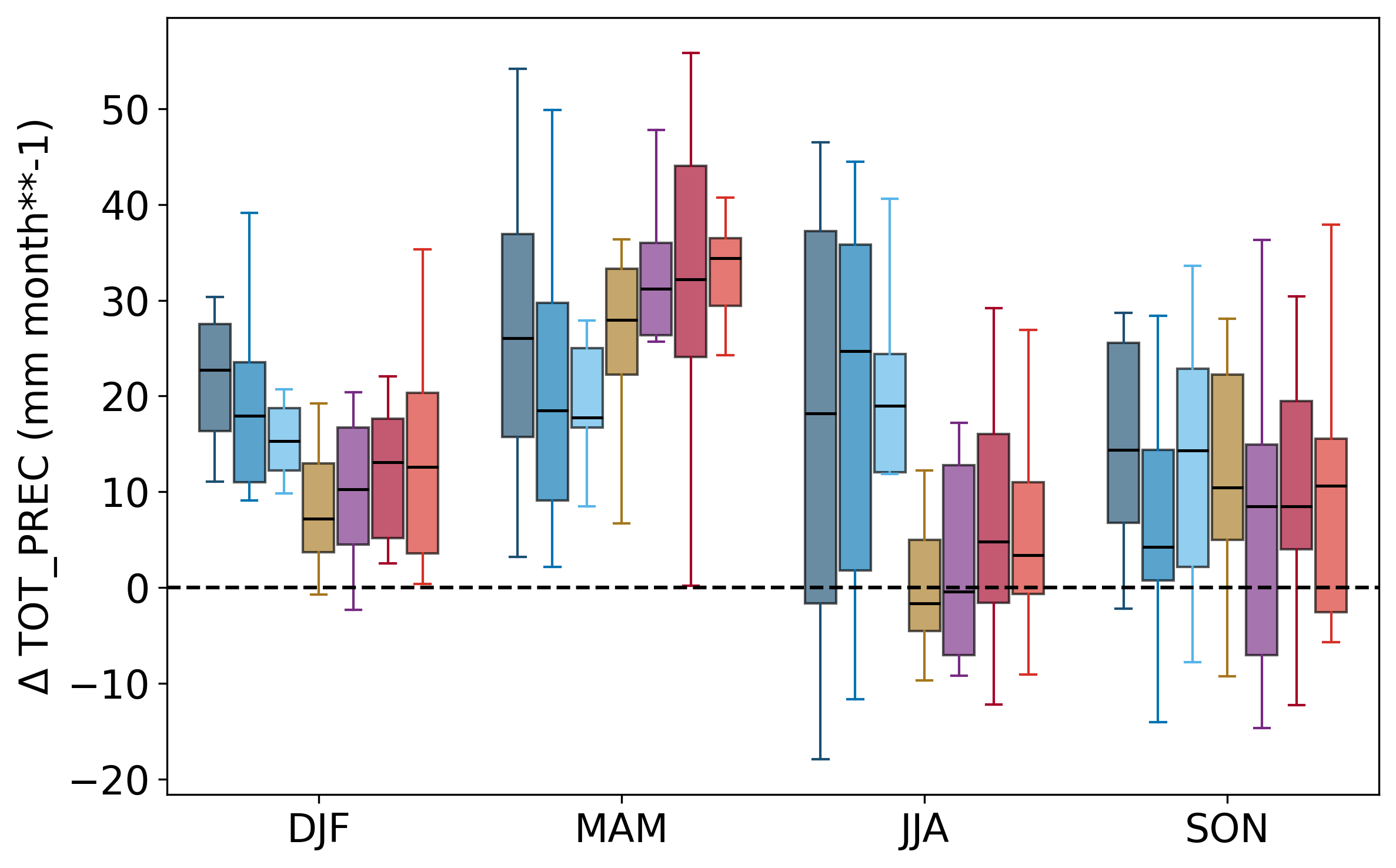}
  \label{fig:3.Po_pp}
\end{subfigure}
\vspace{-9mm}
\medskip
\begin{subfigure}{0.45\textwidth}
  \vspace{-5mm}
  \caption{Rhine}
  \includegraphics[width=\linewidth]{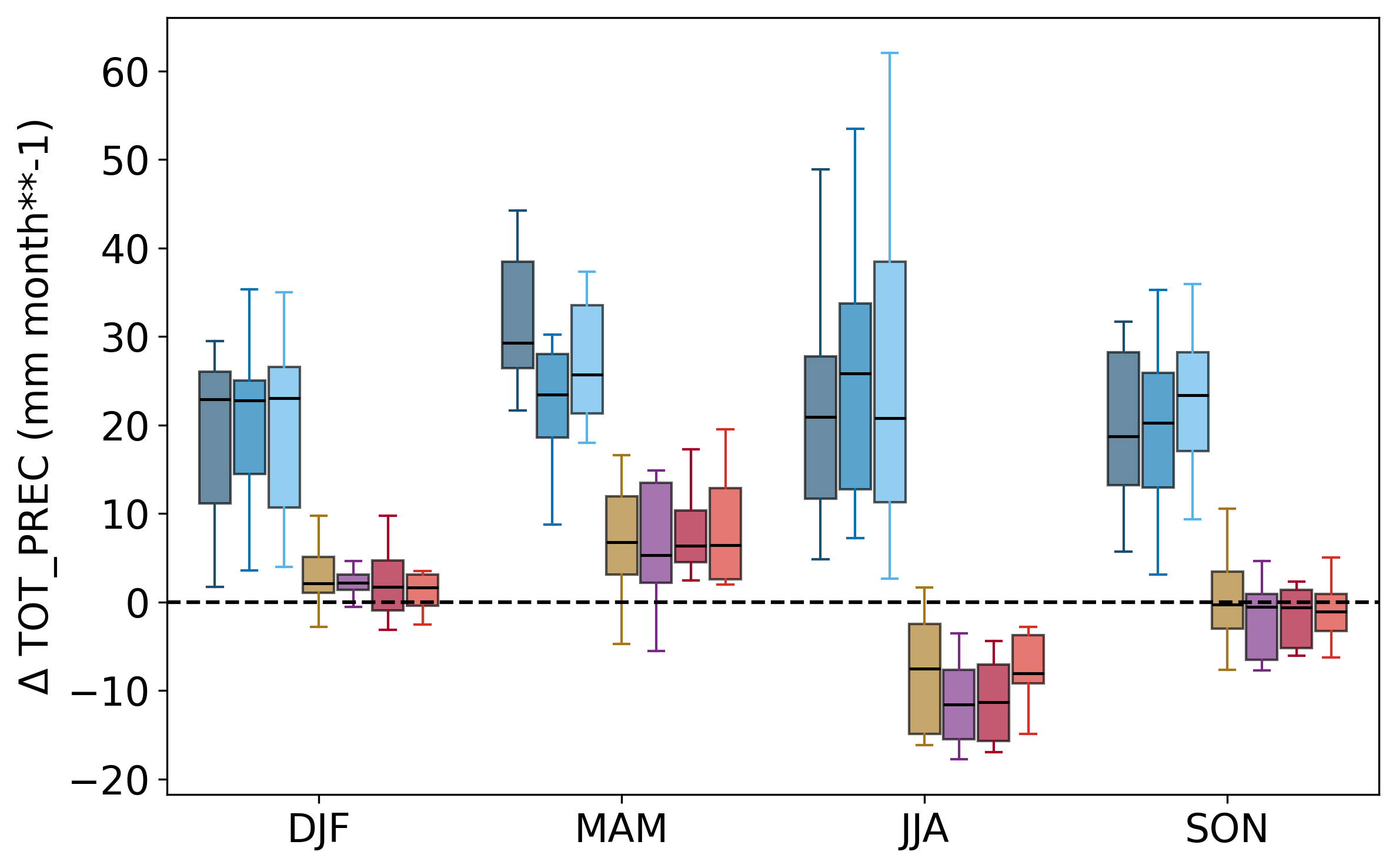}
  \label{fig:4.Rhein_pp}
\end{subfigure}\hspace{1mm} 
\begin{subfigure}{0.45\textwidth}
  \vspace{-5mm}
  \caption{Tisa}
  \includegraphics[width=\linewidth]{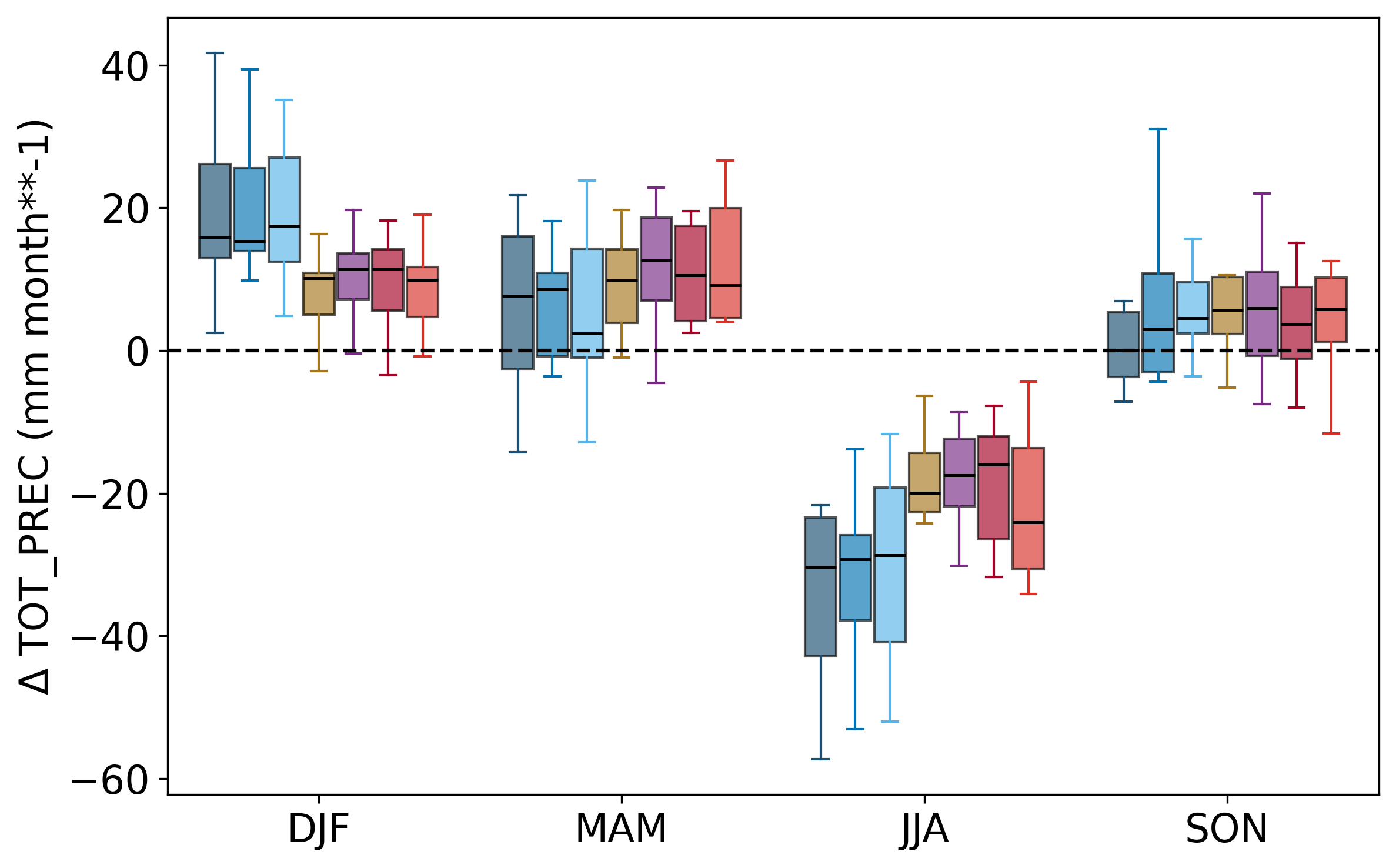}
  \label{fig:5.Tisa_pp}
\end{subfigure}\hspace{1mm}
\vspace{-9mm}
\medskip
\begin{subfigure}{0.45\textwidth}
  \vspace{-21mm}
  \includegraphics[width=\linewidth]{legend_only_2003.png}
\end{subfigure}
\vspace{-19mm}
  \caption{Boxplots of seasonal interannual precipitation biases for different watersheds for the period 2003--2010.}
    \label{TOT_PREC_SEASDIFF_2003}
\end{figure}

Compared to T2m results, seasonal precipitation biases averaged per basin also show differences between ICON and TSMP1 simulations, especially for the Ebro and Rhine river basins during all seasons (Figure \ref{TOT_PREC_SEASDIFF_2003}). 

In DJF, the model performance is similar in the Ebro and Rhine basins. Here, ICON experiments exhibit medians near zero and smaller fluctuations around this quantile. In contrast, all TSMP1 experiments overestimate precipitation, with medians around $\pm$ 20~mm~month$^{-1}$ and a larger bias spread in the Ebro basin. In the other basins, the Po and Tisa, there is a consistent wet bias, with medians between 10 and 20~mm~month$^{-1}$ in the Po basin. In the Tisa basin, ICON has a median bias lower than 10~mm~month$^{-1}$ while TSMP1's median is closer to 20~mm~month$^{-1}$. 

In MAM, most experiments exhibit positive median biases, indicating a tendency toward wet conditions. The precipitation biases of ICON and TSMP1 differ markedly in the Ebro and Rhine basins. ICON shows a wet median bias of approximately $\pm$ 10~mm~month$^{-1}$, whereas TSMP1 exhibits a larger bias of about $\pm$ 20~mm~month$^{-1}$. By contrast, differences between models and experiments are less pronounced in the Po and Tisa basins. 

In JJA, the model behavior varies across basins, but in general two groups can be distinguished again: the Ebro and Rhine basins on one side, and the Po and Tisa basins on the other. In the Ebro and Rhine basins, TSMP1 experiments maintain their characteristic wet bias, while ICON shifts to a dry bias. The spread among TSMP1 experiments is substantially larger than for ICON, and ICON’s medians biases remain much closer to zero, being the -5.10~mm~month$^{-1}$ (ICON\_12\_sst\_mur) the closest to zero in the Ebro. In the Po basin, most experiments continue to show a wide spread. However, ICON median biases remain near zero, while TSMP1 continues to overestimate precipitation. In the Tisa basin, all experiments exhibit a dry bias, with median values between $\pm$ 20 and $\pm$ 30~mm~month$^{-1}$. 

In SON, ICON experiments persist with dry biases but with lower values, whereas TSMP1 maintains the wet biases observed in summer for the Ebro and Rhine basins. In the Po basin, both the spread and median biases remain similar to those of other seasons. In the Tisa basin, most experiments display a wet bias much closer to zero than in the other seasons. Overall, ICON biases are lower in all seasons and basins, however, both models still display a wide spread of biases, which is to be expected. \citet{Pham2021} also reported large ranges of precipitation biases for both ICON-CLM and COSMO-CLM, reflecting the high spatial variability of precipitation. Although they did not identify a clear advantage of one model over the other, our results indicate that coupling COSMO-CLM with different land surface and hydrologic models amplifies the differences in simulated precipitation.

\begin{figure}[htp!]
\centering
\begin{subfigure}{0.45\textwidth}
  \caption{Ebro}
  \includegraphics[width=\linewidth]{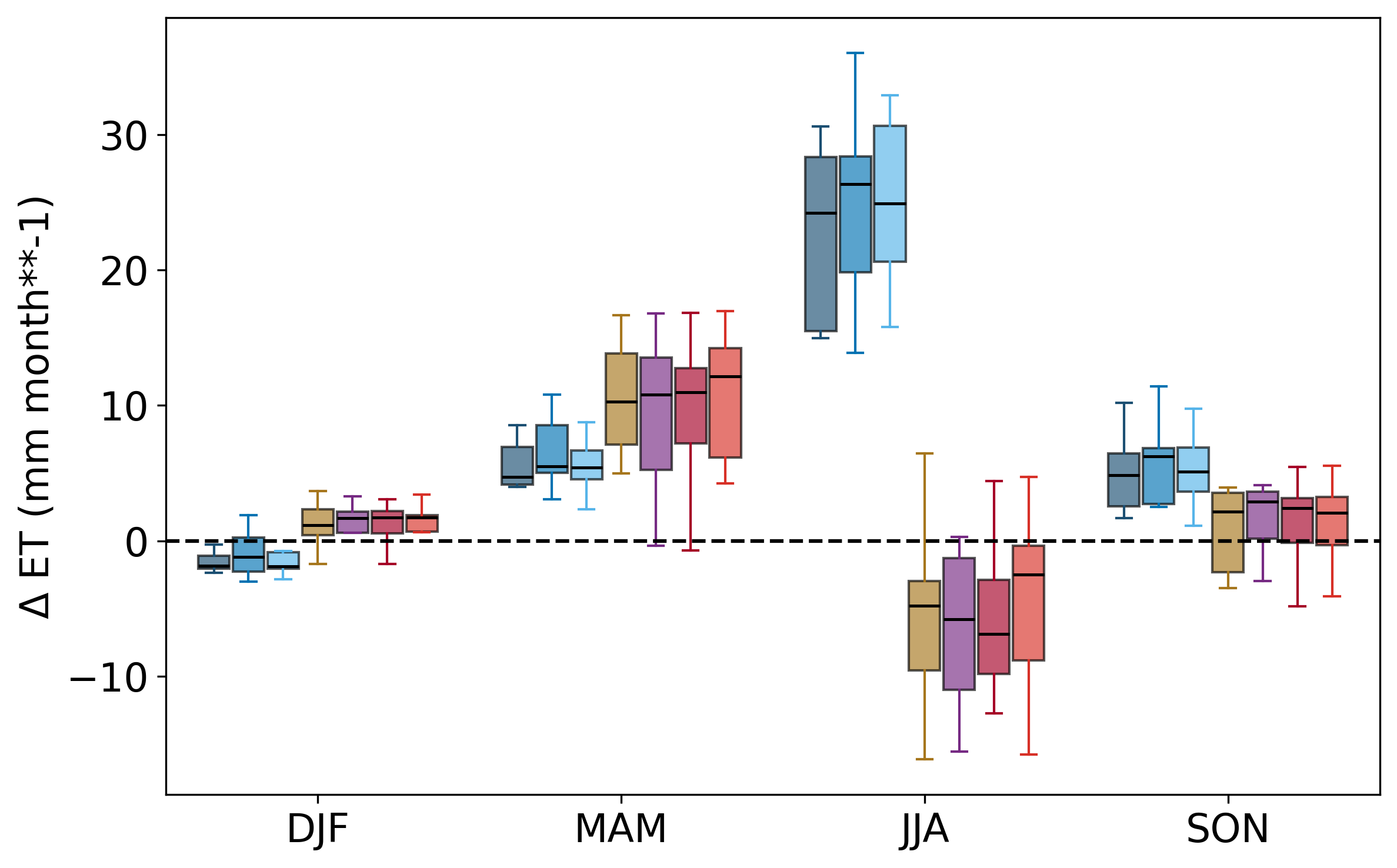}
  \label{fig:2.Ebro}
\end{subfigure}\hspace{1mm} 
\begin{subfigure}{0.45\textwidth}
  \caption{Po}
  \includegraphics[width=\linewidth]{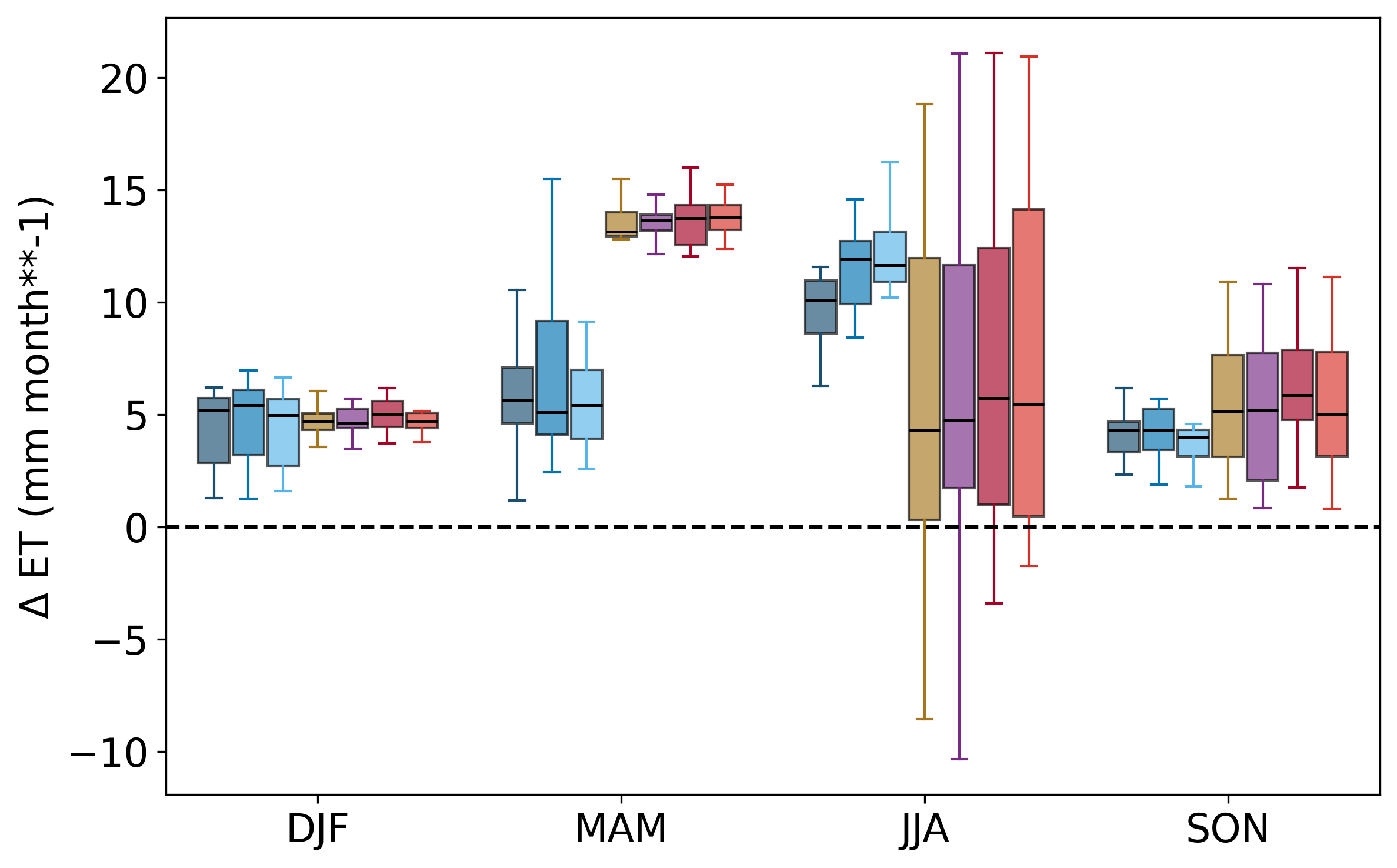}
  \label{fig:3.Po}
\end{subfigure}
\vspace{-9mm}
\medskip
\begin{subfigure}{0.45\textwidth}
  \vspace{-5mm}
  \caption{Rhine}
  \includegraphics[width=\linewidth]{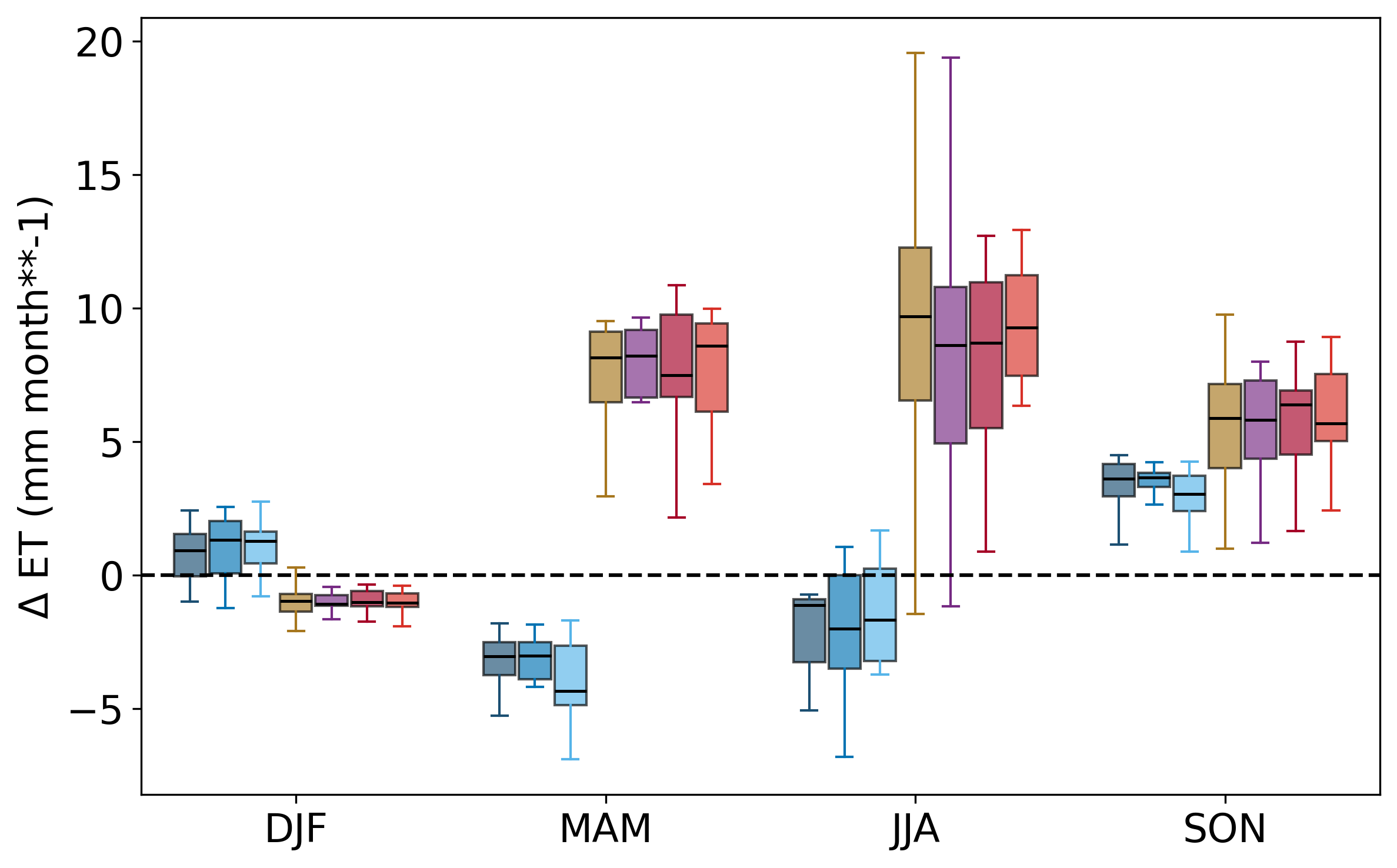}
  \label{fig:4.Rhein}
\end{subfigure}\hspace{1mm} 
\begin{subfigure}{0.45\textwidth}
  \vspace{-5mm}
  \caption{Tisa}
  \includegraphics[width=\linewidth]{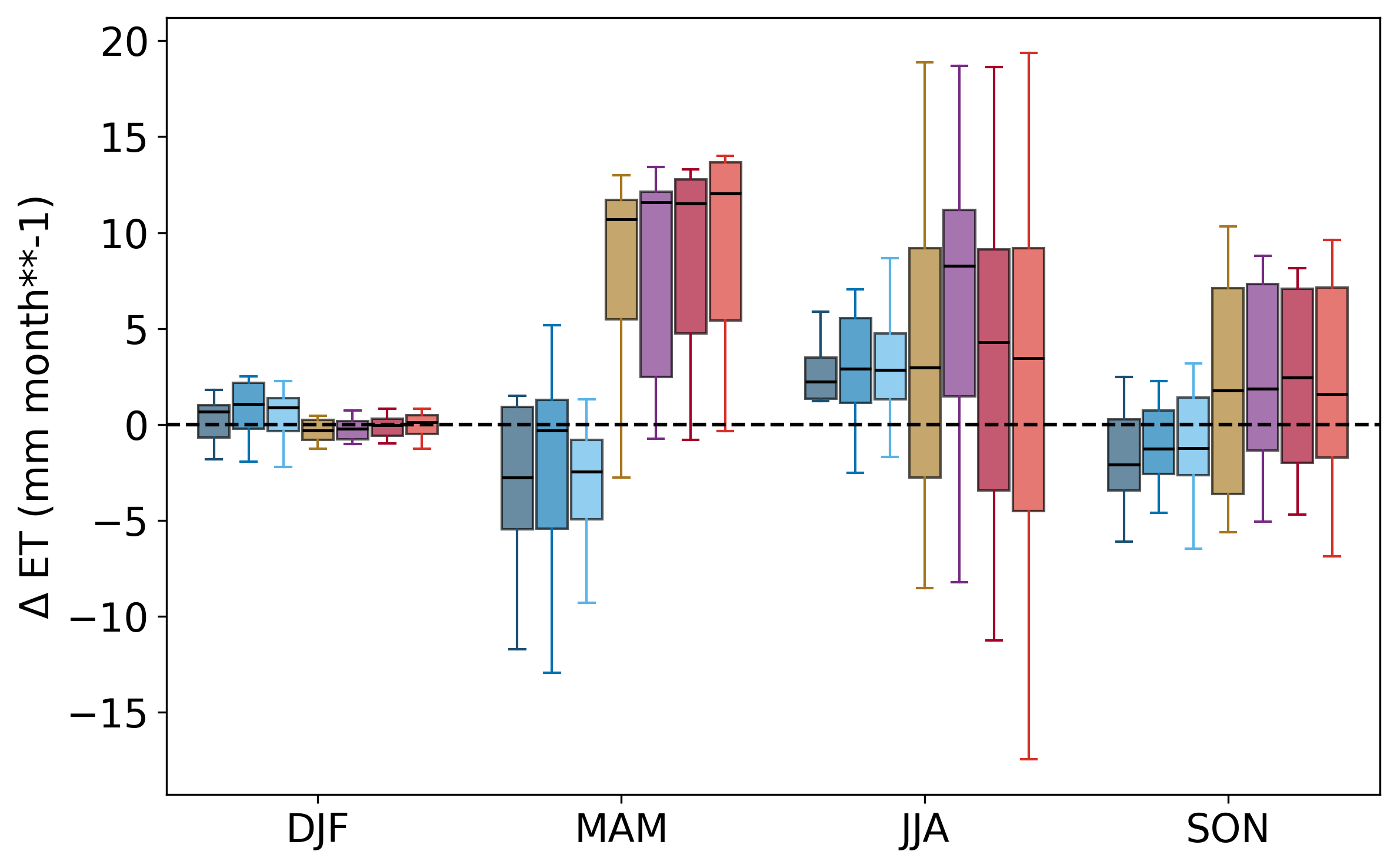}
  \label{fig:5.Tisa}
\end{subfigure}\hspace{1mm} 
\vspace{-9mm}
\medskip
\begin{subfigure}{0.45\textwidth}
  \vspace{-21mm}
  \includegraphics[width=\linewidth]{legend_only_2003.png}
\end{subfigure}
\vspace{-19mm}
  \caption{Boxplots of seasonal interannual ET biases for different watersheds for the period 2003--2010.}
    \label{ET_SEASDIFF_2003}
\end{figure}

Similar to the seasonal distribution of T2m and precipitation biases, the seasonal distribution of ET biases reveals contrasts between the experiments, especially in MAM and JJA(Figure~\ref{ET_SEASDIFF_2003}). 

In DJF, biases are generally small across all simulations and basins, with medians close to zero and limited variability. However, an exception is the Po river basin, as the medians of all experiments are close to 5~mm~month$^{-1}$, showing an overestimation. 

In MAM, most experiments overestimate ET, with ICON simulations showing particularly consistent positive biases across basins. Although the TSMP1 experiments also overestimate ET in the Ebro and the Po basins, their medians are closer to zero compared to the ICON experiments. Moreover, the overall spread of TSMP1 results in the Ebro basin is smaller than that of the ICON experiments. In contrast, TSMP1 experiments slightly underestimate ET in the Rhine and Tisa river basins, often with medians lower than 5~mm~month$^{-1}$. 

In JJA, the models diverge considerably, as both the medians and the spread among experiments vary strongly. Overall, ICON simulations systematically overestimate ET, with median biases well above zero for the Po, Rhine, and Tisa watersheds. In addition, ICON experiments exhibit a substantially larger spread in the Po, and Tisa basins, particularly in Tisa, where values range from about -15 to 20~mm~month$^{-1}$. In contrast, the spread of the bias in the TSMP1 experiments is much less in these basins. In the Ebro basin, ICON experiments perform better than TSMP1 as their median biases underestimate ET but remain closer to zero, whereas TSMP1 consistently overestimates ET, often exceeding 20~mm~month$^{-1}$. This overestimation may be linked to the hydrologic representation in TSMP1. \citet{Naz2022} conducted ParFlow–CLM simulations over the CORDEX European domain at 3 km resolution and reported an ET overestimation in dry regions such as the Iberian Peninsula. Similarly, \citet{Poshyvailo2024} showed that TSMP simulations are particularly sensitive to groundwater representation in the Iberian Peninsula, highlighting this region as susceptible to such biases. 

SON shows relatively moderate biases in most models and basins, with distributions centered close to zero. However, simulations in the Po and Rhine watersheds exhibit a larger positive bias, with medians around 5~mm~month$^{-1}$. The TSMP1 results are consistent with the findings of \citet{Naz2022}, who reported smaller discrepancies between ParFlow-CLM-simulated evapotranspiration and FLUXNET observations during winter, spring, and autumn, but larger differences in summer across most stations. They further suggested that the positive ET bias in summer could be caused by enhanced simulated water availability in ParFlow-CLM, a feature we also observed, particularly in the Ebro and Po basins. Overall, the boxplots highlight a strong seasonality in model performance, with ICON mostly favoring overestimations across different experiments and water basins, while TSMP1 shows variable behavior that depends on the basin and season.

\subsubsection{Yearly interannual variability (Period 2003--2010)}

To provide a concise overview of model performance, Figures \ref{T2m_SEASDIFF_2003_mean}, \ref{PP_SEASDIFF_2003_mean} and \ref{ET_SEASDIFF_2003_mean} present both the seasonal biases and the yearly Mean Absolute Error (MAE) for T2m, precipitation, and ET across all analyzed basins for the current period. The results indicate that within the TSMP1 experiment group, the control simulation (TSMP1\_12\_ctrl) exhibits the lowest MAE for all three variables in the Ebro and Tisa basins, demonstrating the best overall mean agreement with the reference datasets. 

Regarding the interannual variability in the Ebro basin, the Taylor diagram indicates that the setup TSMP1\_12\_irri\_w most accurately reproduces the observed interannual precipitation variability, whereas the control simulation continues to perform best for ET and T2m (Figure \ref{Taylor_Ebro}). 
In the Tisa basin, the control simulation attains again the most favorable combination of RMS, variance and correlation across all variables, confirming its overall superior performance (Figure \ref{Taylor_Tisa}).
In the Po basin, TSMP1\_12\_ctrl maintains the lowest MAE for ET, whereas the TSMP1\_12\_irri\_s experiment performs best for precipitation and T2m. This improved performance in the irrigation experiment highlights the importance of representing irrigation processes in this basin, which contains one of the largest irrigated areas in Europe. 
Similarly, in the Rhine basin, the TSMP1 simulations that include irrigation achieve the best performance. Specifically, TSMP1\_12\_irri\_w produces the lowest MAE for ET, while TSMP1\_12\_irri\_s performs best for precipitation and T2m. Interannual Taylor diagrams further support these findings. In the Po basin, TSMP1\_12\_irri\_w outperforms other configurations for ET and precipitation, whereas TSMP1\_12\_irri\_s provides the best representation of T2m (Figure \ref{Taylor_Po}). In the Rhine basin, TSMP1\_12\_irri\_w achieves the best agreement for ET and T2m, while the control simulation (TSMP1\_12\_ctrl) performs best for precipitation (Figure \ref{Taylor_Rhein}). Although irrigation is less extensive in the Rhine basin compared to the Po, these results suggest that even median to small irrigated areas can influence the surface–atmosphere exchanges. 

The evaluation of ICON\_12 simulations across the analyzed basins reveals different sensitivities to SST boundary conditions. In the Ebro basin, the control simulation (ICON\_12\_ctrl) achieves the lowest MAE for ET, while ICON\_12\_sst\_mur and ICON\_12\_sst\_era5/ICON\_12\_sst\_ostia perform best for precipitation and T2m respectively. This pattern indicates that SST boundary forcing might play a relevant role in modulating precipitation and temperature fields, emphasizing the importance of using observation-based SST datasets as lower boundary in free simulations. Similarly, in the Po basin, ICON\_12\_sst\_era5 obtains the lowest MAE for ET and T2m, whereas the ICON\_12\_ctrl simulation performs best for precipitation. Although ICON\_12\_ctrl and ICON\_12\_sst\_era5 best capture the mean state biases in the Ebro and Po basins for different variables, the interannual variability illustrated in the Taylor diagrams indicates that no single experiment consistently outperforms the others. For instance, ICON\_12\_sst\_mur shows the highest agreement for ET in the Ebro basin (Figure \ref{Taylor_Ebro}), while ICON\_12\_sst\_era5 provides the best representation of precipitation. In the Po basin, although ICON\_12\_sst\_mur obtains the lowest RMS for ET, the correlation is negative (Figure \ref{Taylor_Po}). In the Rhine basin, all SST experiments outperform the control simulation, with ICON\_12\_sst\_era5 achieving the lowest MAE for ET and ICON\_12\_sst\_ostia for precipitation and T2m. This suggests that, in a basin characterized by strong synoptic influence, the representation of SST conditions significantly influences the simulated surface fluxes and temperature. In the Tisa basin, results are partly consistent with those of the Ebro and Po basins since the ICON\_12\_ctrl simulation shows the best performance for ET and T2m, while ICON\_12\_sst\_mur exhibits the lowest MAE for precipitation. Taylor diagram analysis indicates that in the Rhine, no single configuration consistently outperforms the others across all variables (Figure \ref{Taylor_Rhein}). However, in the Tisa basin  ICON\_12\_sst\_mur exhibits the highest skill for precipitation and T2m, while the control run best reproduces ET, matching with the MAE results. The consistent better representation of precipitation by ICON\_12\_sst\_mur in this basin might be related to the high resolution of MURSST dataset (1~km), however, this hypothesis still needs some testing. The outcomes indicate that, while SST boundary conditions affect atmospheric variables to varying degrees, their influence is region-specific. 

\subsubsection{Seasonal interannual variability (Period 2011--2020)}

\begin{figure}[htp!]
\centering
\begin{subfigure}{0.45\textwidth}
  \caption{Ebro}
  \includegraphics[width=\linewidth]{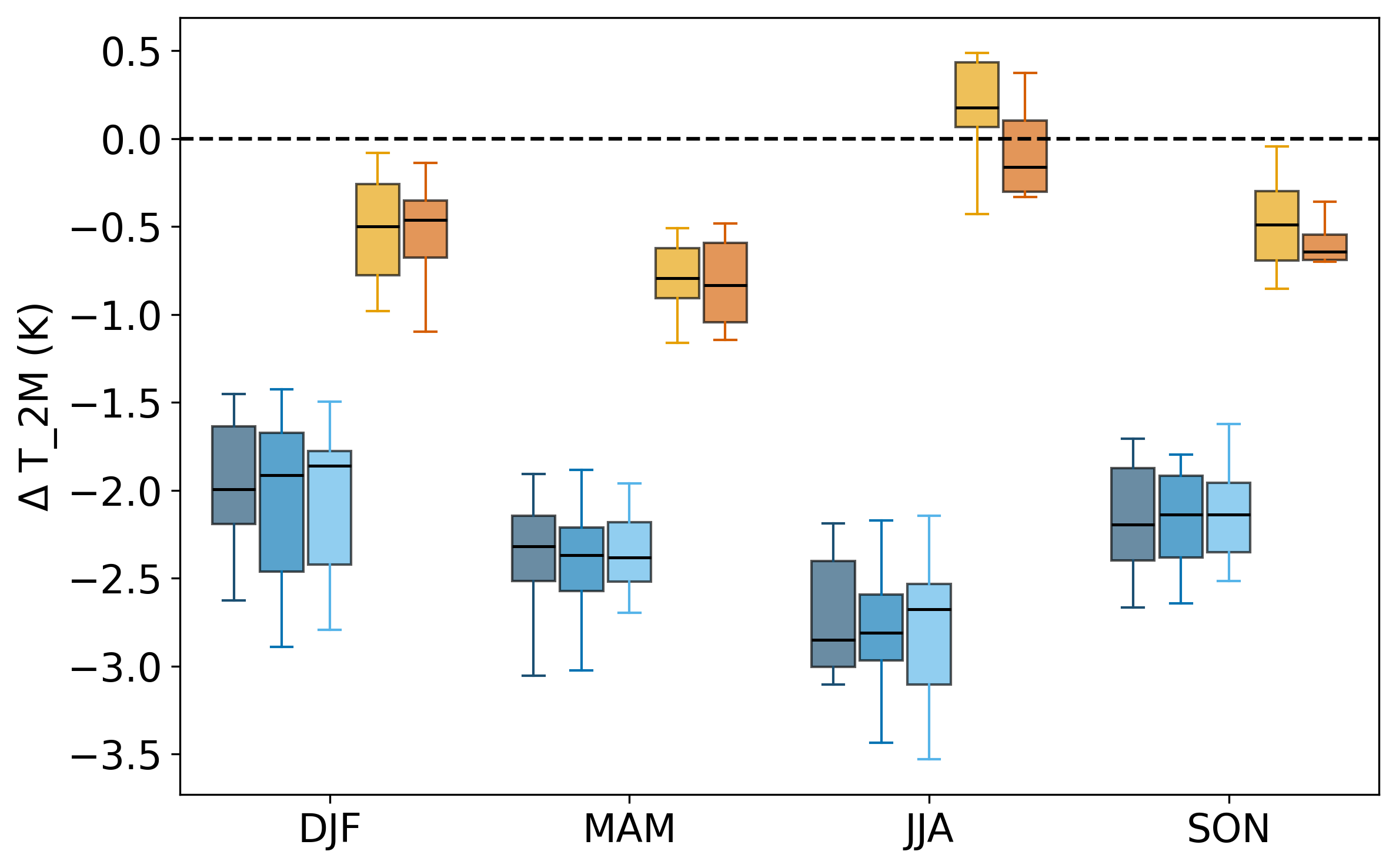}
  \label{fig:2.Ebro_2011_t2m}
\end{subfigure}\hspace{1mm} 
\begin{subfigure}{0.45\textwidth}
  \caption{Po}
  \includegraphics[width=\linewidth]{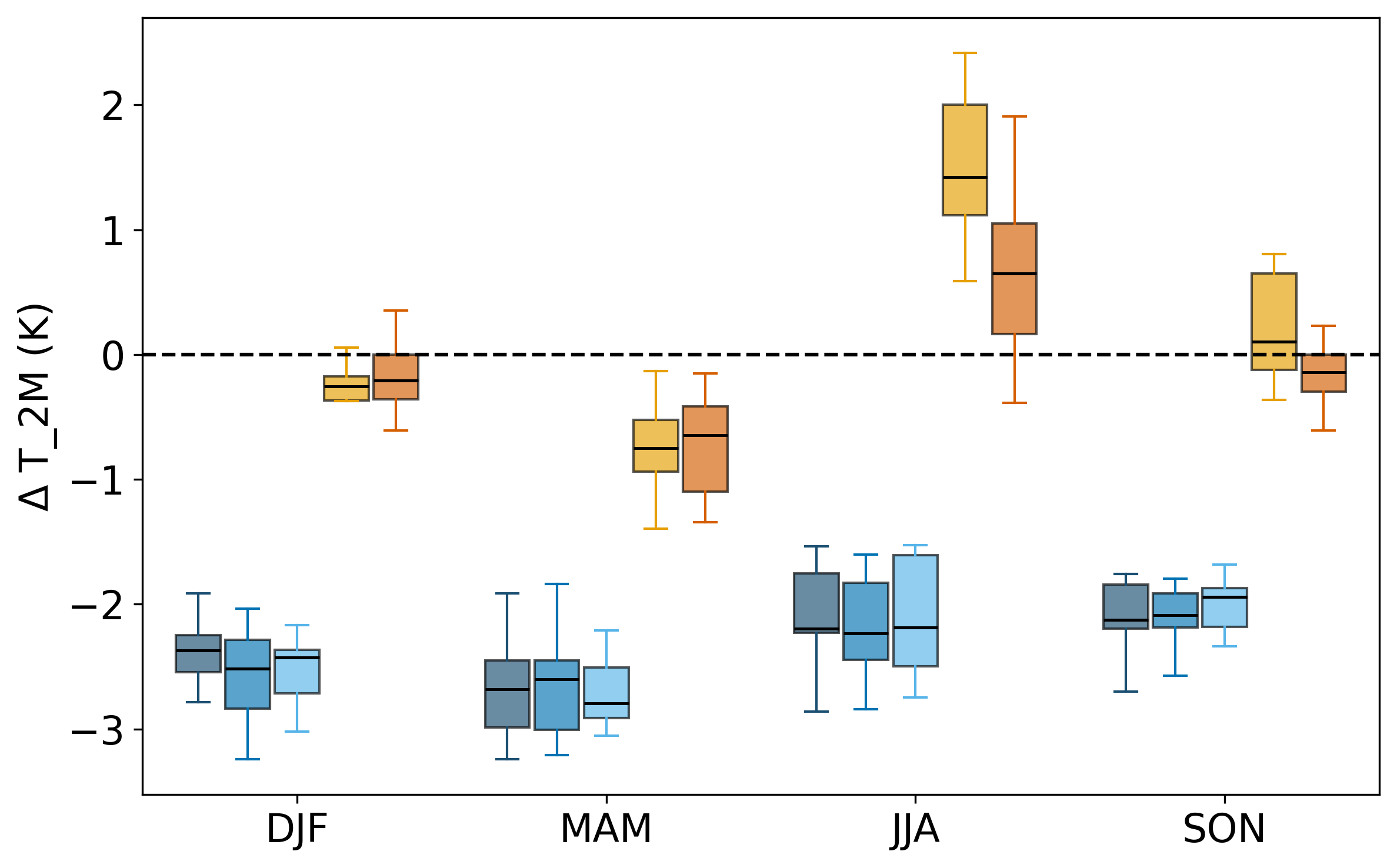}
  \label{fig:3.Po_2011_t2m}
\end{subfigure}
\vspace{-9mm}
\medskip
\begin{subfigure}{0.45\textwidth}
  \vspace{-5mm}
  \caption{Rhine}
  \includegraphics[width=\linewidth]{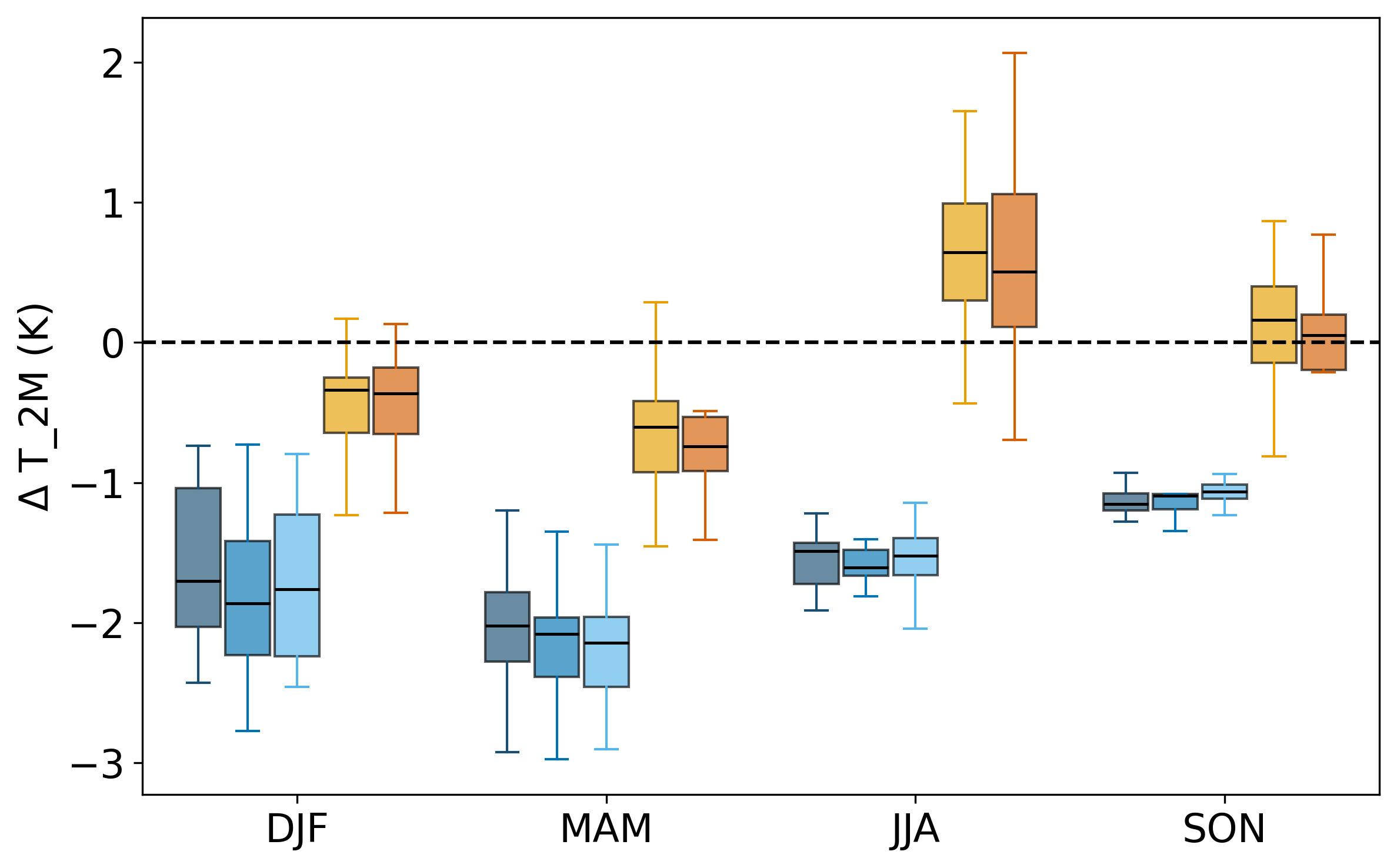}
  \label{fig:4.Rhein_2011_t2m}
\end{subfigure}\hspace{1mm} 
\begin{subfigure}{0.45\textwidth}
  \vspace{-5mm}
  \caption{Tisa}
  \includegraphics[width=\linewidth]{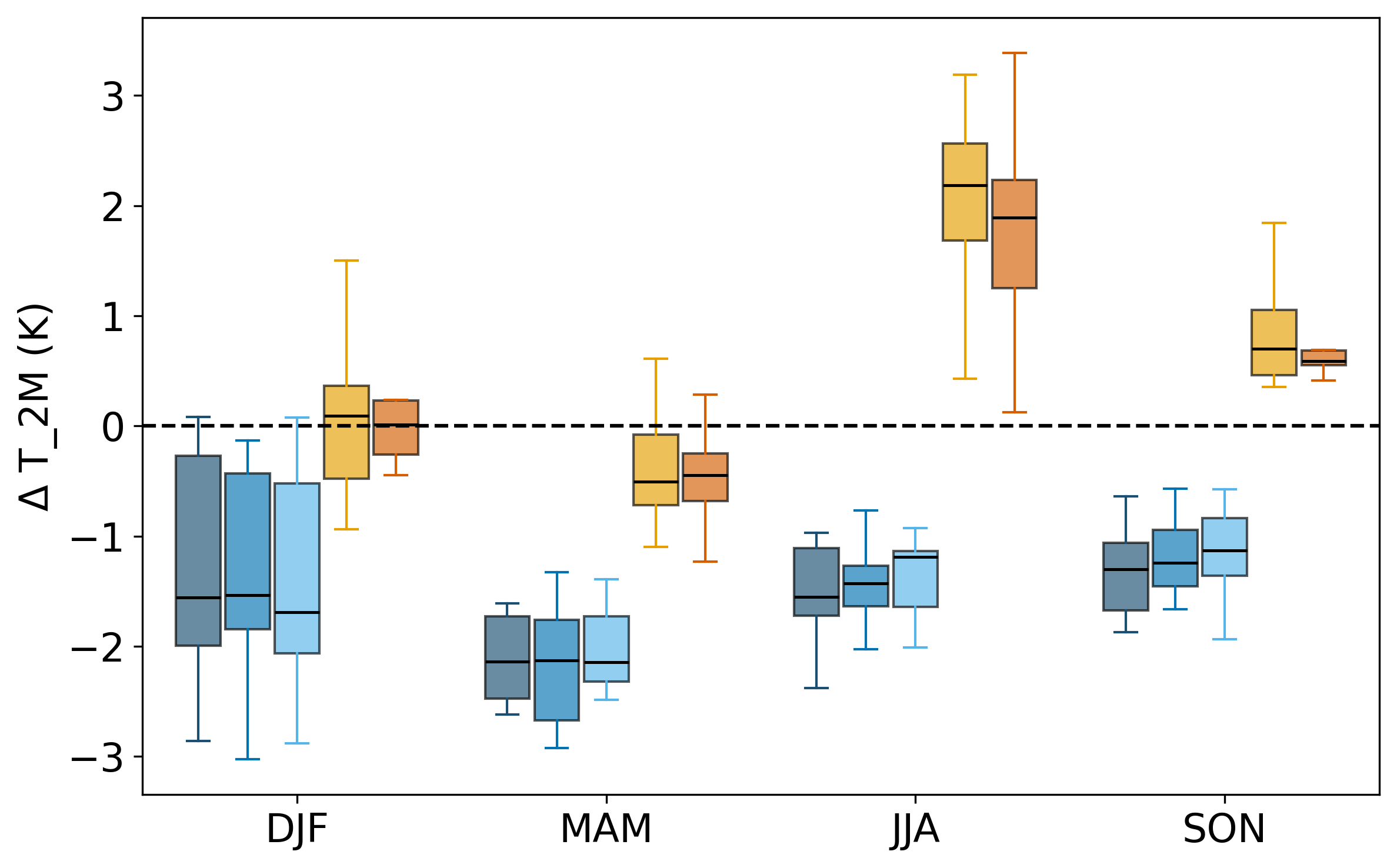}
  \label{fig:5.Tisa_2011_t2m}
\end{subfigure}\hspace{1mm} 
\begin{subfigure}{0.45\textwidth}
  \vspace{-26mm}
  \includegraphics[width=\linewidth]{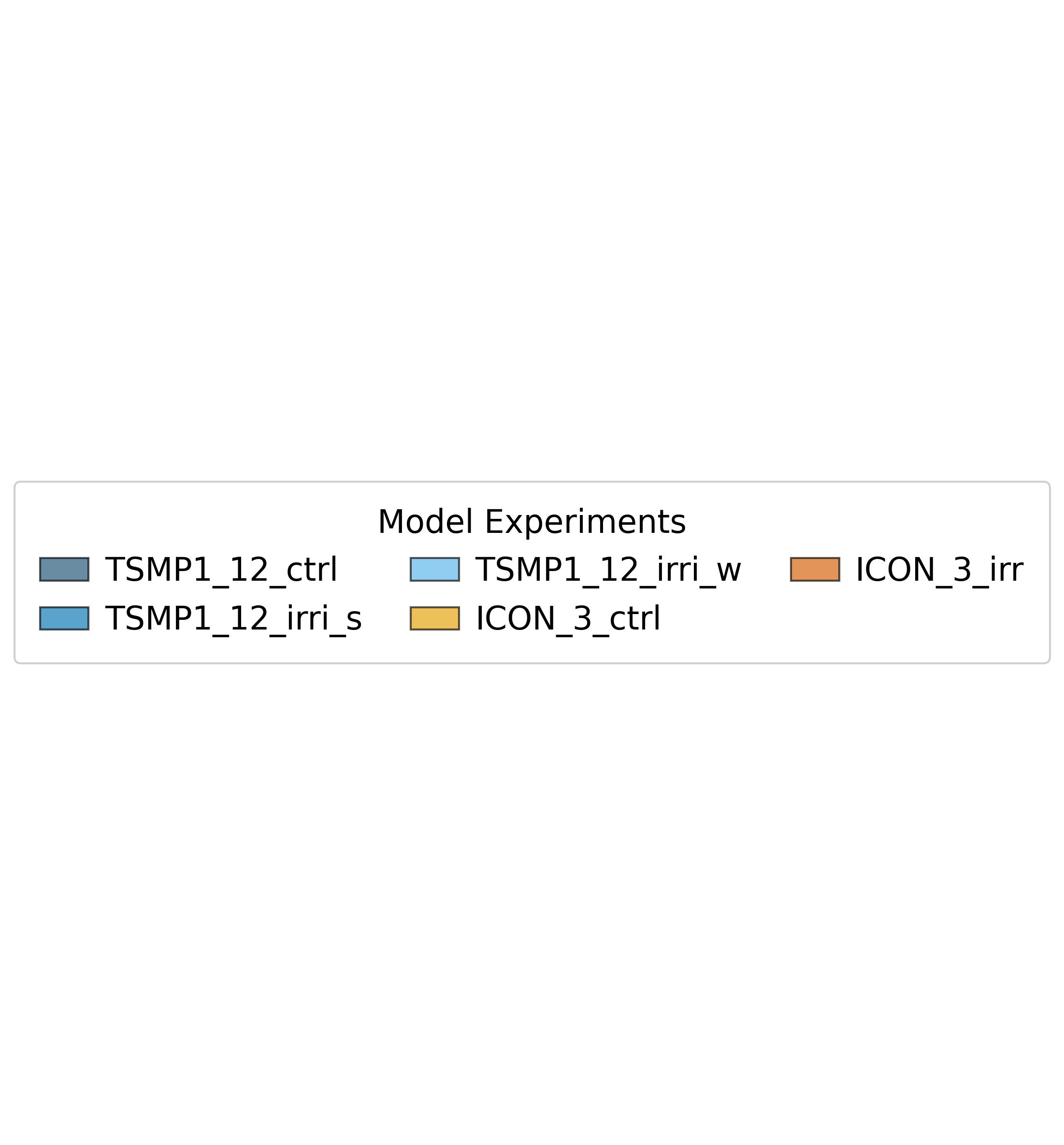}
\end{subfigure}
\vspace{-31mm}
  \caption{Boxplots of seasonal interannual T2m biases for different watersheds for the period 2011--2020.}
    \label{T2M_SEASDIFF_2011}
\end{figure}

For the period 2011--2020, we also compare ICON and TSMP1 experiments, however, ICON simulations are only available at 3 km resolution. Similar to the period 2003-2010, the bias distribution for T2m shows pronounced differences between TSMP1 and ICON experiments for all basins and seasons (Figure \ref{T2M_SEASDIFF_2011}). 
In DJF, both models exhibit a cold bias in the Ebro, Po, and Rhine basins. On average, ICON experiments show relatively small biases, with a median cold bias of less than 1 K, whereas TSMP1 simulations are systematically colder, with 50\% of the data below -1.5 K. In the Tisa basin, ICON cold bias is distributed around zero, while TSMP1 retains a median cold bias below -1 K. 
In MAM, both models maintain a cold bias as well, although ICON medians remain consistently closer to zero than TSMP1. 
Similarly, JJA is the season when the differences between models increase. In the Ebro basin, ICON simulations perform notably better, specifically ICON\_3\_irr experiment which has distributions centered near zero, while TSMP1 biases are grouped around –2.5 K. The data spread is comparable for both models. For the Po, Rhine, and Tisa basins, ICON exhibits a warm bias in about 50\% of its data, with median values of +1.5 K, +0.5 K, and +2 K, respectively. \citet{Iriza2023} also found that ICON-LAM and COSMO-LAM at 2.8 km resolution tend to overestimate summer forecasted values in Romania compared to observations, but only during nighttime. While our analysis focused on monthly means, this discrepancy may be related. In contrast, TSMP1 continues to show cold biases with medians around –2 K, –1.5 K, and –1.5 K, and generally smaller spreads than ICON. In SON, ICON also outperforms TSMP1, as its bias distribution is centered around zero in the Po and Rhine, while in the Ebro the bias is centered around -0.5 K. The Tisa basin remains an exception, where ICON continues to show a warm bias of about +0.5 K. Here, TSMP1 biases are consistently colder with a bias spread comparable to ICON. In general, ICON experiments show smaller and more balanced seasonal biases, while TSMP1 consistently maintains a stronger cold bias across all basins and seasons. These results are consistent with \citet{Rieger2022}, who found that ICON-LAM outperforms COSMO-LAM in representing surface variables such as T2m and temperature profiles across different regional configurations (e.g., ICON-IT vs. COSMO-IT, ICON-RO vs. COSMO-RO).

While the seasonal bias analysis (2011–2020) excludes the reanalysis (ICON\_3\_rea) dataset due to its shorter simulation period (2018–2020), insights from the corresponding time series analysis (Figure \ref{Time_T_2M_bias_2018} and Table \ref{table_P_metrics_2018}, Appendix) provide additional context for ICON simulations. ICON\_3\_rea results show consistently lower RMSE values than ICON\_3 and TSMP1 experiments in the Po, Rhine, and Tisa basins, further supporting its better short-term performance in reproducing T2m patterns. Therefore, assimilating T2m and relative humidity through the DA routine allows to improve these scores.

\begin{figure}[htp!]
\centering
\begin{subfigure}{0.45\textwidth}
  \caption{Ebro}
  \includegraphics[width=\linewidth]{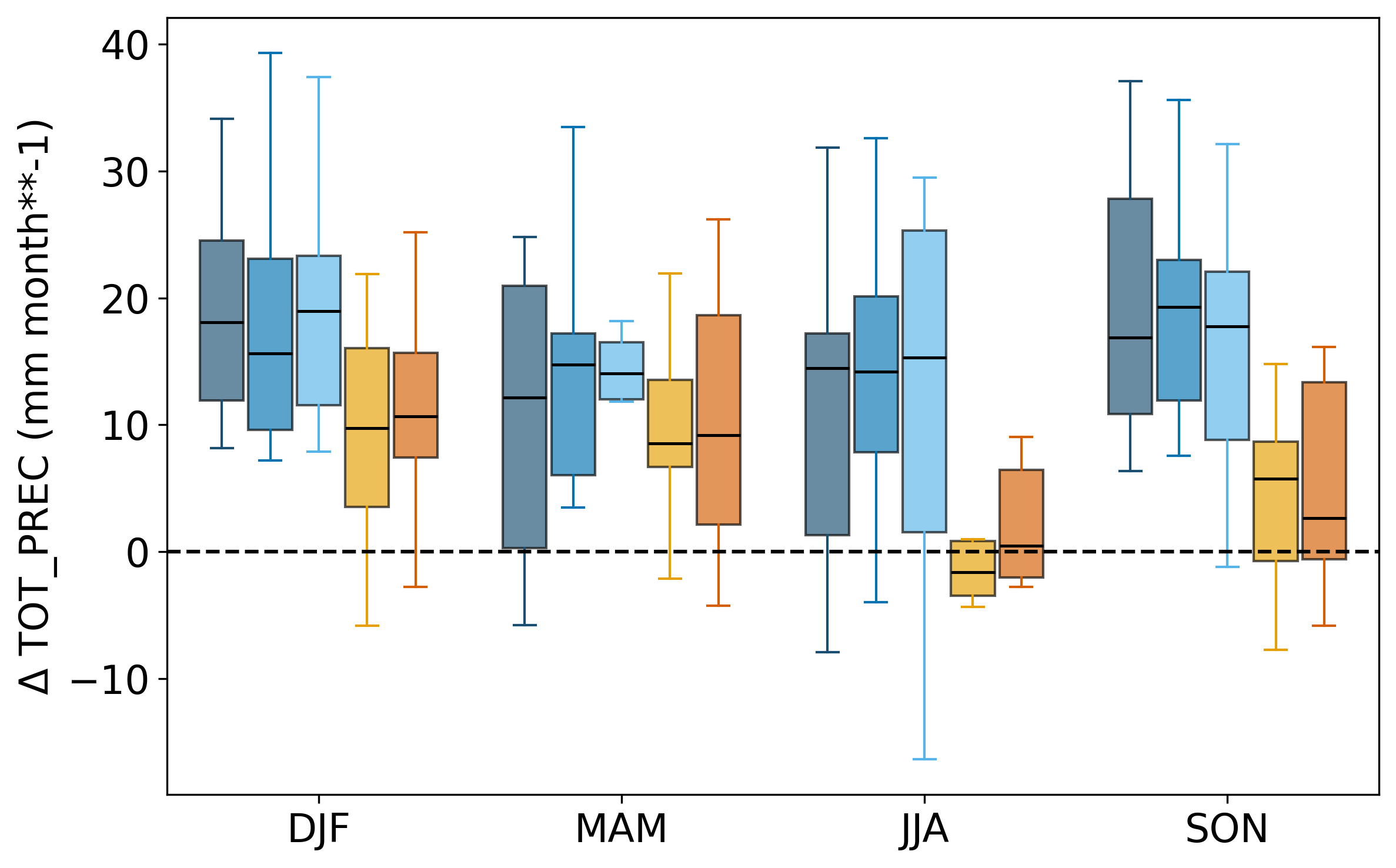}
  \label{fig:2.Ebro_2011_pp}
\end{subfigure}\hspace{1mm} 
\begin{subfigure}{0.45\textwidth}
  \caption{Po}
  \includegraphics[width=\linewidth]{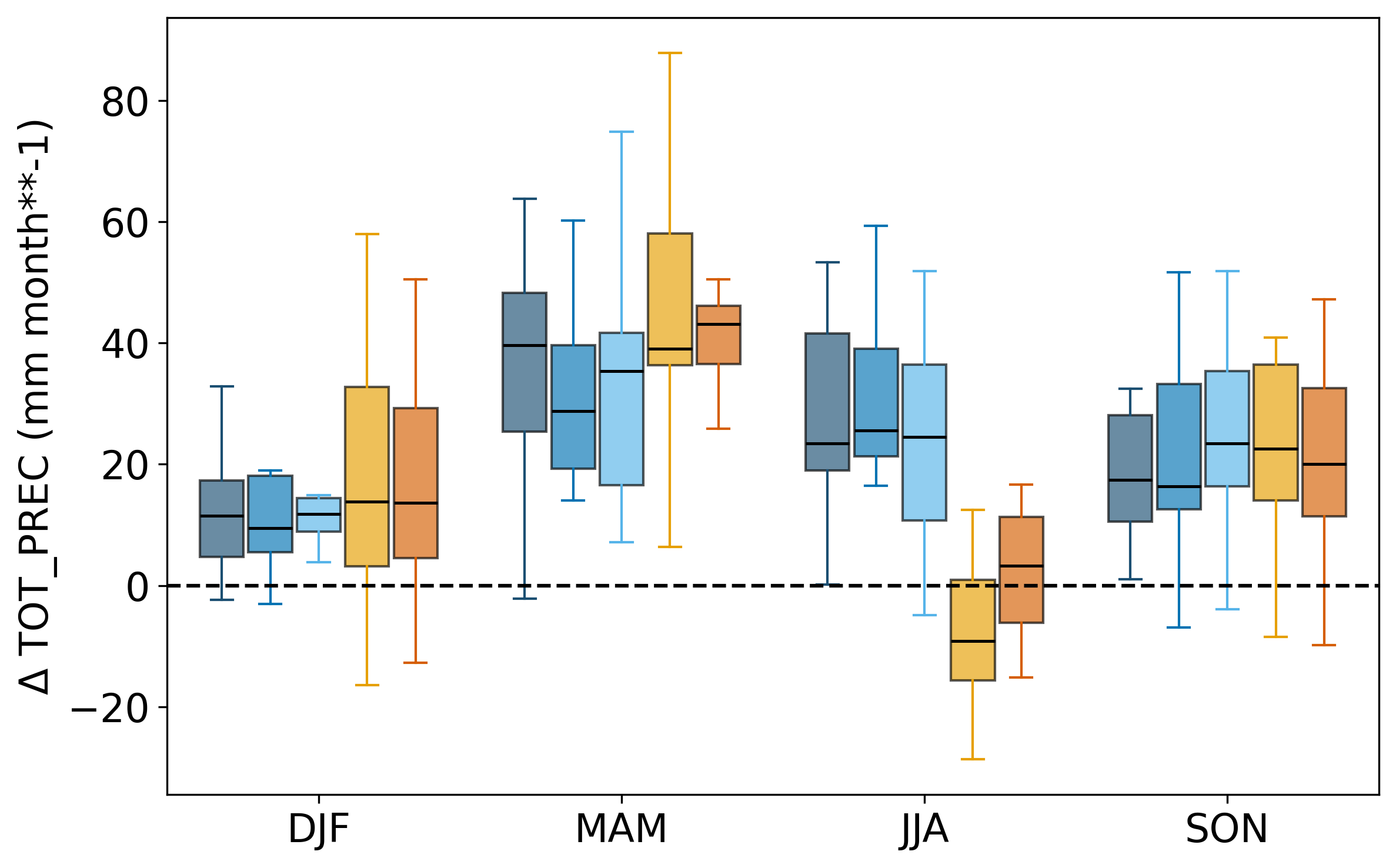}
  \label{fig:3.Po_2011_pp}
\end{subfigure}
\vspace{-9mm}
\medskip
\begin{subfigure}{0.45\textwidth}
  \vspace{-5mm}
  \caption{Rhine}
  \includegraphics[width=\linewidth]{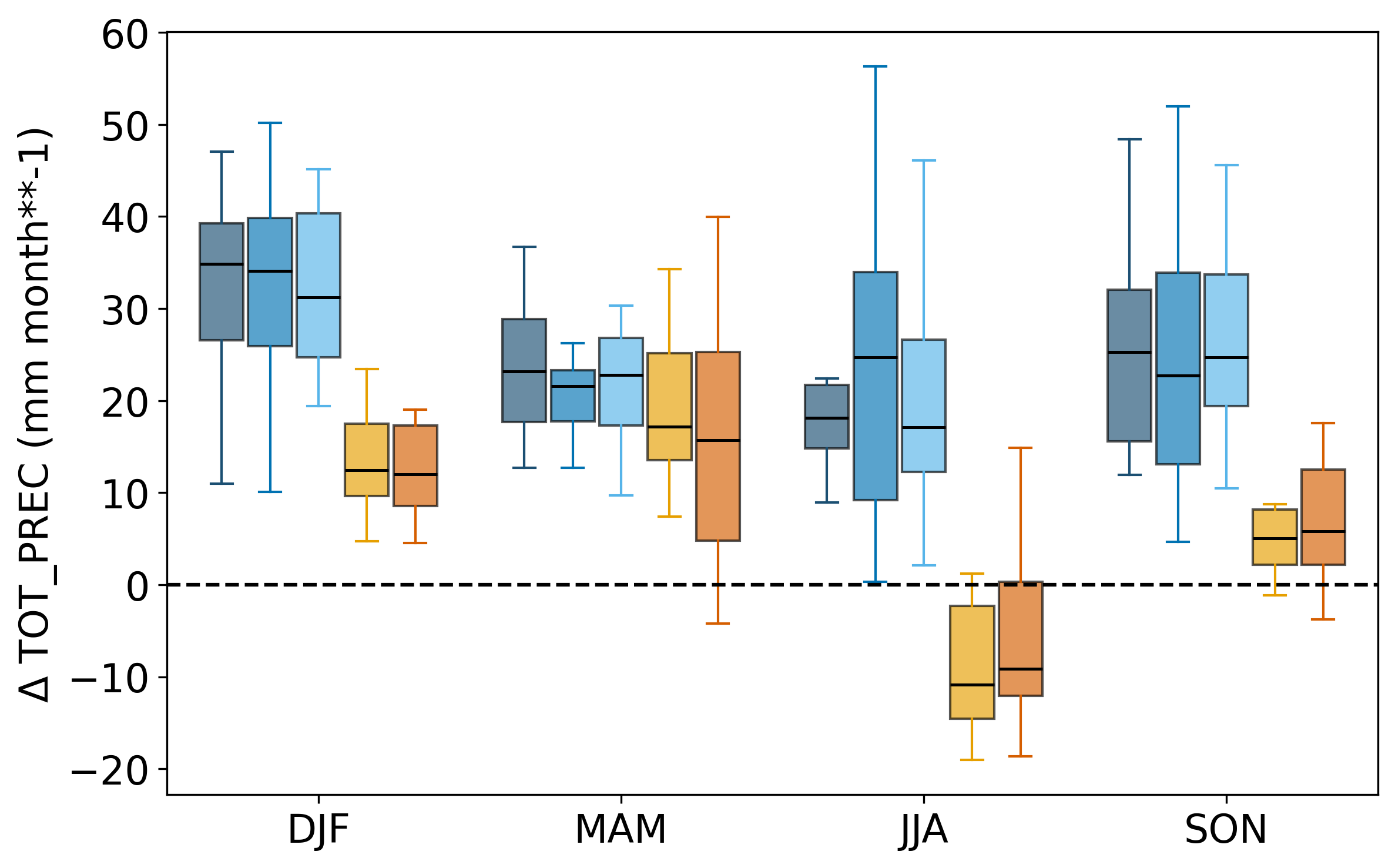}
  \label{fig:4.Rhein_2011_pp}
\end{subfigure}\hspace{1mm} 
\begin{subfigure}{0.45\textwidth}
  \vspace{-5mm}
  \caption{Tisa}
  \includegraphics[width=\linewidth]{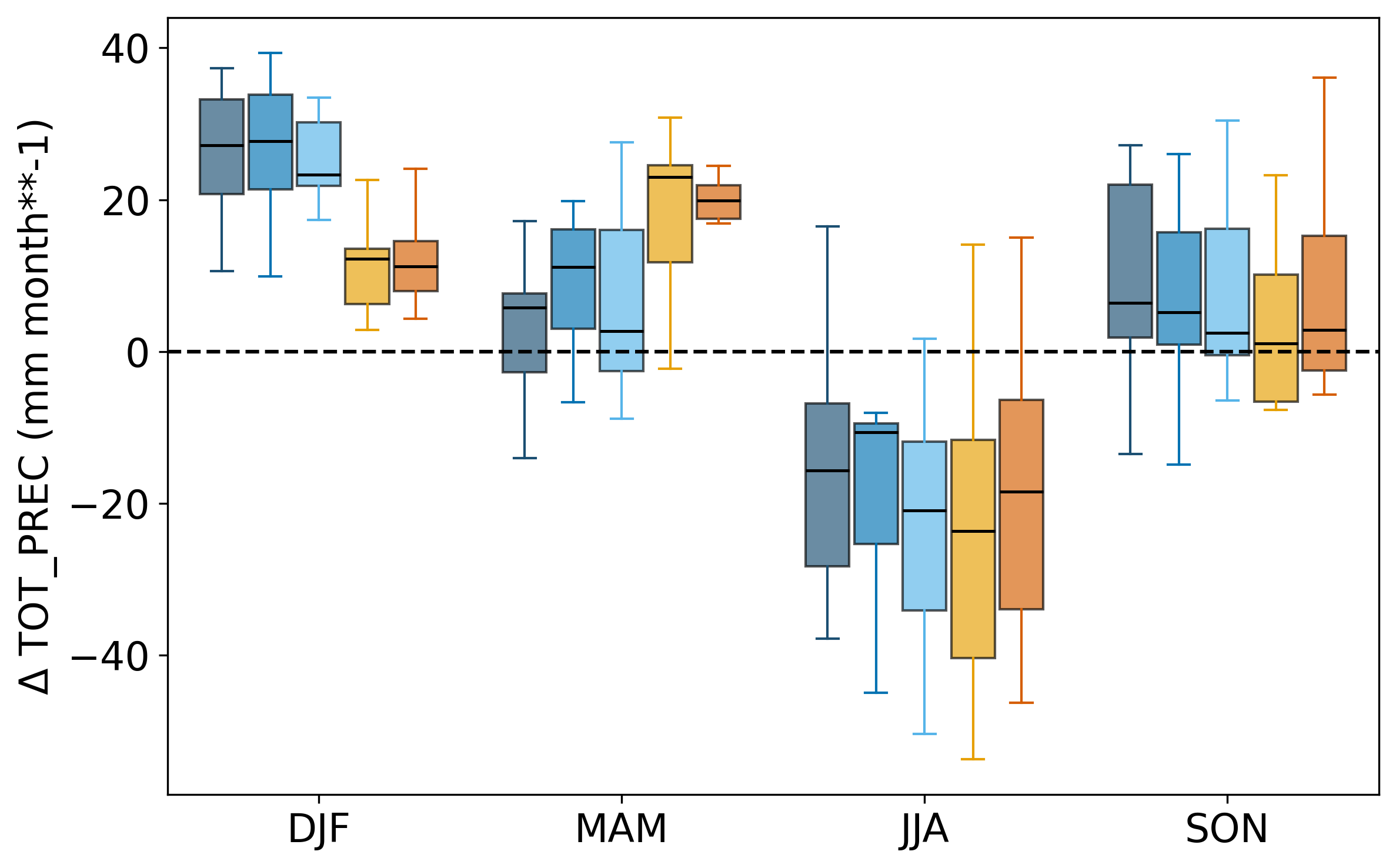}
  \label{fig:5.Tisa_2011_pp}
\end{subfigure}\hspace{1mm} 
\begin{subfigure}{0.45\textwidth}
  \vspace{-26mm}
  \includegraphics[width=\linewidth]{legend_only_2011.png}
\end{subfigure}
\vspace{-31mm}
  \caption{Boxplots of seasonal interannual precipitation biases for different watersheds.}
    \label{TOT_PREC_SEASDIFF_2011}
\end{figure}

Unlike T2m, the seasonal bias distribution for precipitation shows mainly a moderate difference between ICON and TSMP1 simulations (Figure \ref{TOT_PREC_SEASDIFF_2011}), as the bias distribution is comparable across most basins and seasons. During DJF, all experiments exhibit a wet bias in approximately 50\% of the data, with the strongest overestimation found in the Rhine and Tisa basins for the TSMP1 simulations. In both the Rhine and Tisa basins, ICON outperforms TSMP1, showing narrower bias spreads and medians below 20~mm~month$^{-1}$. Other high-resolution simulations (2.8 km) over Romania (within the Tisa basin) with ICON-LAM and COSMO-LAM also indicate that both models overestimate the magnitude and spatial extent of intense winter precipitation \citep{Iriza2023}. Similarly, in MAM, all experiments systematically overestimate precipitation, with consistent results in all basins. The largest differences between TSMP1 and ICON occur in JJA, particularly in the Ebro, Po, and Rhine basins. In Ebro and Po, ICON biases are centered close to zero, indicating good performance, while TSMP1 strongly overestimates precipitation, with median biases exceeding 10~mm~month$^{-1}$. In contrast, the Rhine shows opposite tendencies between models as TSMP1 simulations are predominantly wet-biased and ICON simulations are predominantly dry-biased. The generally better representation of precipitation by ICON in this season and across these basins may be linked to its treatment of shallow and deep convection. For instance, \citet{DeLucia} showed that ICON simulations at 2.5 km resolution over Italy during summer performed better when deep convection was switched off compared to runs where it was enabled. In the Tisa basin, the bias distribution is similar across all experiments, with about half of the data consistently showing a dry bias. In SON, the bias distribution is again comparable between experiments in the Ebro, Po, and Tisa basins. The Tisa basin has generally the best representation, as 50\% of the distribution of all experiments centers around zero. In the Rhine, ICON performs better, with median biases closer to zero and smaller spreads than TSMP1. Overall, ICON tends to reproduce precipitation more accurately in summer and autumn, whereas both models systematically overestimate precipitation in winter and spring.

Additional insights from the reanalysis dataset for the period 2018–2020 (Figure \ref{Time_PP_bias_2018} and Table \ref{table_P_metrics_2018}) confirm ICON\_3 tendencies. ICON\_3\_rea simulation achieved lower RMSE values than both ICON\_3 and TSMP1 across all basins, indicating again improved short-term skill in reproducing precipitation variability. 

\begin{figure}[htp!]
\centering
\begin{subfigure}{0.45\textwidth}
  \caption{Ebro}
  \includegraphics[width=\linewidth]{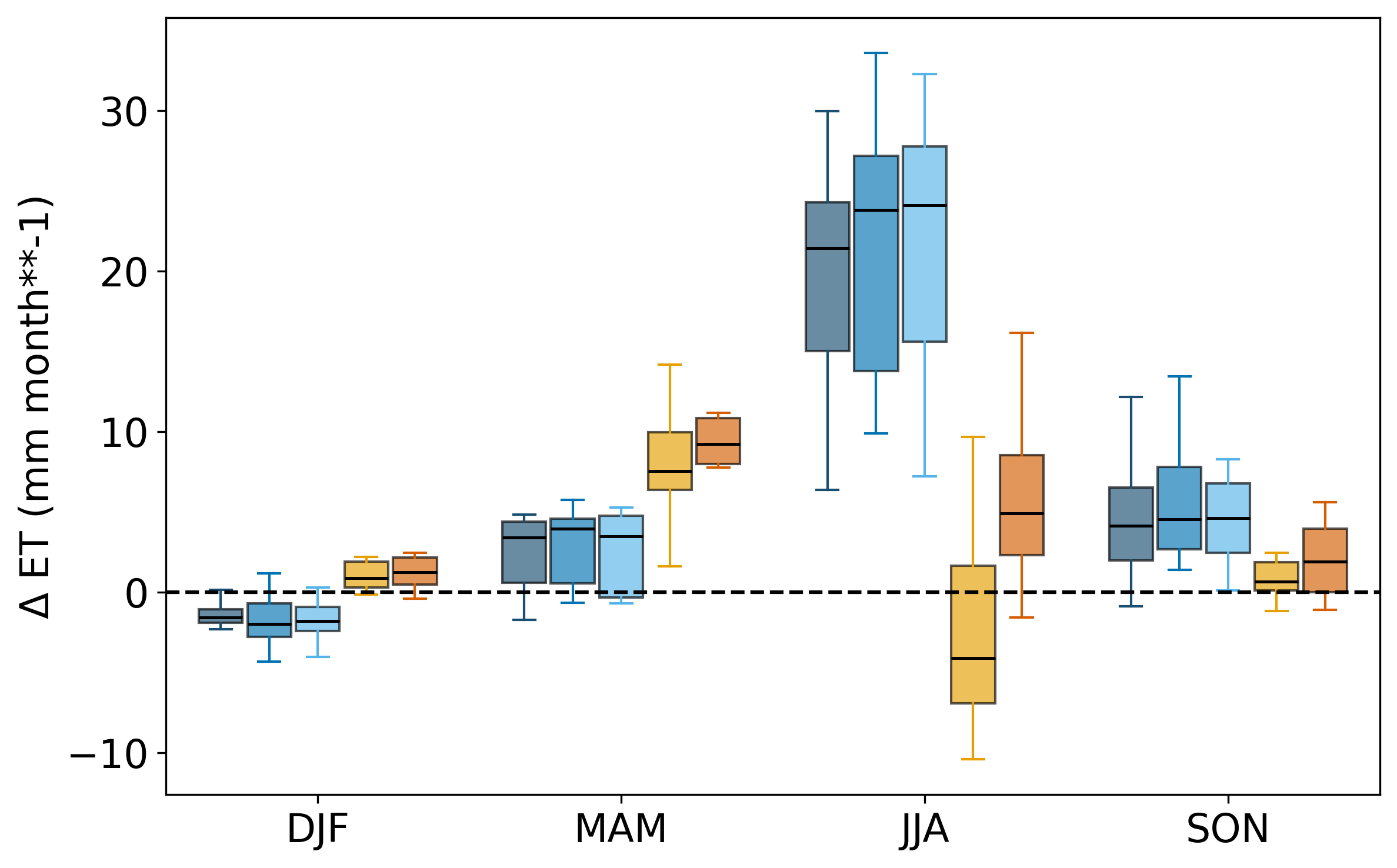}
  \label{fig:2.Ebro_2011}
\end{subfigure}\hspace{1mm} 
\begin{subfigure}{0.45\textwidth}
  \caption{Po}
  \includegraphics[width=\linewidth]{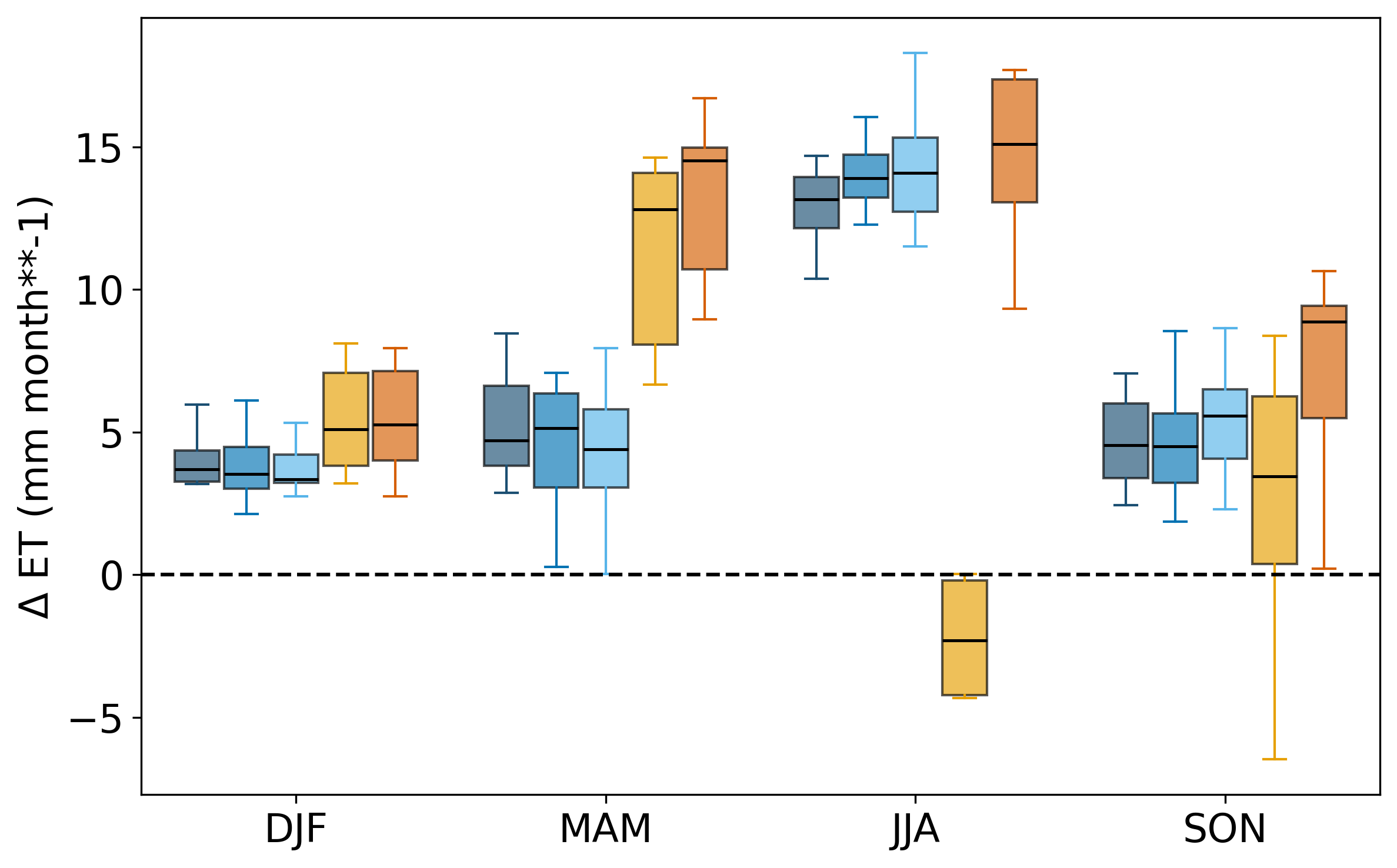}
  \label{fig:3.Po_2011}
\end{subfigure}
\vspace{-9mm}
\medskip
\begin{subfigure}{0.45\textwidth}
  \vspace{-5mm}
  \caption{Rhine}
  \includegraphics[width=\linewidth]{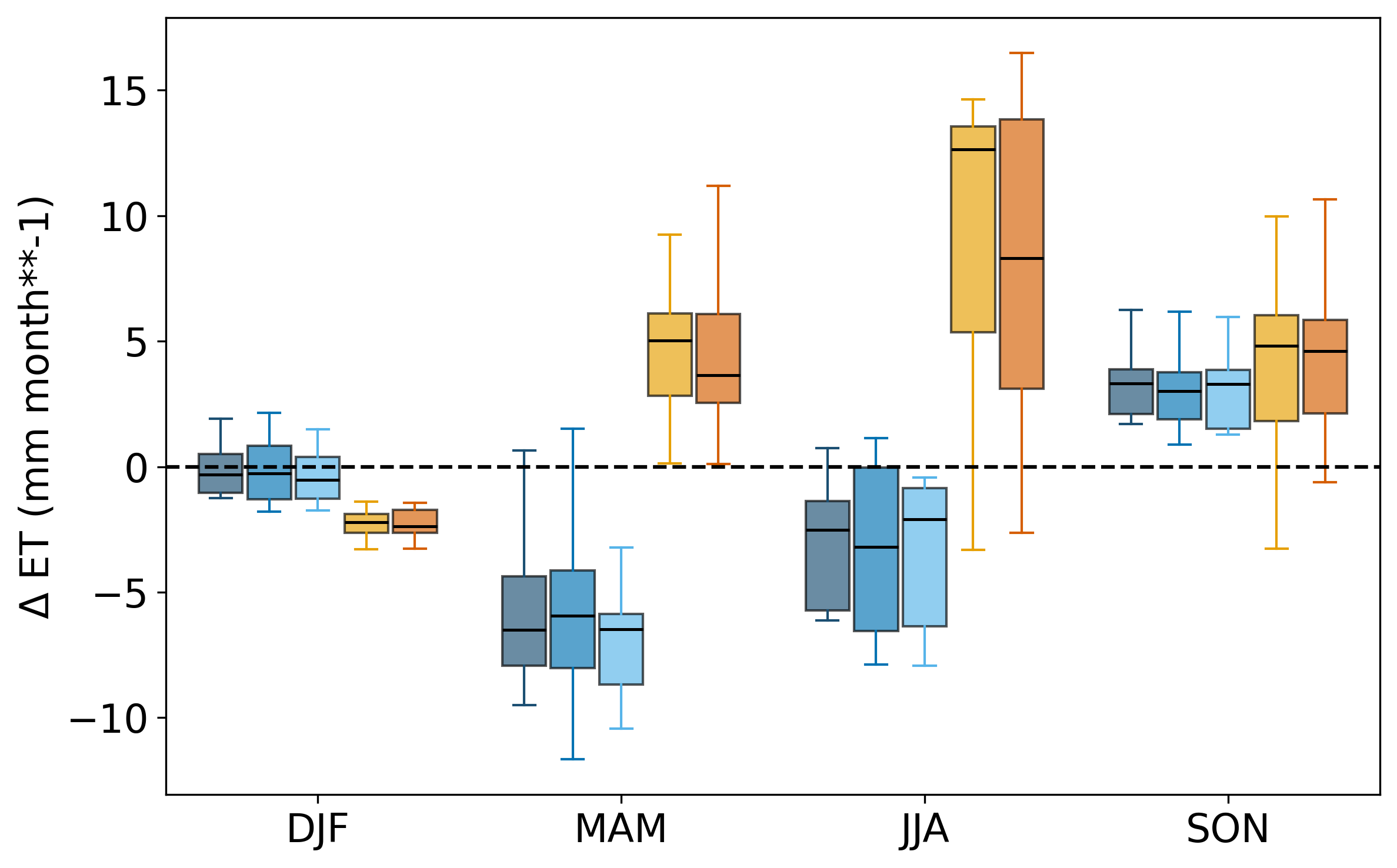}
  \label{fig:4.Rhein_2011}
\end{subfigure}\hspace{1mm} 
\begin{subfigure}{0.45\textwidth}
  \vspace{-5mm}
  \caption{Tisa}
  \includegraphics[width=\linewidth]{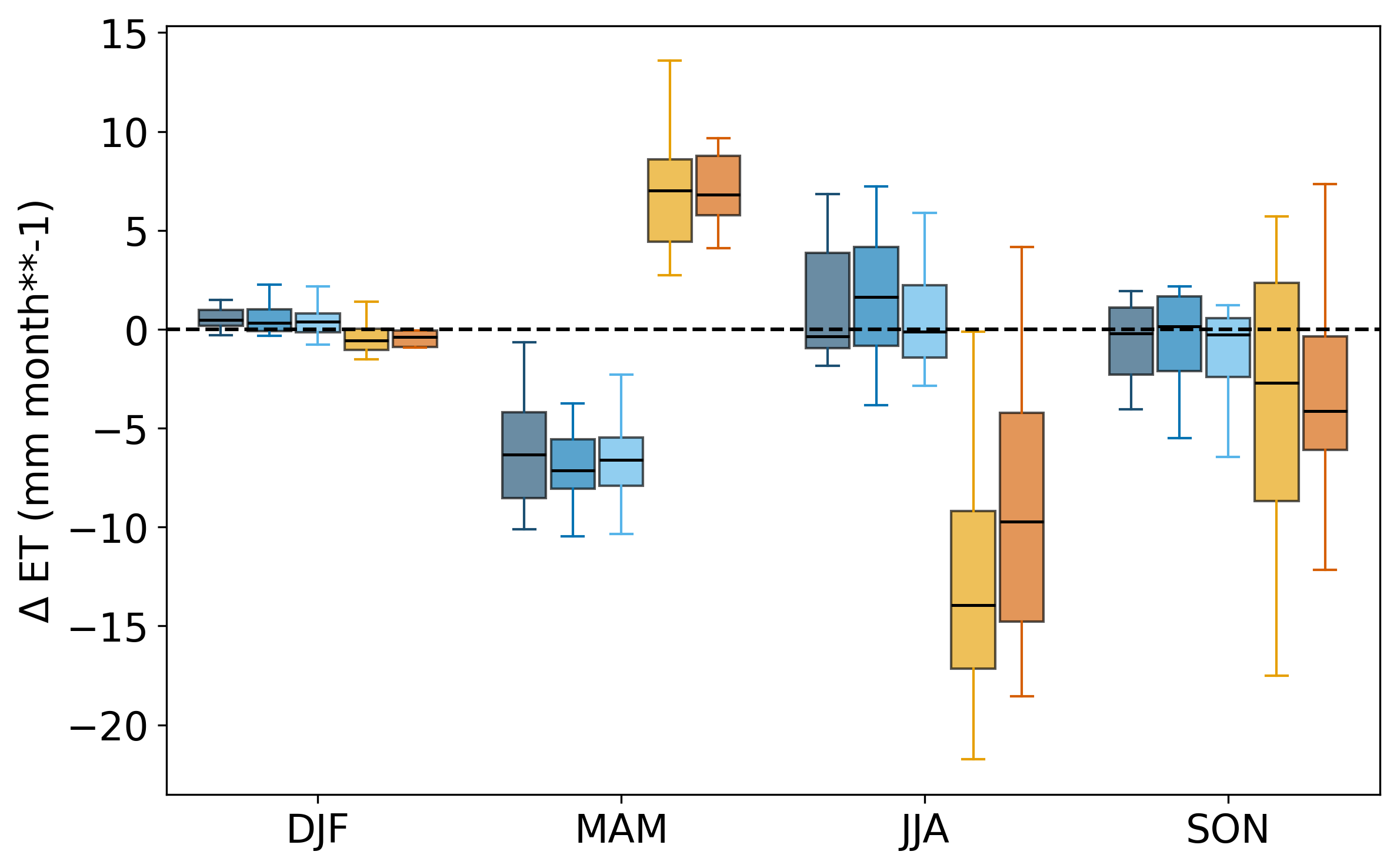}
  \label{fig:5.Tisa_2011}
\end{subfigure}\hspace{1mm} 
\begin{subfigure}{0.45\textwidth}
  \vspace{-26mm}
  \includegraphics[width=\linewidth]{legend_only_2011.png}
\end{subfigure}
\vspace{-31mm}
  \caption{Boxplots of seasonal interannual ET biases for different watersheds for the period 2011--2020.}
    \label{ET_SEASDIFF_2011}
\end{figure}

Figure \ref{ET_SEASDIFF_2011} presents the distribution of seasonal ET biases averaged per basin. DJF generally shows small biases, which are close to zero for most experiments and basins. However, Po river basin is an exception, as the medians are located around 5 mm month\textsuperscript{–1}. In MAM, the bias spread increases across all experiments, although the direction of the bias varies per basin. In the Ebro and Po basins, both models overestimate ET, with TSMP1 experiments medians nearer to zero than ICON. In the case of the Rhine and Tisa basins, ICON continues overestimating ET while TSMP1 underestimates it. In general, in both basins, the biases are of similar magnitude in opposite directions. During JJA, ICON\_3\_ctrl exhibits median biases closer to zero in the Ebro and Po basins compared to TSMP1. In the Rhine and Tisa basins, both the spread and median biases of TSMP1 are lower compared to ICON. SON shows a moderate bias in most experiments and basins, generally reflecting ET overestimation. In the Ebro and Tisa basins, bias distributions are closer to zero for both models, whereas in the Po and Rhine basins ICON experiments show larger spreads. In general, the boxplots show that the TSMP1 experiments generally exhibit smaller spreads and biases than ICON, particularly in the Rhine and Tisa basins, while ICON performs better in reducing median biases in the Ebro basin.

Complementary results of ICON\_3\_rea, period 2018–2020, (Figure \ref{Time_ET_bias_2018} and Table \ref{table_ET_metrics_2018}) provide additional context for ICON simulations. During this overlapping period, the reanalysis exhibits lower RMSE values than the ICON\_3\_ctrl and ICON\_3\_irr simulations in the Ebro, Rhine and Tisa basins, suggesting a better monthly representation of evapotranspiration. So, even though evapotranspiration is not assimilated, its representation improves. However, in the Po basin, the highest bias and RMSE correspond to ICON\_3\_rea. These results might indicate a poor representation of the observed variability on a monthly scale.

\subsubsection{Yearly interannual variability (Period 2011--2020)}

The mean absolute error (MAE) for the 2011--2020 period is shown in Figures \ref{ET_SEASDIFF_2011_mean}, \ref{PP_SEASDIFF_2011_mean} and \ref{T2m_SEASDIFF_2011_mean}. They present seasonal biases and the yearly Mean Absolute Error (MAE) for ET, precipitation, and T2m across all analyzed basins. Here we focus on the comparison within two modeling groups, TSMP1 and ICON. Similar to the previous period, results indicate that within the TSMP1 experiment group, the control simulation (TSMP1\_12\_ctrl) exhibits the lowest MAE for ET and T2m in the Ebro basin, demonstrating the best overall agreement with the reference datasets. Taylor diagrams support the fact that TSMP1\_12\_ctrl achieves the best results for ET and T2m, while TSMP1\_12\_irri\_s obtains the lowest RMS and the highest correlation for precipitation (Figure \ref{Taylor_Ebro_2011}).
In the Po basin, TSMP1\_12\_irri\_s again achieve the best performance in precipitation with the lowest MAE for ET, while TSMP1\_12\_ctrl performs best for ET. T2m has TSMP1\_12\_ctrl and TSMP1\_12\_irri\_w very close. Interannual Taylor diagrams further support that TMSP1 irrigation experiments always outperform in all variables, specifically TSMP1\_12\_irri\_w (Figure \ref{Taylor_Po_2011}). This consistent improvement in the irrigation experiments highlights the continued importance of representing irrigation processes in this basin, which is among the most heavily irrigated regions in Europe. 
For the Rhine basin, both the control simulation and TSMP1\_12\_irri\_s achieve the lowest MAE for ET, indicating comparable skill in reproducing surface fluxes. Precipitation is best represented by TSMP1\_12\_irri\_w, whereas the control run performs best for T2m. Likewise, the interannual variability of ET is best captured by TSMP1\_12\_ctrl, while the representation of interannual variability in precipitation and T2m is superior in TSMP1\_12\_irri\_w (Figure \ref{Taylor_Rhein_2011}). This differs from the previous period, where irrigation configurations consistently outperformed the control run. These results might suggest that model performance in this basin is highly sensitive to interannual climatic variability. In the Tisa basin, the results are again mixed, with different experiments performing best for different variables. TSMP1\_12\_irri\_s shows the lowest MAE for ET, the control simulation the lowest for precipitation, and TSMP1\_12\_irri\_w for T2m. When representing the interannual variability in this basin, TSMP1\_12\_irri\_w outperforms the other experiments in ET and T2m, however, the control run obtains better results for precipitation (Figure \ref{Taylor_Tisa_2011}). These findings emphasize that the model skill at 12 km for TSMP1 varies by variable and basin, reflecting differences in land-surface processes, and even in the influence of irrigation on surface energy and water fluxes. 

The ICON experiments ICON\_3\_ctrl and ICON\_3\_irr focus on the impact of irrigation for the period 2011–2020. In the Ebro and Tisa basins, the control simulation achieves the lowest MAE for ET, precipitation and T2m. Figure \ref{Taylor_Ebro_2011} supports these results since ICON\_3\_ctrl RMSE is the lowest for all three variables in the Ebro basin and for precipitation and ET in the Tisa basin. In the case of T2m, ICON\_3\_irr obtained a slightly lower RMSE than the control. The better performance of ICON\_3\_ctrl for ET might be attributed to the TERRA land surface scheme, which is already well-tunned within the ICON modeling framework. In the Po Basin, ICON\_3\_ctrl shows the lowest annual MAE for ET and T2m, while ICON\_3\_irr performs best for precipitation. Although ICON\_3\_ctrl performs slightly better annually (0.23 vs. 0.25 MAE), ICON\_3\_irr improves summer results, reducing the biases from 1.48 K to 0.65 K. In addition, ICON\_3\_irr obtains a marginally lower RMSE than the control for T2m when representing the interannual variability (Figure \ref{Taylor_Po_2011}). This experiment also obtains the lowest RMS errors for ET, although it exhibits a negative correlation. The same T2m trend holds for the Rhine in relation to the MAE, since ICON\_3\_ctrl shows only a slightly better improvement for T2m  (0.26 vs. 0.28 MAE). In the seasonal comparison, ICON\_3\_irr exhibits the best agreement in JJA. This irrigation experiment also improves the MAE for ET. The interannual variability for the Rhine shows that ICON\_3\_irr provides the best representation of T2m, characterized by positive correlations and the lowest RMSE. This experiment also achieves the lowest RMSE with a negative correlation for ET. ICON\_3\_ctrl outperforms for precipitation with lower RMSE, higher correlation and similar STD than GPCC. These findings imply that irrigation is indeed an anthopogenic factor that influences T2m, an effect widely recognized within the Earth system community \citep{avalm2023}.
In the current experiments, irrigation also appears to improve the representation of precipitation in the Po basin, a variable that remains subject to considerable uncertainties regarding its interaction with irrigation \citep{McDermid2023}.
In any case, more analysis is needed to confirm this improvement. Overall, the results highlight again the importance of accounting for irrigation in weather and climate simulations, especially for the summer season. However, further parameter tuning and reduction of irrigation related uncertainties remain necessary for the rest of the year.

\section{SUMMARY AND CONCLUSIONS} 
The present study presents a multi-model regional climate model ensemble dataset of the CRC 1502 - DETECT, designed for water cycle process analyses. We compare the performance of the atmospheric simulations in representing key aspects of the regional climate over the period 1990--2020 across four representative basins within the EURO-CORDEX domain. Our analysis focuses on the variables 2 m air temperature (T2m), precipitation (P) and evapotranspiration (ET), evaluated using monthly values spatially-averaged over the Ebro, Po, Rhine, and Tisa basins. To assess the model performance against the reference observational datasets GLEAM, GPCC, and E-OBS, we employ common statistical metrics as well as Taylor diagrams to analyze the interannual variability of seasonal and yearly averages or accumulations. This study provides a baseline analysis of the simulated weather and climate in the DETECT project pan-European analysis domain and may inform ensuing analyses on water cycle processes, such as land-atmosphere interactions, and can be used to evaluate future model and simulation experiment improvements. The dataset will become publicly available also within the course of the DETECT project. 

Our study shows that all experiments systematically underestimate T2m across most seasons and basins, although the magnitude of the bias differs between ICON and TSMP1. In general, TSMP1 shows higher variability and stronger cold biases often exceeding -2~K, whereas ICON performs particularly better, with generally smaller biases. Moreover, the largest model differences for this variable are observed in summer. Regarding P and ET, we found that all simulations across both periods (2003--2010, 2011--2020) exhibit a comparable bias distribution in winter; while the largest differences between model performance occur in spring and, most prominently, in summer. For instance, substantial contrasts between ICON and TSMP1 simulations are evident for P, particularly over the Ebro and Rhine river basins in all seasons, with the largest differences occurring in summer. In contrast, the Tisa basin shows the smallest inter-model differences across all experiments. Overall, ICON exhibits smaller biases across all seasons and basins. However, both models display considerable spread in their P bias, which is expected given the inherent difficulty of simulating P \citep{giorgi2019, tapiador2019}. 
In the case of ET, ICON simulations systematically overestimate this variable and show a larger spread, whereas the TSMP1 experiments exhibit a narrower bias distribution. In general, TSMP1 outperforms ICON in the Rhine and Tisa basins, whereas ICON performs better in the Ebro basin. These results highlight that model performance varies substantially by basin and variable and may be clearly attributed to model structural differences.

When analyzing the interannual variability across all variables and basins per simulation group, we found that both TSMP1 and ICON\_3 irrigation experiments tend to perform better in the Po basin, mainly for T2m and P. ET also shows a yearly improved representation in TSMP1 simulations. These results are supported by lower MAE and RMS error values compared with their respective control simulation. Given that the Po basin is among the most intensively irrigated regions in Europe, these findings emphasize the importance of accurately representing irrigation processes to improve the model performance in this basin.

The SST sensitivity experiments conducted with the ICON\_12 simulations reveal that SST influences T2m and P simulations within the EURO-CORDEX domain. This is especially the case when analyzing results from the Rhine basin, where all SST experiments outperformed the control run across all assessed variables. Among the configurations, ICON\_12\_sst\_era5 presented better results in simulated T2m for three of the four analyzed basins, while ICON\_12\_sst\_mur stood out in two of them. Using foundation SSTs instead of skin SSTs, \cite{daSilvaLopes2025} found an increase of about 2.4\% in annual continental precipitation and a corresponding rise in surface freshwater flux, at finer spatial scales, precipitation differences were locally up to five times larger than the domain average. These results highlight the importance of accurate SST specification in constraining regional climate model boundary conditions and improving simulated land-atmosphere interactions.

This study provides a comprehensive, general intercomparison of regional atmospheric simulations for essential climate variables, with a focus on water cycle investigations. Despite the dedicated ensemble, we acknowledge some limitations. First, the simulation period analyzed was relatively short, especially when divided into two sub-periods. Second, all simulations have different specific goals, and their model configuration reflects those distinct objectives. Third, as mentioned in the Methods section, the reference datasets contain inherent uncertainties, particularly in regions with sparse station coverage, such as E-OBS. Finally, we focused on monthly averages for specific basins, omitting daily, diurnal, and spatial variability. In the future, more specific events that require shorter timescales can be analyzed, such as heatwaves or heavy precipitation events. Nonetheless, this study lays the basis for future intercomparisons and integrated analyses across the diverse simulation packages and modeling components developed within CRC 1502 - DETECT and can be used as a baseline for future model and simulation experiment improvements.

\section*{Acknowledgements} 

Funded by the Deutsche Forschungsgemeinschaft (DFG, German Research Foundation) – SFB 1502/1–2022 - Projektnummer: 450058266. \\
We would like to thank members of the CRC 1502 - DETECT for their supervision, discussion, and their support and contributions to model development, data generation, data provisioning, as well as input to the set-up and operation of simulations: Stefan Siebert (B05), Stefan Kollet and Michael Schindelegger (D02); support was also provided by Niklas Wagner. \\
The authors gratefully acknowledge the Gauss Centre for Supercomputing e.V. (www.gauss-centre.eu) for funding this project by providing computing time on the GCS supercomputer JUWELS \citep{JUWELS} and on the supercomputer JURECA \citep{JURECA} under grant no. cjjsc39, detectrea, both at the Jülich Supercomputing Centre (JSC) at Forschungszentrum Jülich. \\ 
The authors thankfully acknowledge the Deutsche Wetterdienst (DWD - German Weather Service) for providing computing resources in the operational High Performance Computing system. \\ 
We thank the research data infrastructure and services provided by the Jülich Supercomputing Centre, Germany, as part of data project 'detectdata'. \\
We thank the "Global Runoff Data Centre at the Federal Institute of Hydrology (BfG), 56068 Koblenz, Germany" for providing HydroSHEDS-based watershed boundaries. \\
We acknowledge the E-OBS dataset from the EU-FP6 project UERRA (http://www.uerra.eu) and the Copernicus Climate Change Service, and the data providers in the ECA\&D project (https://www.ecad.eu).

\section*{conflict of interest} 
The authors declare no conflict of interest.

\printendnotes

\bibliography{references_alphabetic}

\appendix
\section{Appendix} 
\renewcommand{\thefigure}{S\arabic{figure}}
\renewcommand{\thetable}{S\arabic{table}}
\setcounter{figure}{0}
\setcounter{table}{0}

\begin{figure}[H]
\centering
\begin{subfigure}{0.45\textwidth}
  \caption{Ebro}
  \includegraphics[width=\linewidth]{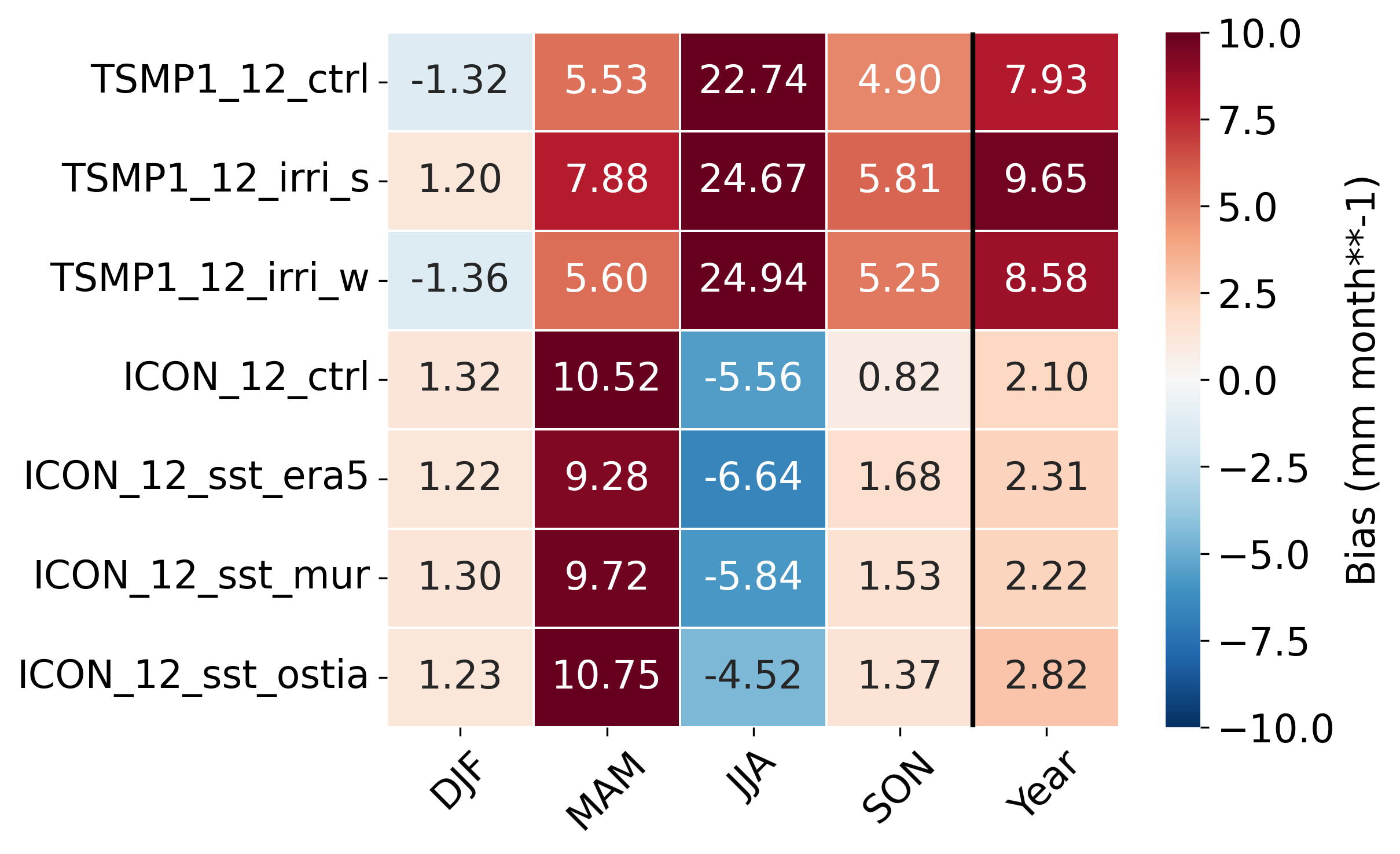}
  \label{et_Ebro}
\end{subfigure}\hspace{1mm} 
\begin{subfigure}{0.45\textwidth}
  \caption{Po}
  \includegraphics[width=\linewidth]{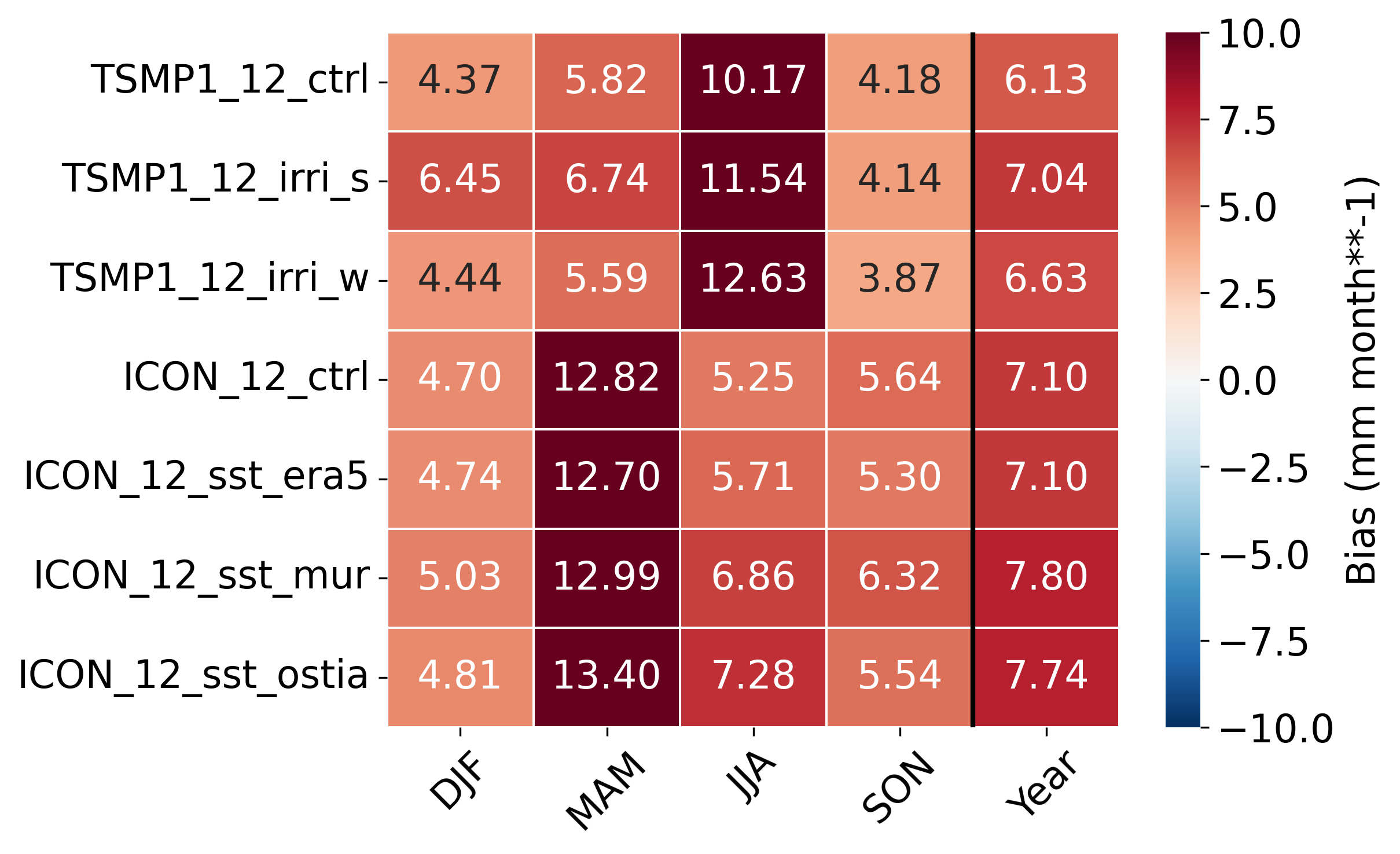}
  \label{et_Po}
\end{subfigure}
\vspace{-9mm}
\medskip
\begin{subfigure}{0.45\textwidth}
  \vspace{-5mm}
  \caption{Rhine}
  \includegraphics[width=\linewidth]{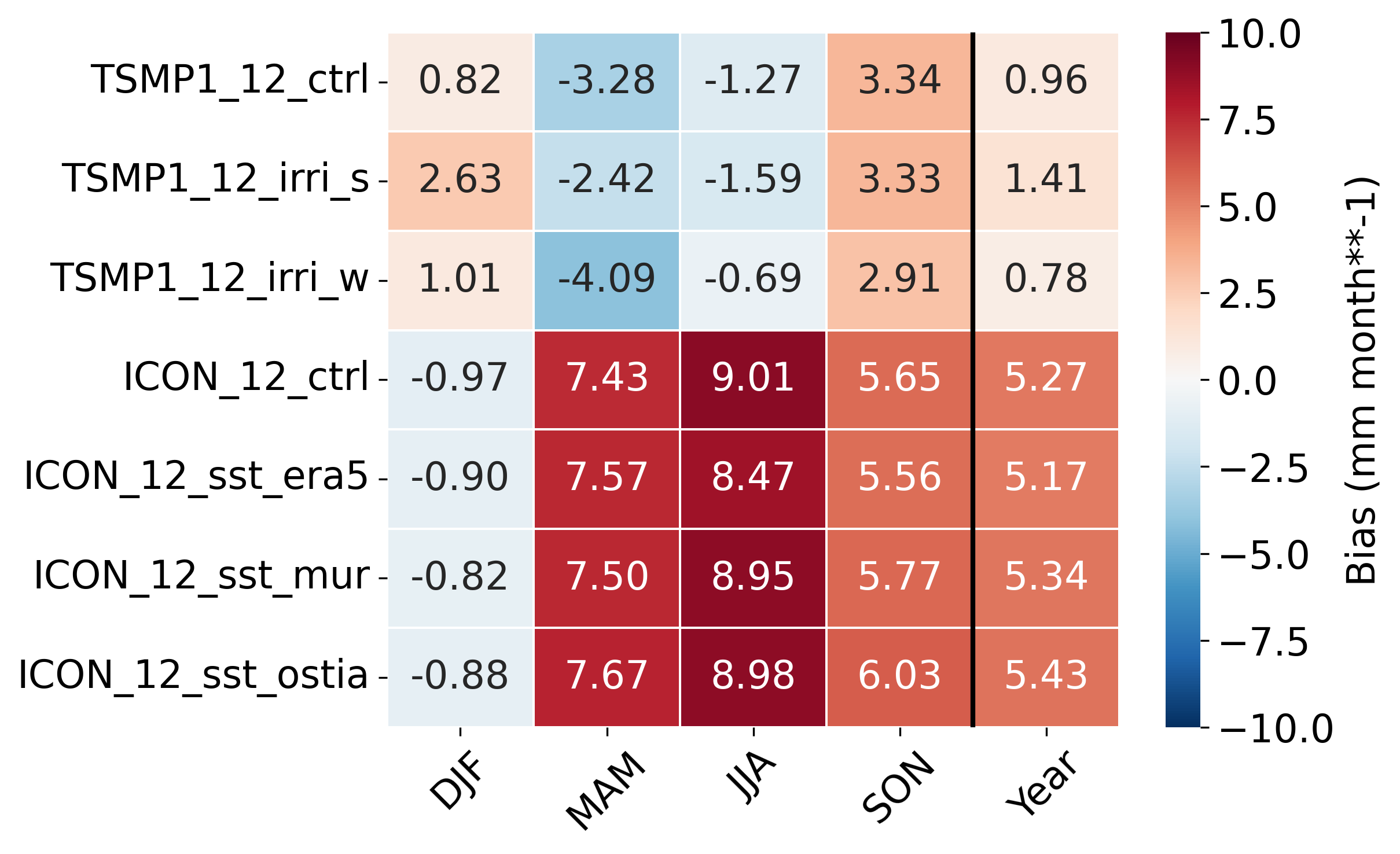}
  \label{et_Rhein}
\end{subfigure}\hspace{1mm} 
\begin{subfigure}{0.45\textwidth}
  \vspace{-5mm}
  \caption{Tisa}
  \includegraphics[width=\linewidth]{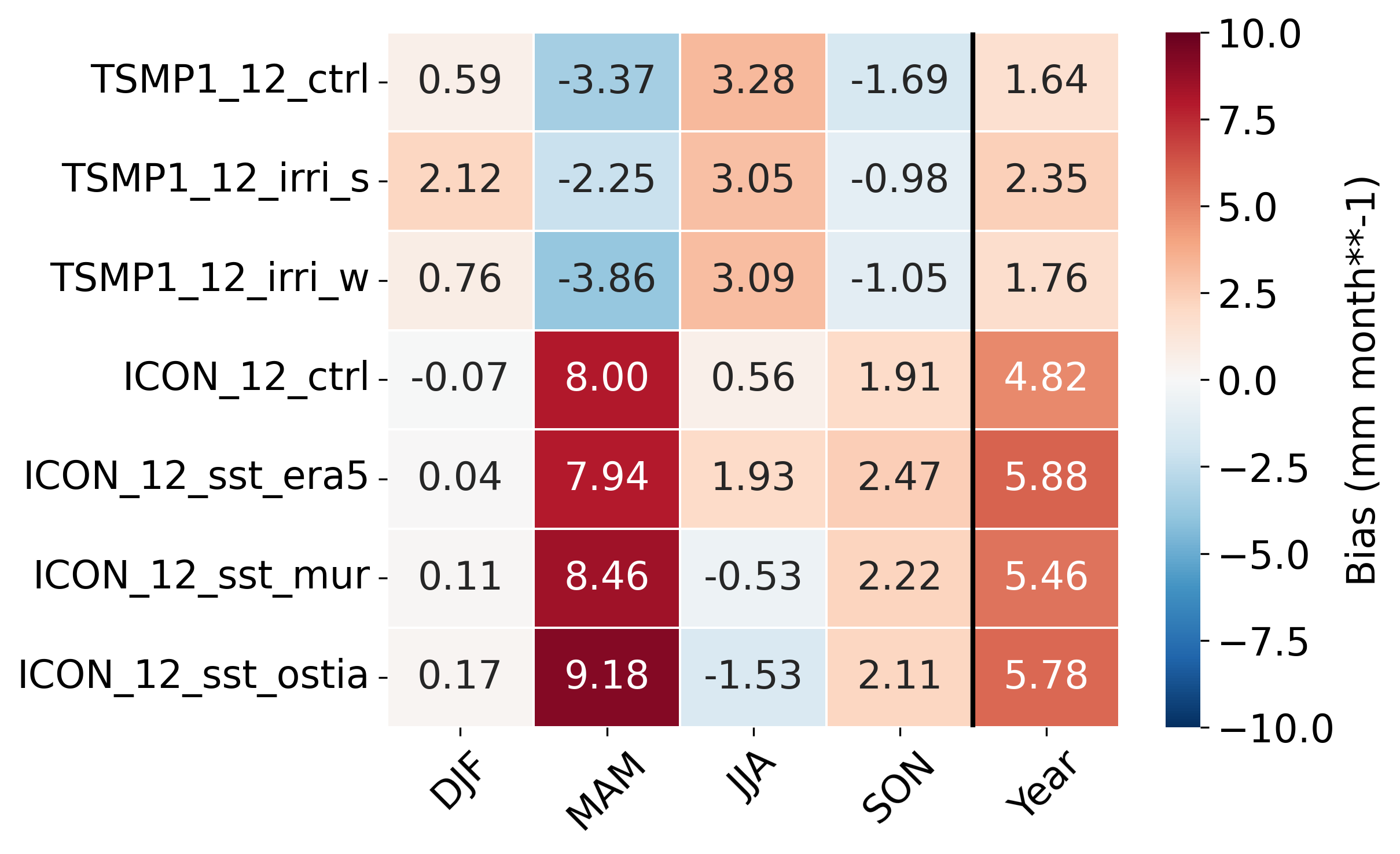}
  \label{et_Tisa}
\end{subfigure}\hspace{1mm} 
  \caption{Mean seasonal ET biases for different watersheds for the period 2003-2010. The column "Year", shows the yearly MAE for the whole period.}
    \label{ET_SEASDIFF_2003_mean}
\end{figure}

\begin{figure}[H]
\centering
\begin{subfigure}{0.45\textwidth}
  \caption{Ebro}
  \includegraphics[width=\linewidth]{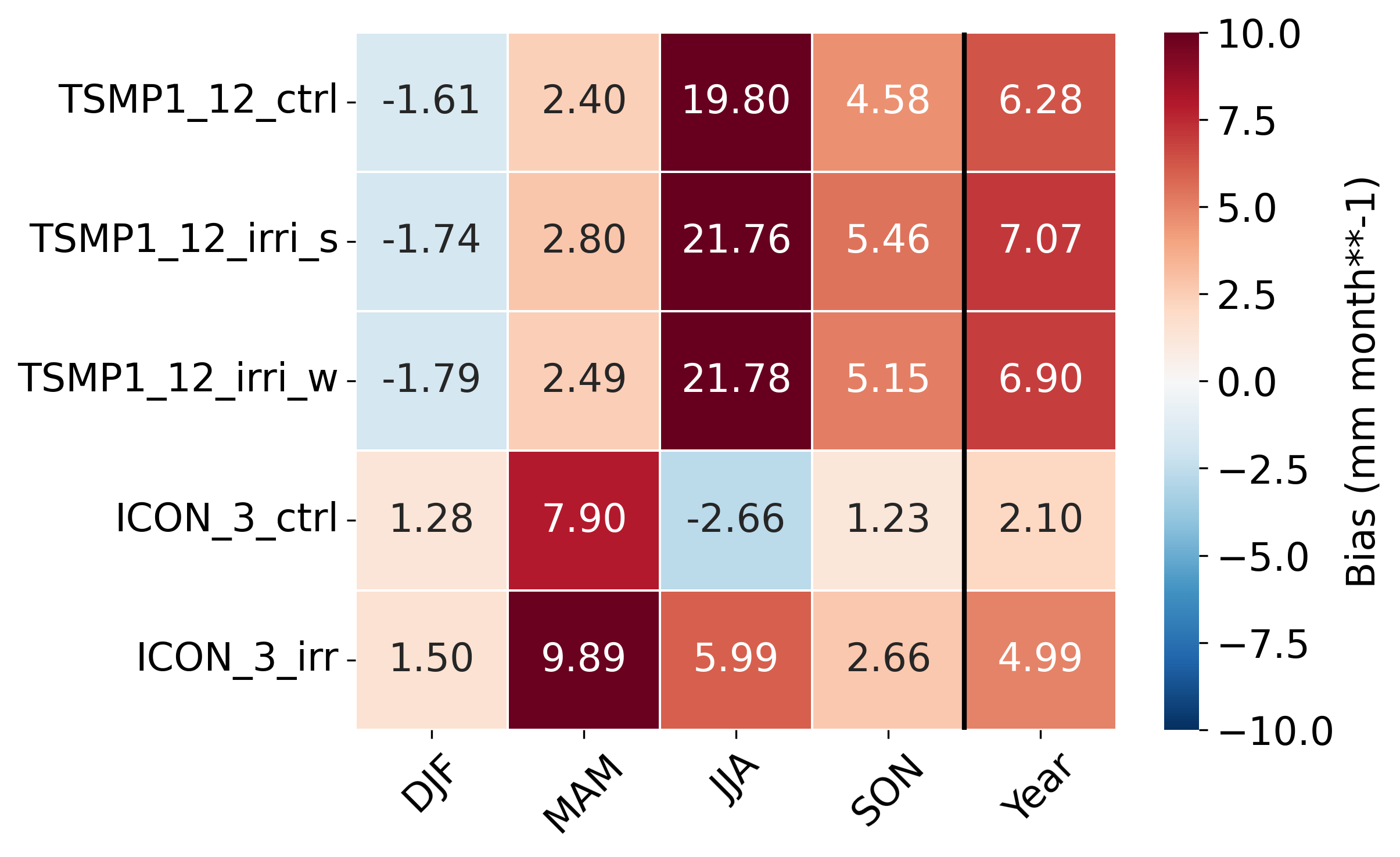}
  \label{et_Ebro_11}
\end{subfigure}\hspace{1mm} 
\begin{subfigure}{0.45\textwidth}
  \caption{Po}
  \includegraphics[width=\linewidth]{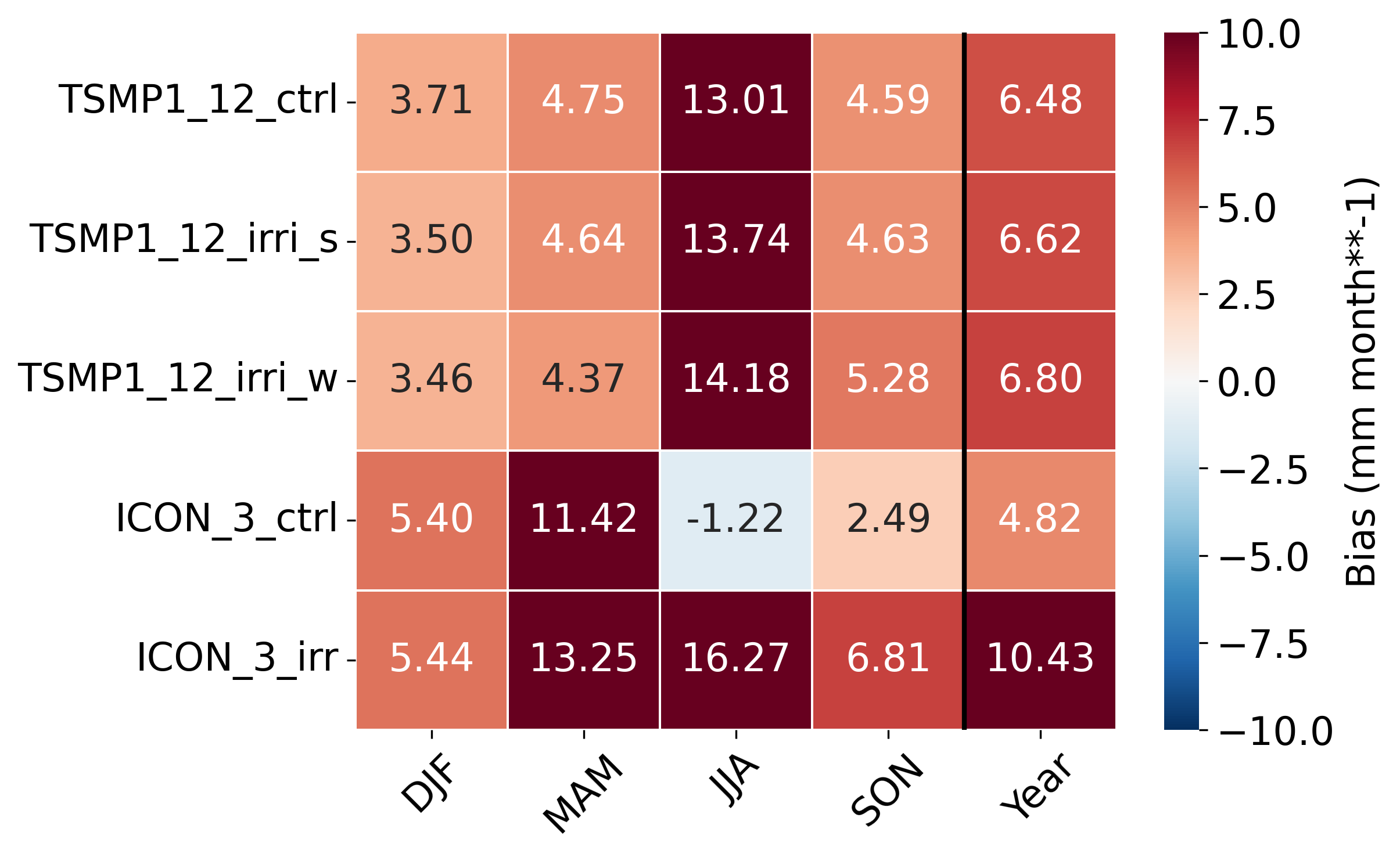}
  \label{et_Po_11}
\end{subfigure}
\vspace{-9mm}
\medskip
\begin{subfigure}{0.45\textwidth}
  \vspace{-5mm}
  \caption{Rhine}
  \includegraphics[width=\linewidth]{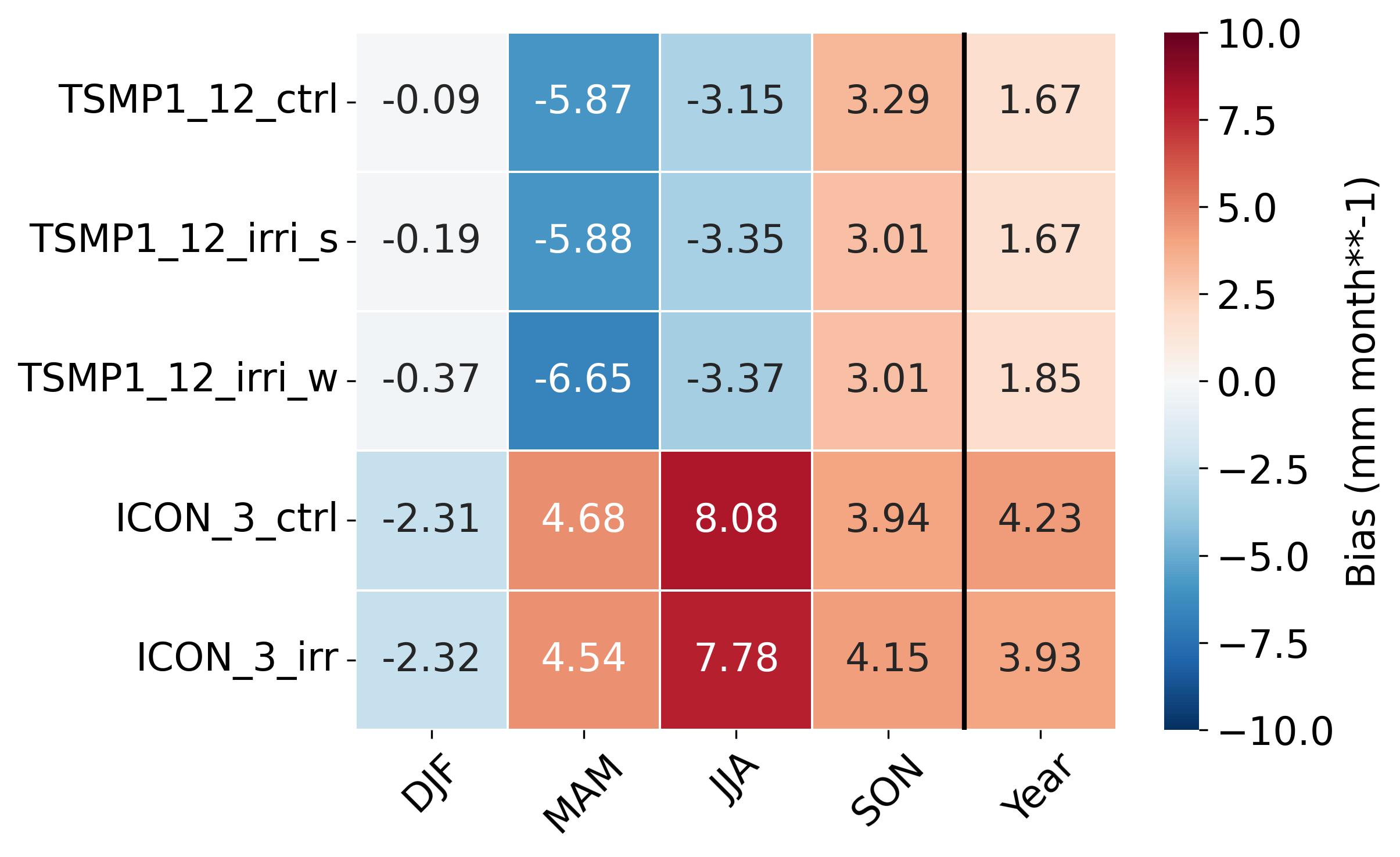}
  \label{et_Rhein_11}
\end{subfigure}\hspace{1mm} 
\begin{subfigure}{0.45\textwidth}
  \vspace{-5mm}
  \caption{Tisa}
  \includegraphics[width=\linewidth]{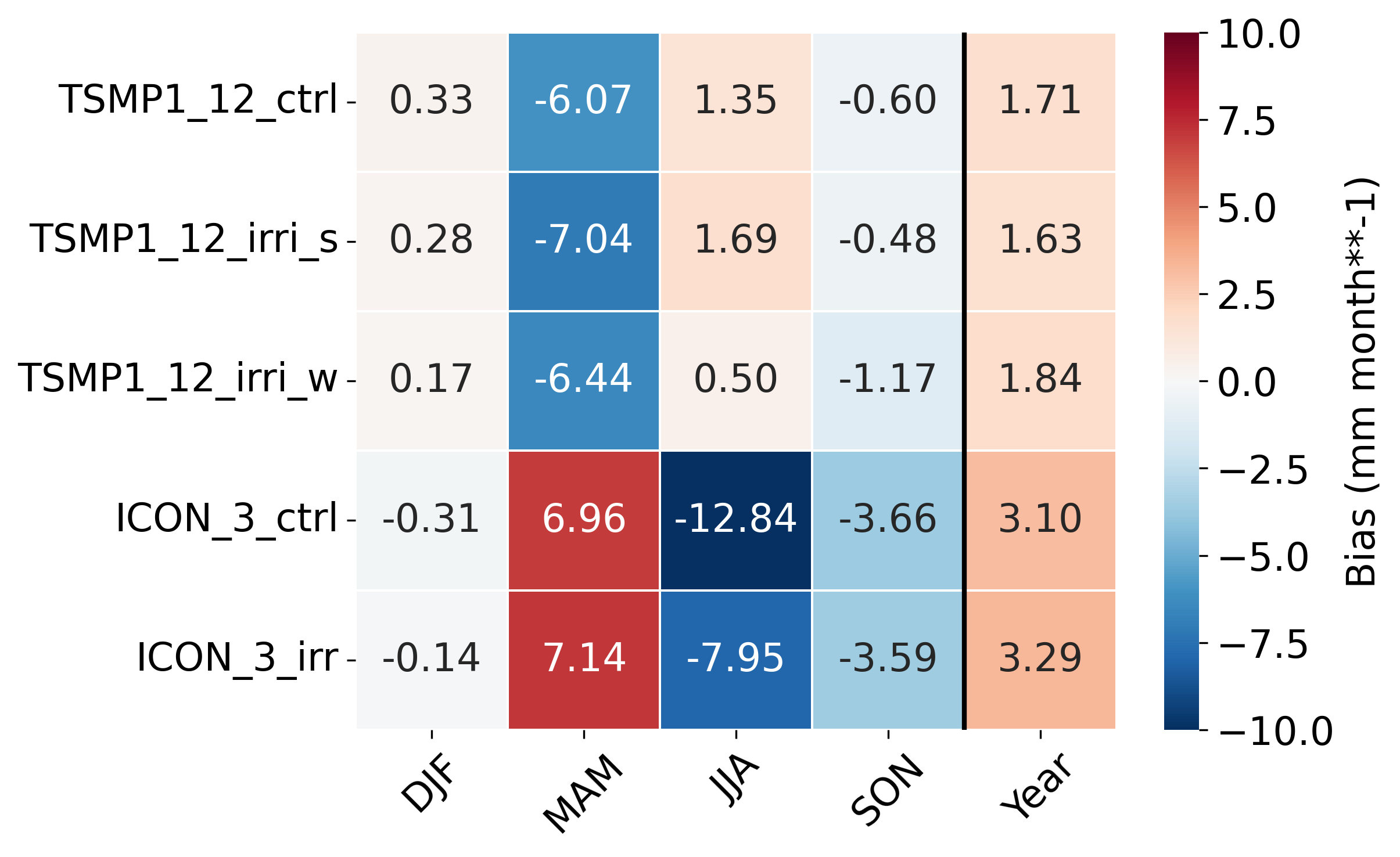}
  \label{et_Tisa_11}
\end{subfigure}\hspace{1mm} 
  \caption{Mean seasonal ET biases for different watersheds for the period 2011-2020. The column "Year", shows the yearly MAE for the whole period.}
    \label{ET_SEASDIFF_2011_mean}
\end{figure}

\begin{figure}[H]
\centering
\begin{subfigure}{0.45\textwidth}
  \caption{Ebro}
  \includegraphics[width=\linewidth]{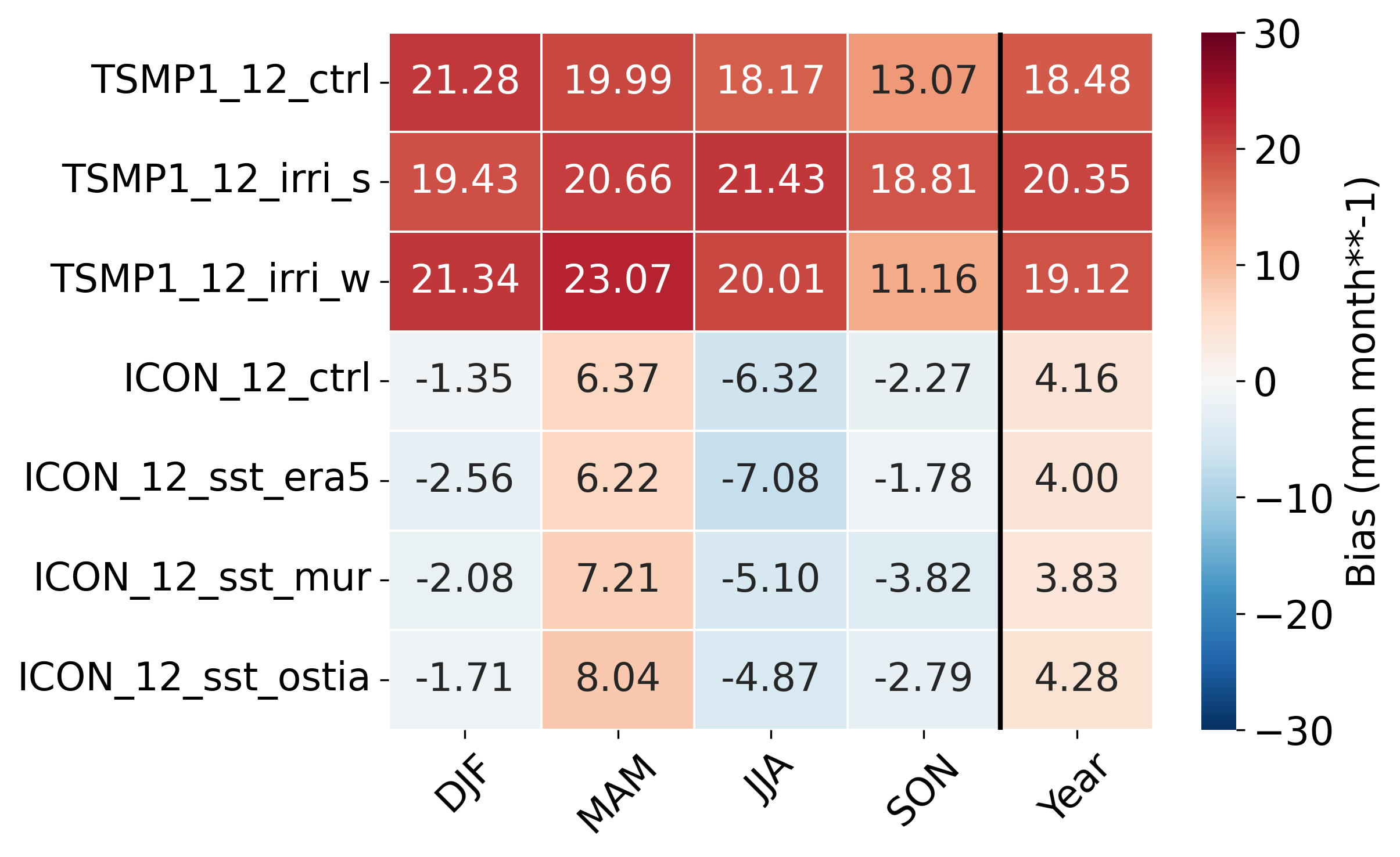}
  \label{pp_Ebro}
\end{subfigure}\hspace{1mm} 
\begin{subfigure}{0.45\textwidth}
  \caption{Po}
  \includegraphics[width=\linewidth]{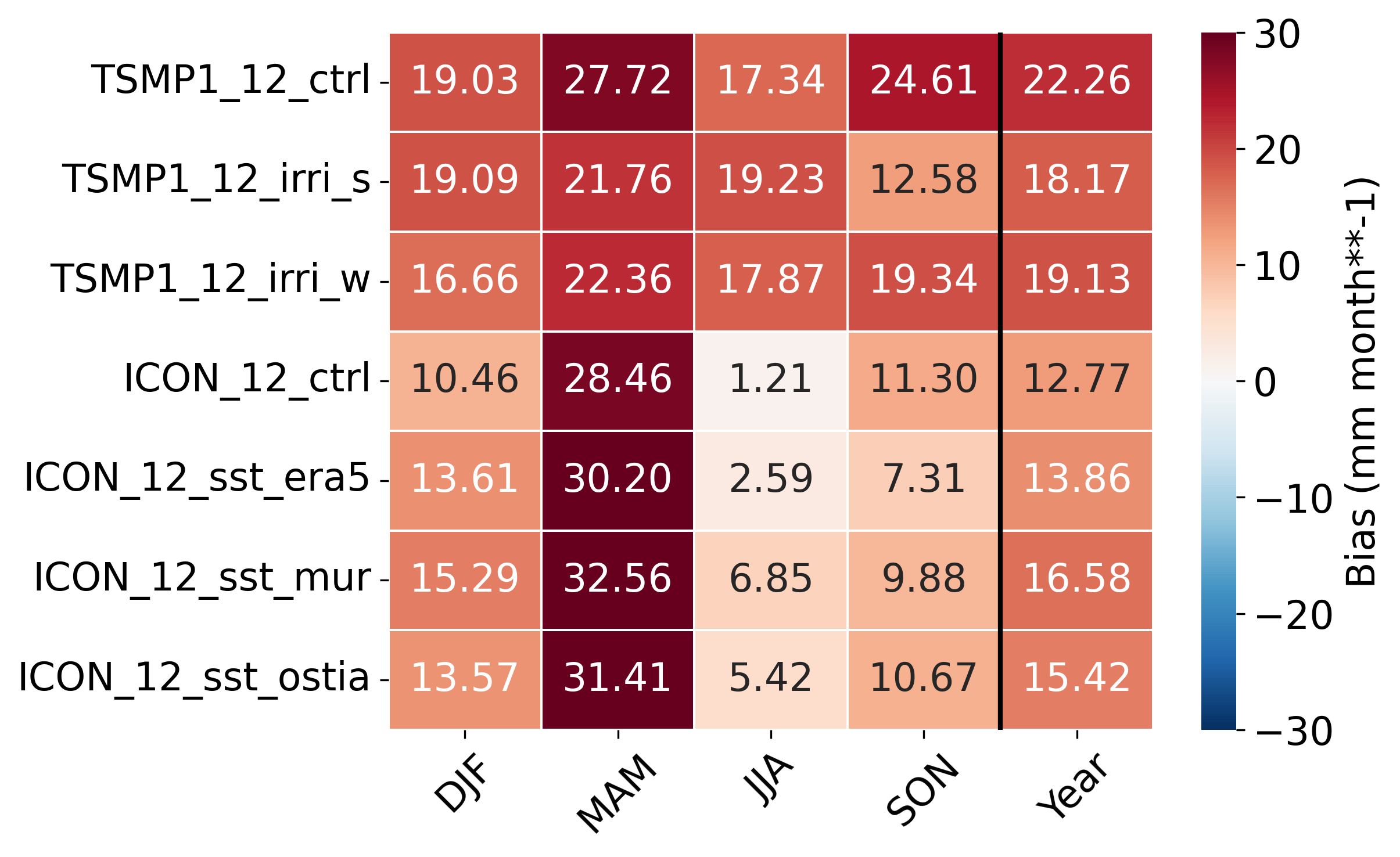}
  \label{pp_Po}
\end{subfigure}
\vspace{-9mm}
\medskip
\begin{subfigure}{0.45\textwidth}
  \vspace{-5mm}
  \caption{Rhine}
  \includegraphics[width=\linewidth]{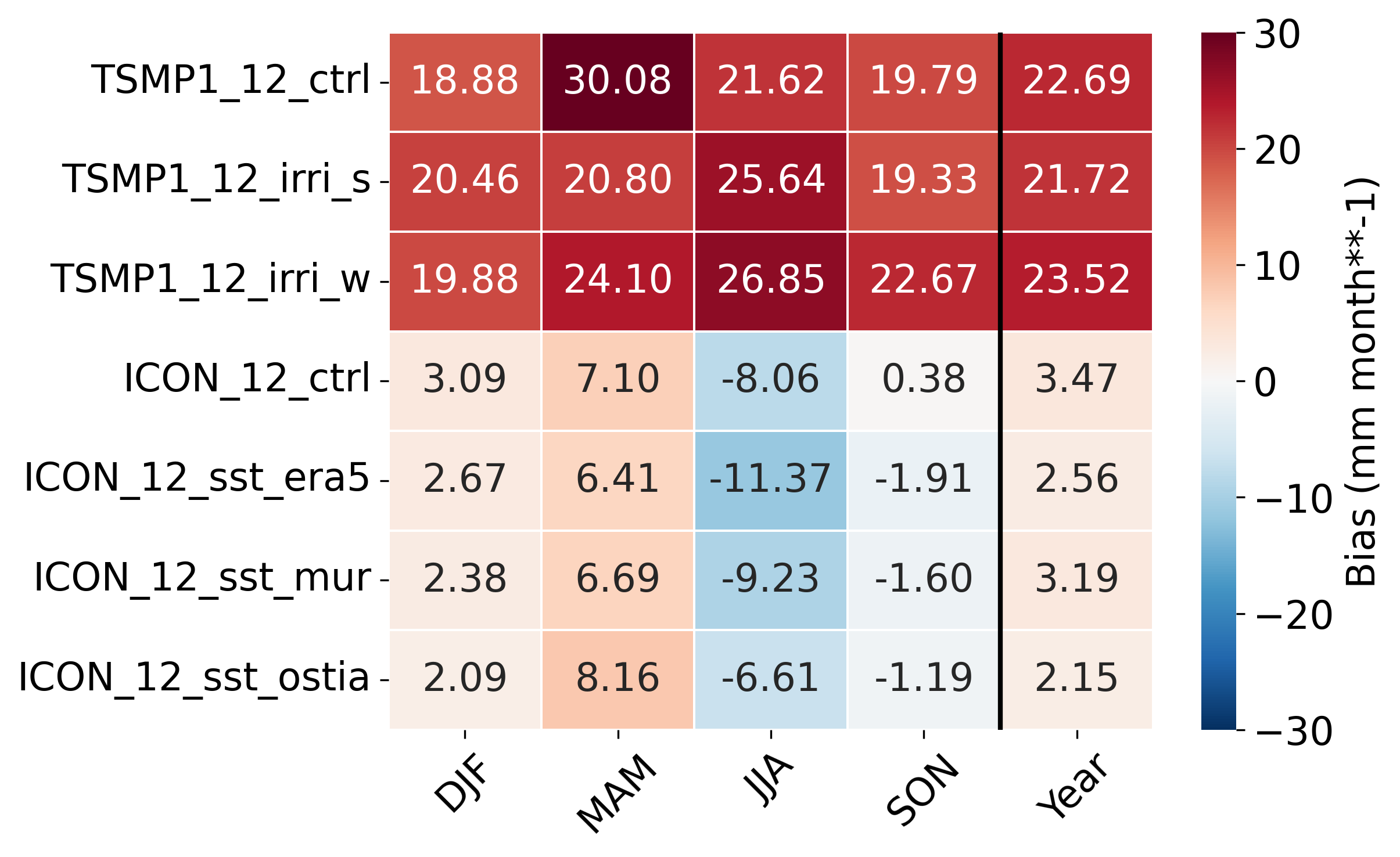}
  \label{pp_Rhein}
\end{subfigure}\hspace{1mm} 
\begin{subfigure}{0.45\textwidth}
  \vspace{-5mm}
  \caption{Tisa}
  \includegraphics[width=\linewidth]{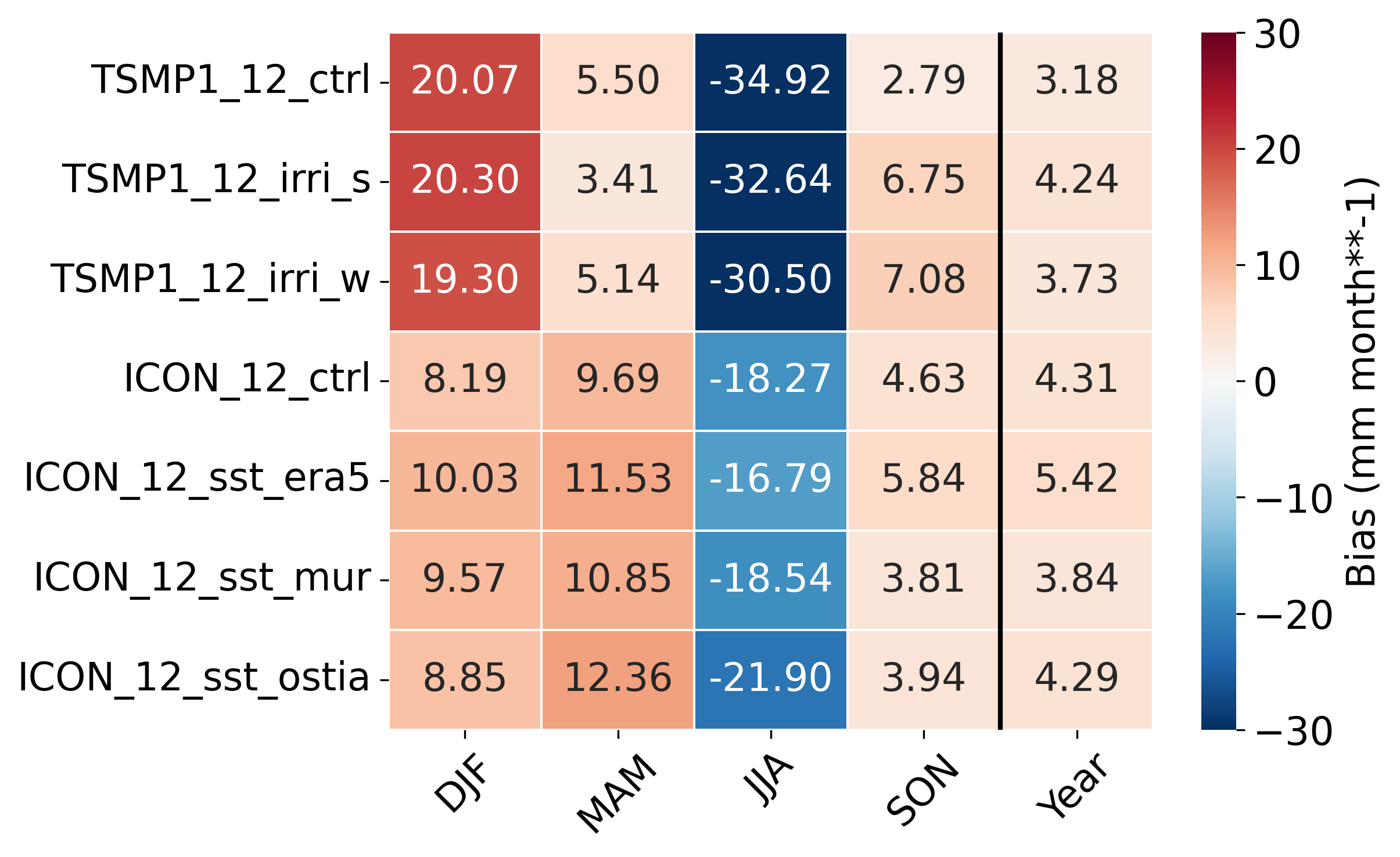}
  \label{pp_Tisa}
\end{subfigure}\hspace{1mm} 
  \caption{Mean seasonal precipitation biases for different watersheds for the period 2003-2010. The column "Year", shows the yearly MAE for the whole period.}
    \label{PP_SEASDIFF_2003_mean}
\end{figure}

\begin{figure}[H]
\centering
\begin{subfigure}{0.45\textwidth}
  \caption{Ebro}
  \includegraphics[width=\linewidth]{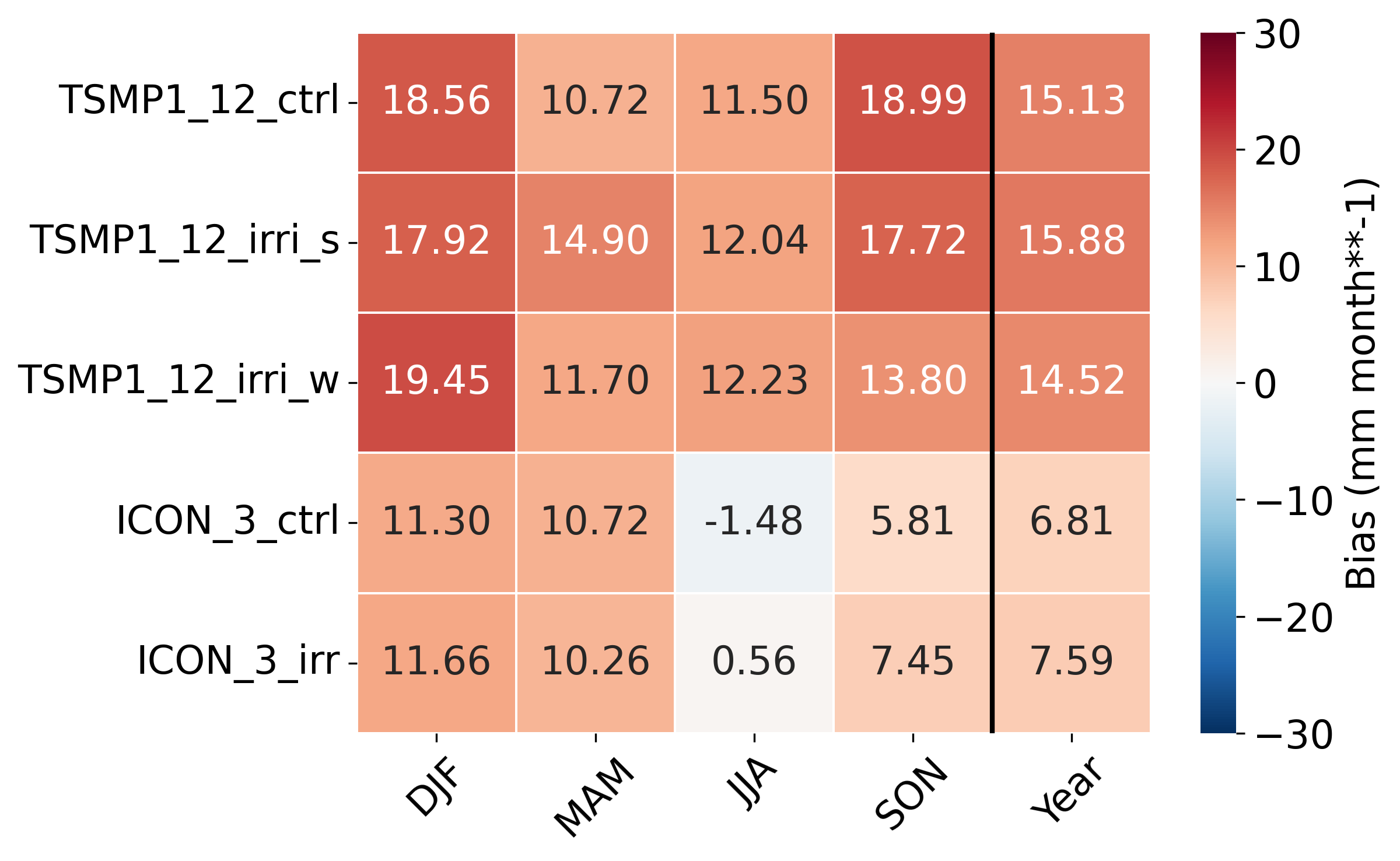}
  \label{pp_Ebro_11}
\end{subfigure}\hspace{1mm} 
\begin{subfigure}{0.45\textwidth}
  \caption{Po}
  \includegraphics[width=\linewidth]{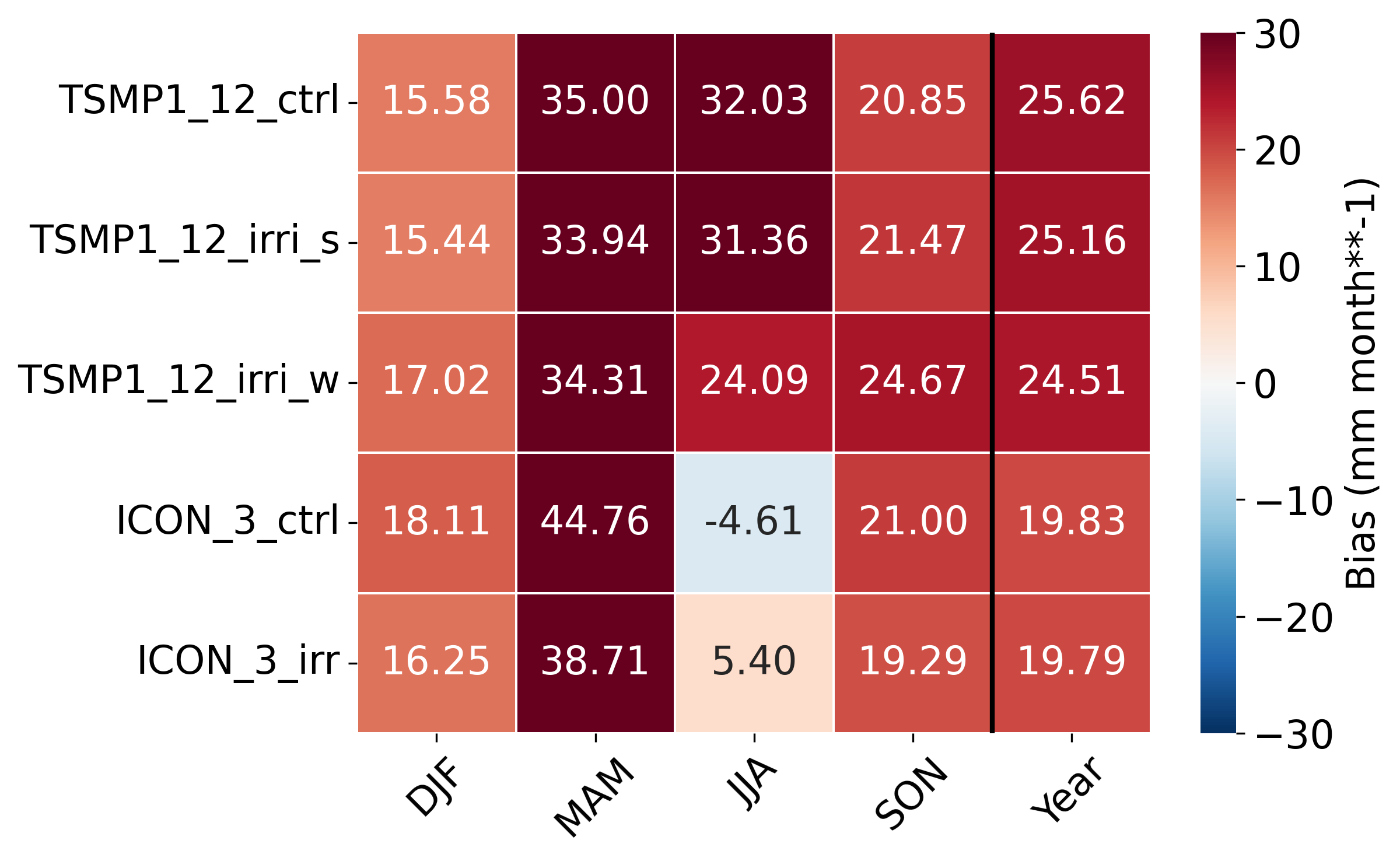}
  \label{pp_Po_11}
\end{subfigure}
\vspace{-9mm}
\medskip
\begin{subfigure}{0.45\textwidth}
  \vspace{-5mm}
  \caption{Rhine}
  \includegraphics[width=\linewidth]{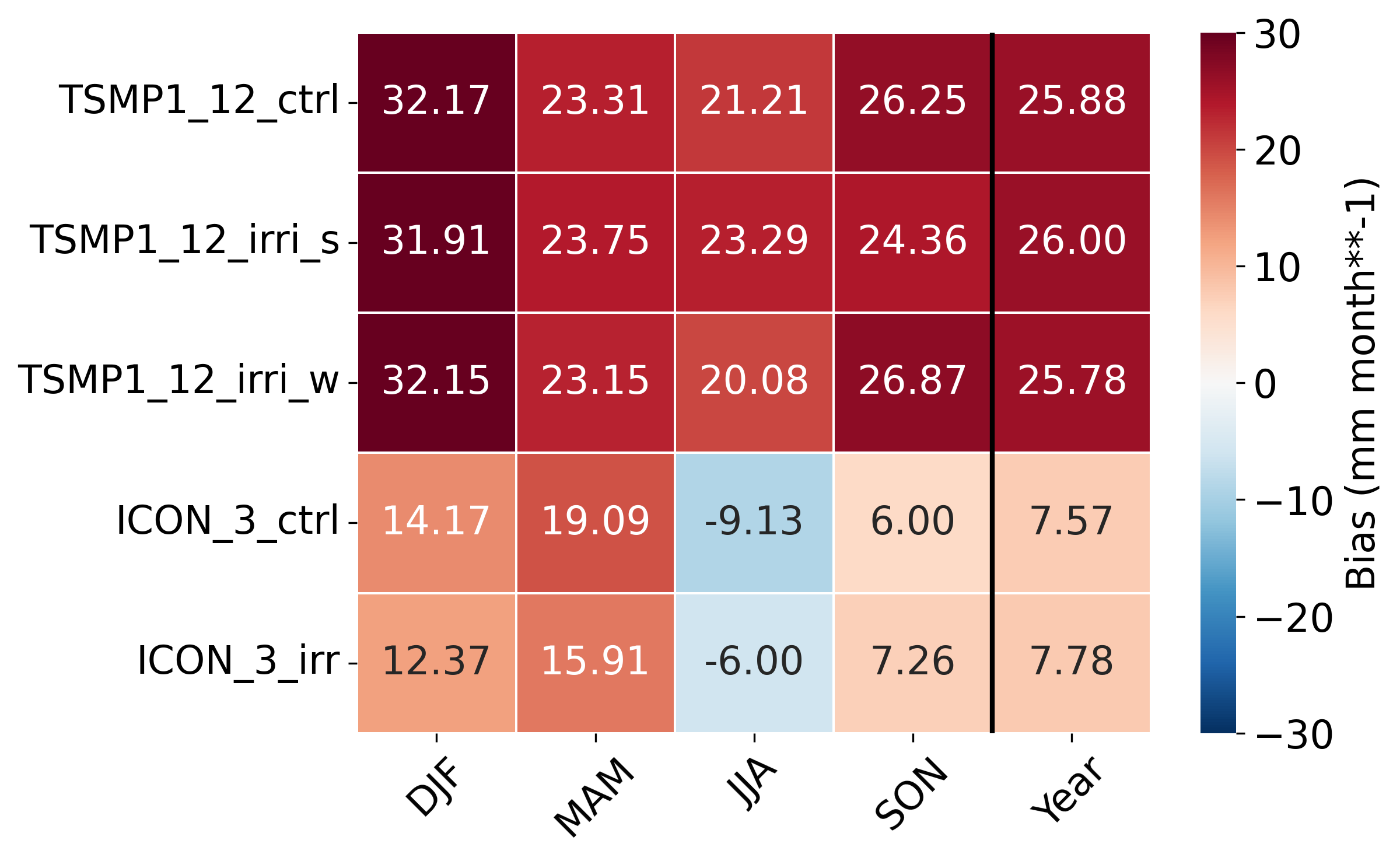}
  \label{pp_Rhein_11}
\end{subfigure}\hspace{1mm} 
\begin{subfigure}{0.45\textwidth}
  \vspace{-5mm}
  \caption{Tisa}
  \includegraphics[width=\linewidth]{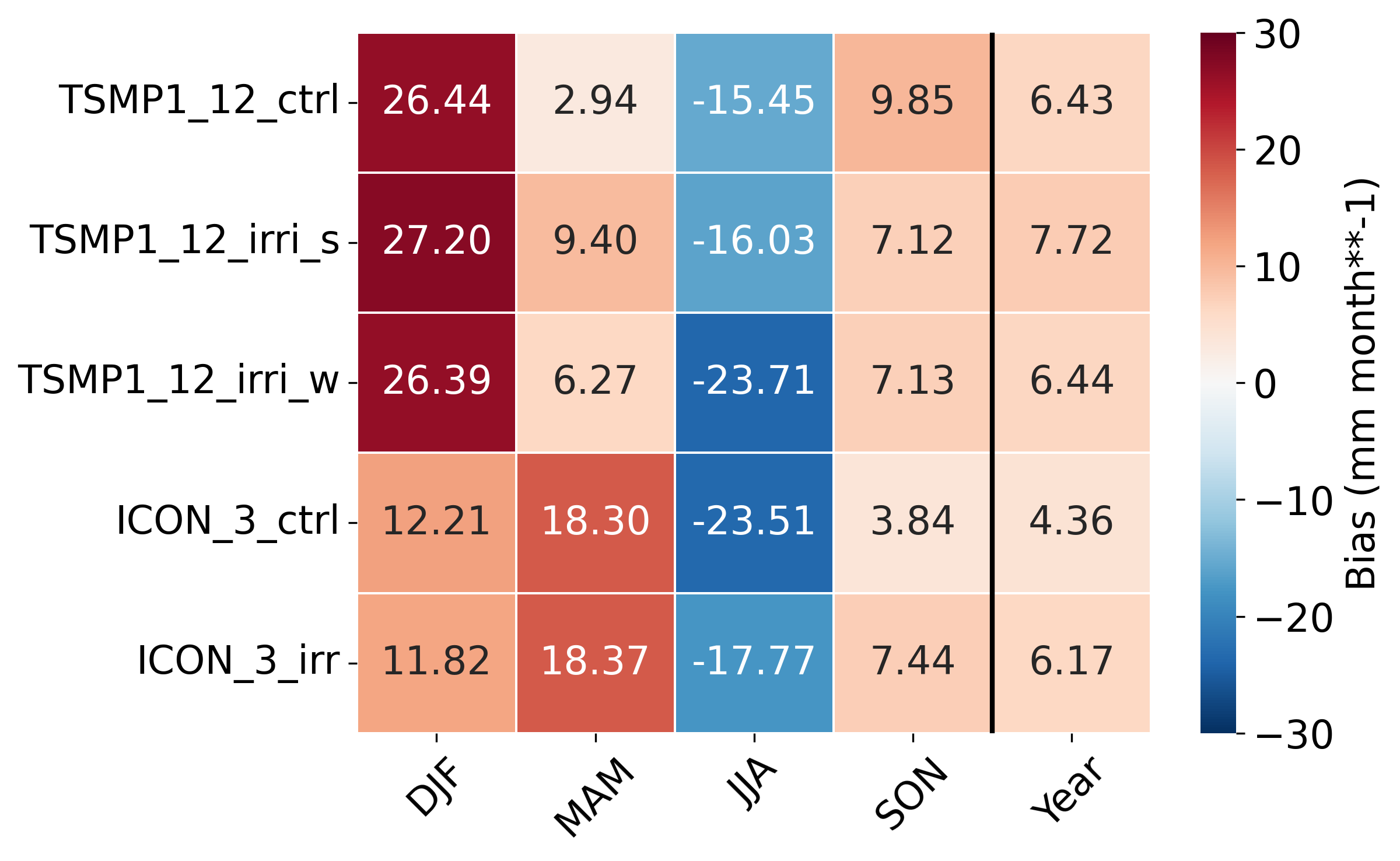}
  \label{pp_Tisa_11}
\end{subfigure}\hspace{1mm} 
  \caption{Mean seasonal precipitation biases for different watersheds for the period 2011-2020. The column "Year", shows the yearly MAE for the whole period.}
    \label{PP_SEASDIFF_2011_mean}
\end{figure}

\begin{figure}[H]
\centering
\begin{subfigure}{0.45\textwidth}
  \caption{Ebro}
  \includegraphics[width=\linewidth]{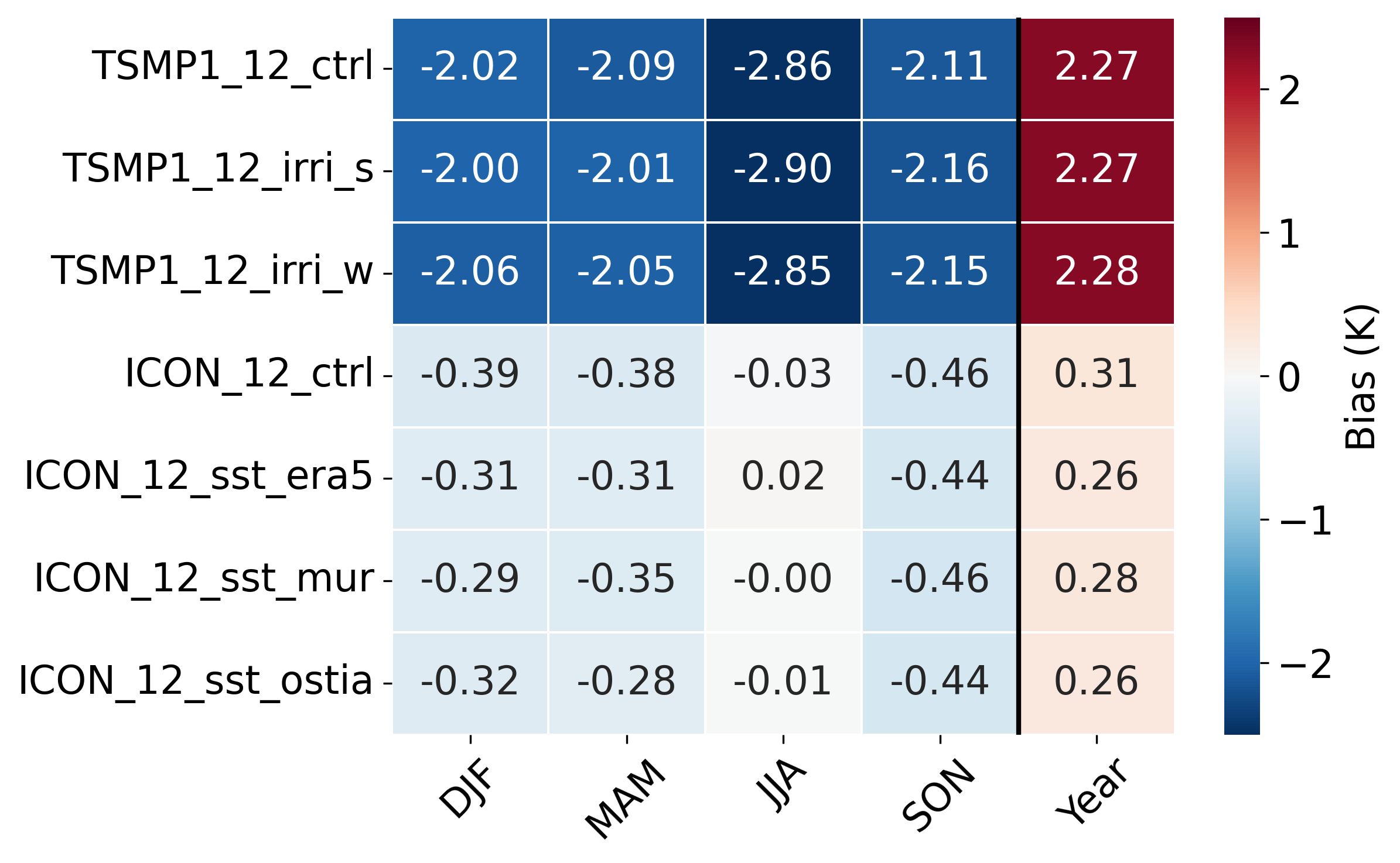}
  \label{tm2_Ebro}
\end{subfigure}\hspace{1mm} 
\begin{subfigure}{0.45\textwidth}
  \caption{Po}
  \includegraphics[width=\linewidth]{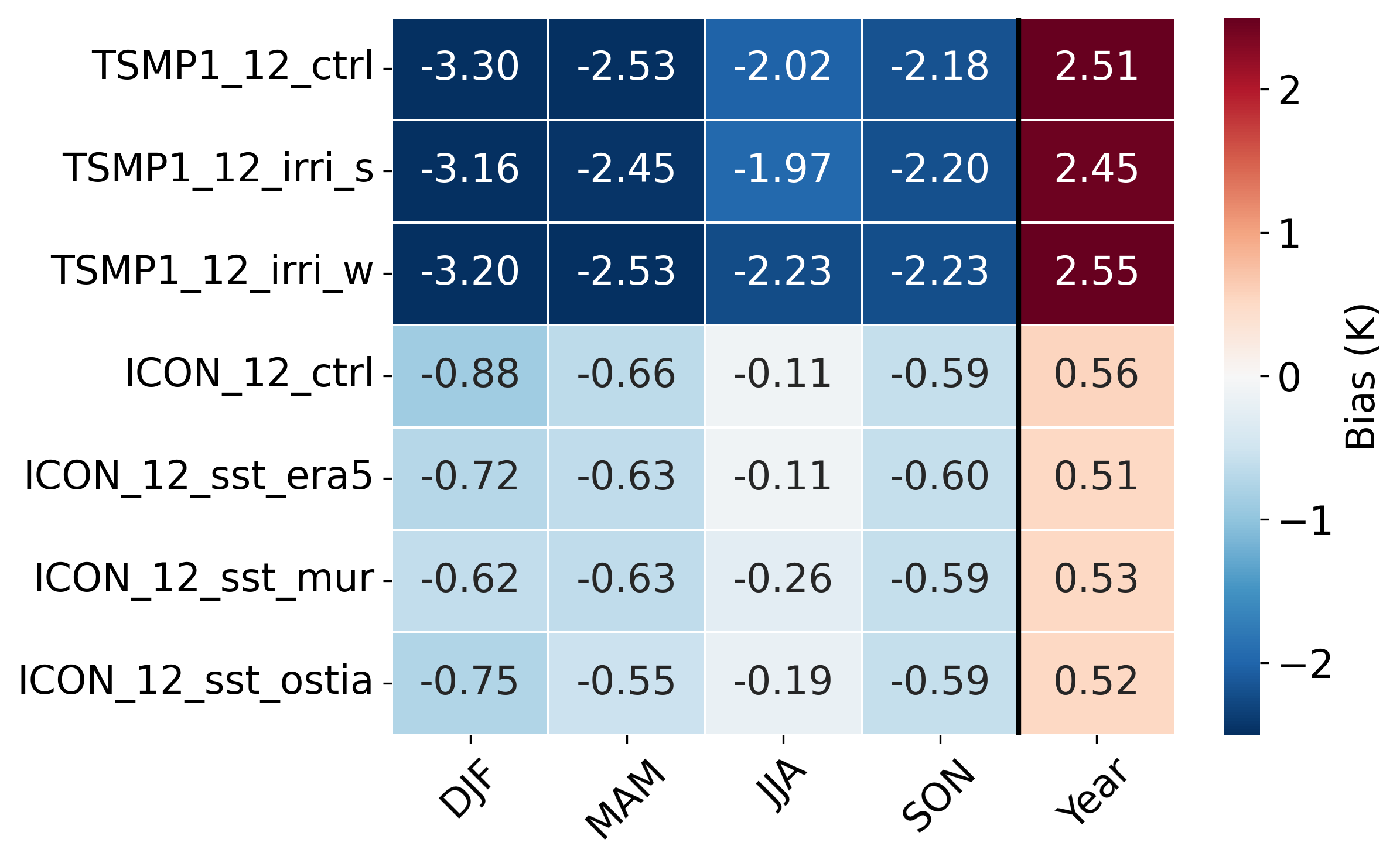}
  \label{tm2_Po}
\end{subfigure}
\vspace{-9mm}
\medskip
\begin{subfigure}{0.45\textwidth}
  \vspace{-5mm}
  \caption{Rhine}
  \includegraphics[width=\linewidth]{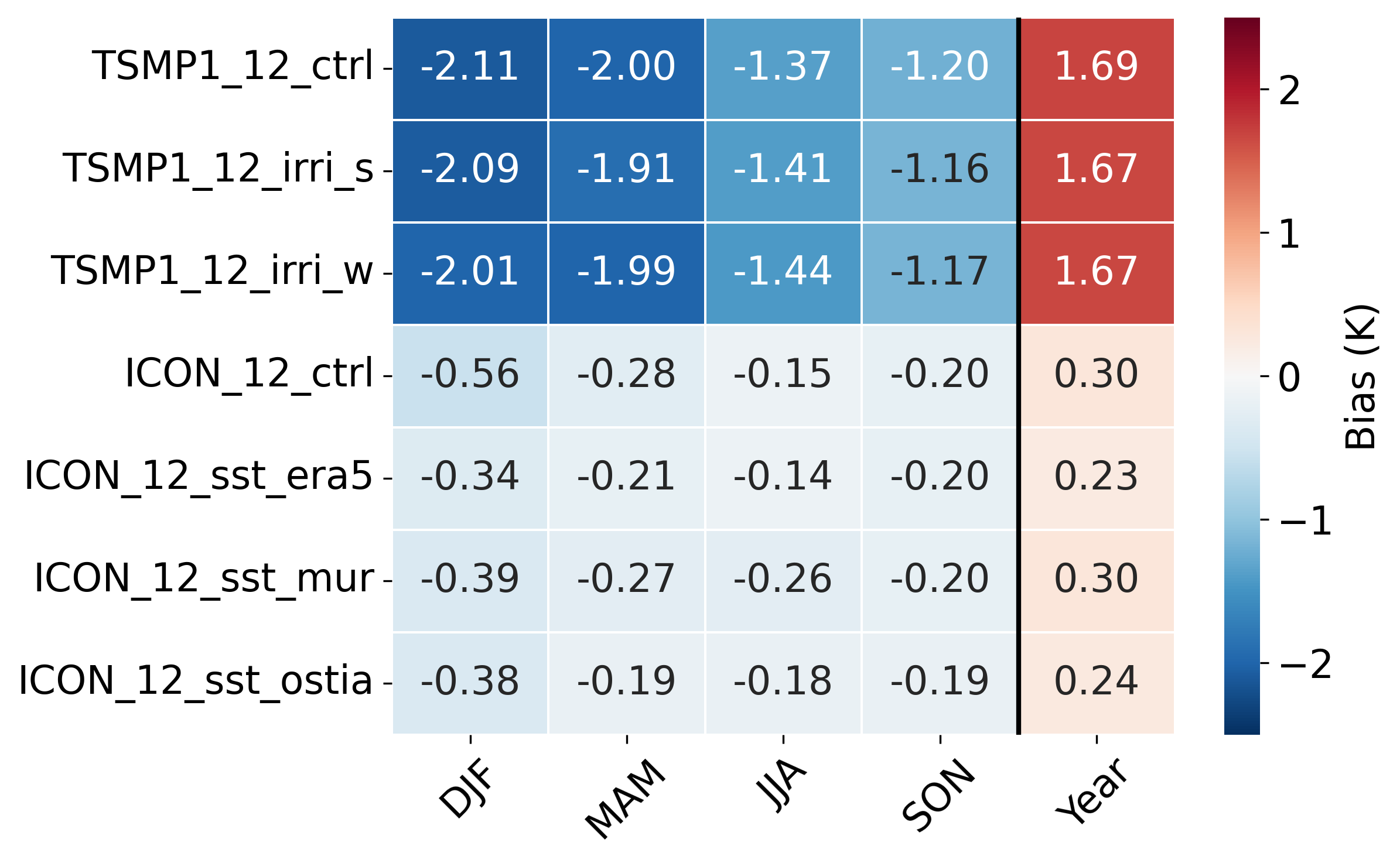}
  \label{tm2_Rhein}
\end{subfigure}\hspace{1mm} 
\begin{subfigure}{0.45\textwidth}
  \vspace{-5mm}
  \caption{Tisa}
  \includegraphics[width=\linewidth]{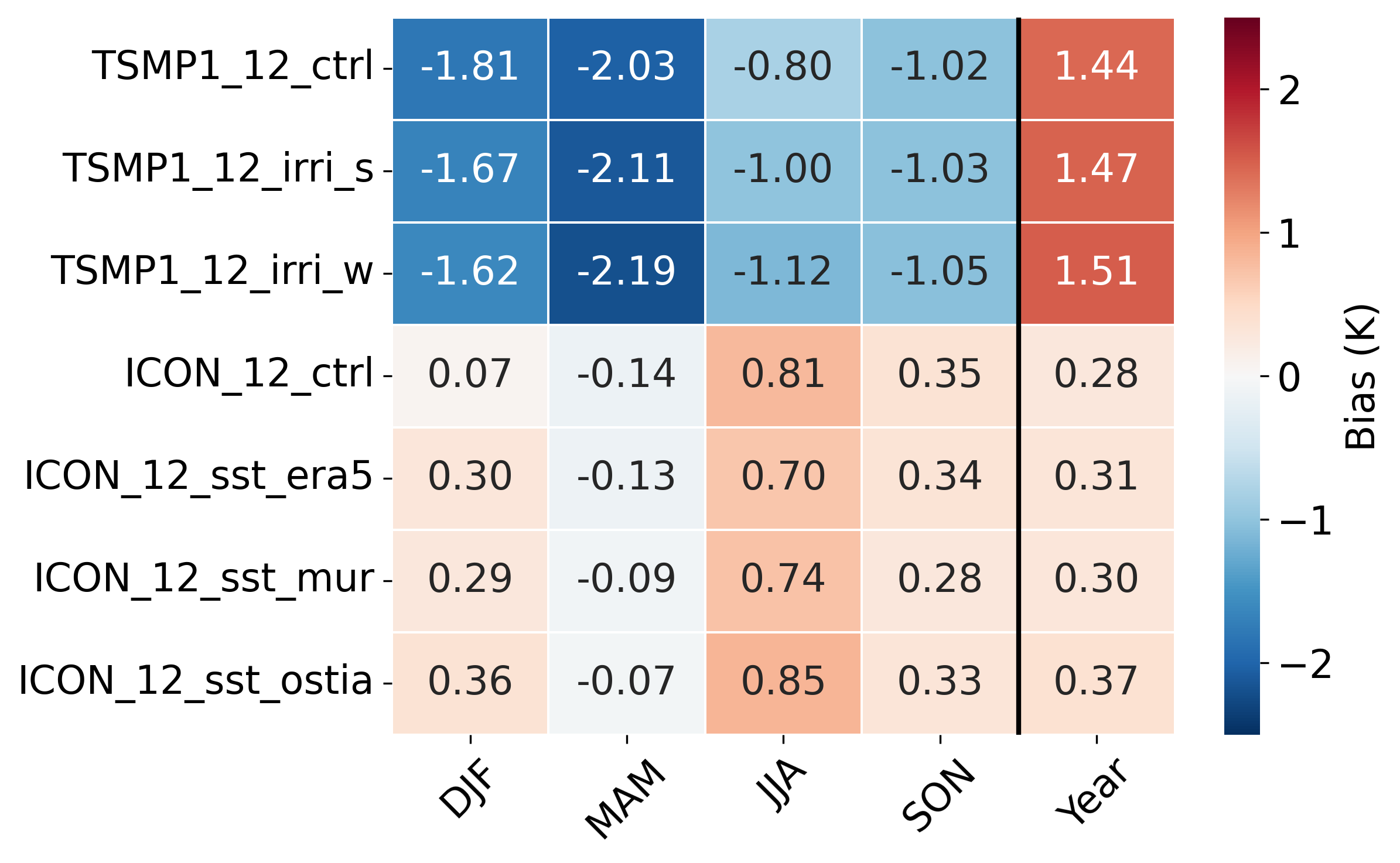}
  \label{tm2_Tisa}
\end{subfigure}\hspace{1mm} 
  \caption{Mean seasonal T2m for different watersheds for the period 2003-2010. The column "Year", shows the yearly MAE for the whole period.}
    \label{T2m_SEASDIFF_2003_mean}
\end{figure}

\begin{figure}[H]
\centering
\begin{subfigure}{0.45\textwidth}
  \caption{Ebro}
  \includegraphics[width=\linewidth]{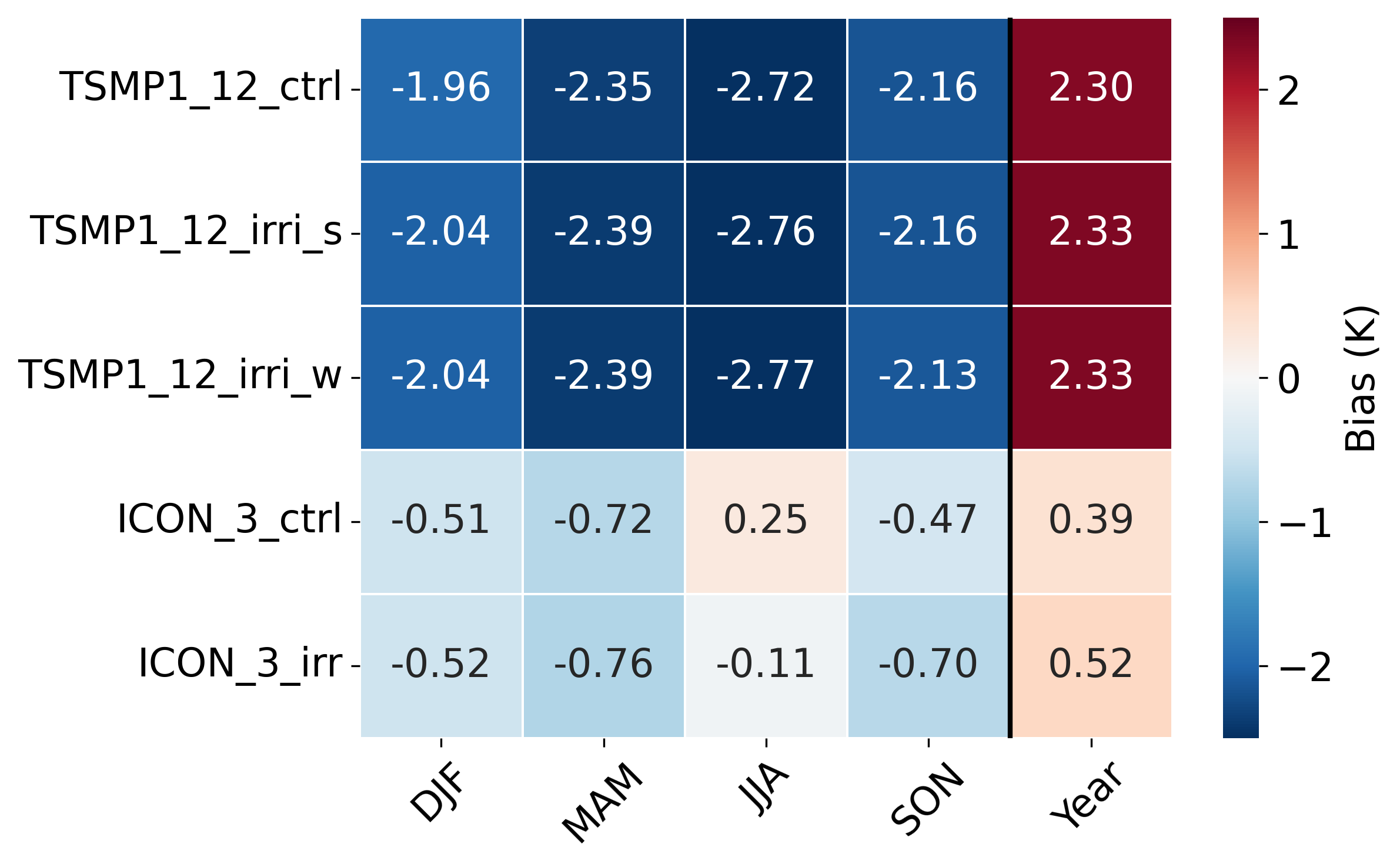}
  \label{tm2_Ebro_11}
\end{subfigure}\hspace{1mm} 
\begin{subfigure}{0.45\textwidth}
  \caption{Po}
  \includegraphics[width=\linewidth]{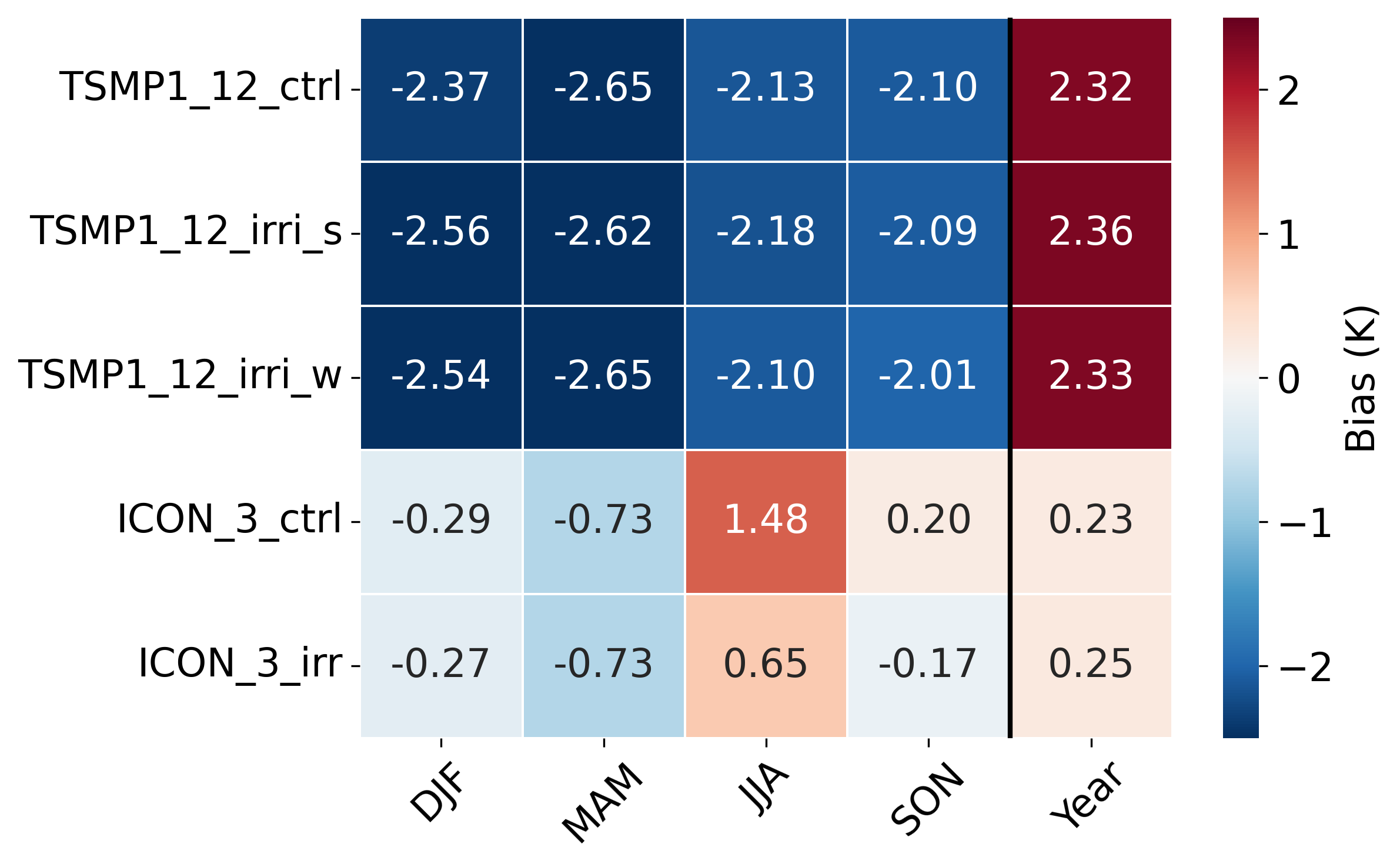}
  \label{tm2_Po_11}
\end{subfigure}
\vspace{-9mm}
\medskip
\begin{subfigure}{0.45\textwidth}
  \vspace{-5mm}
  \caption{Rhine}
  \includegraphics[width=\linewidth]{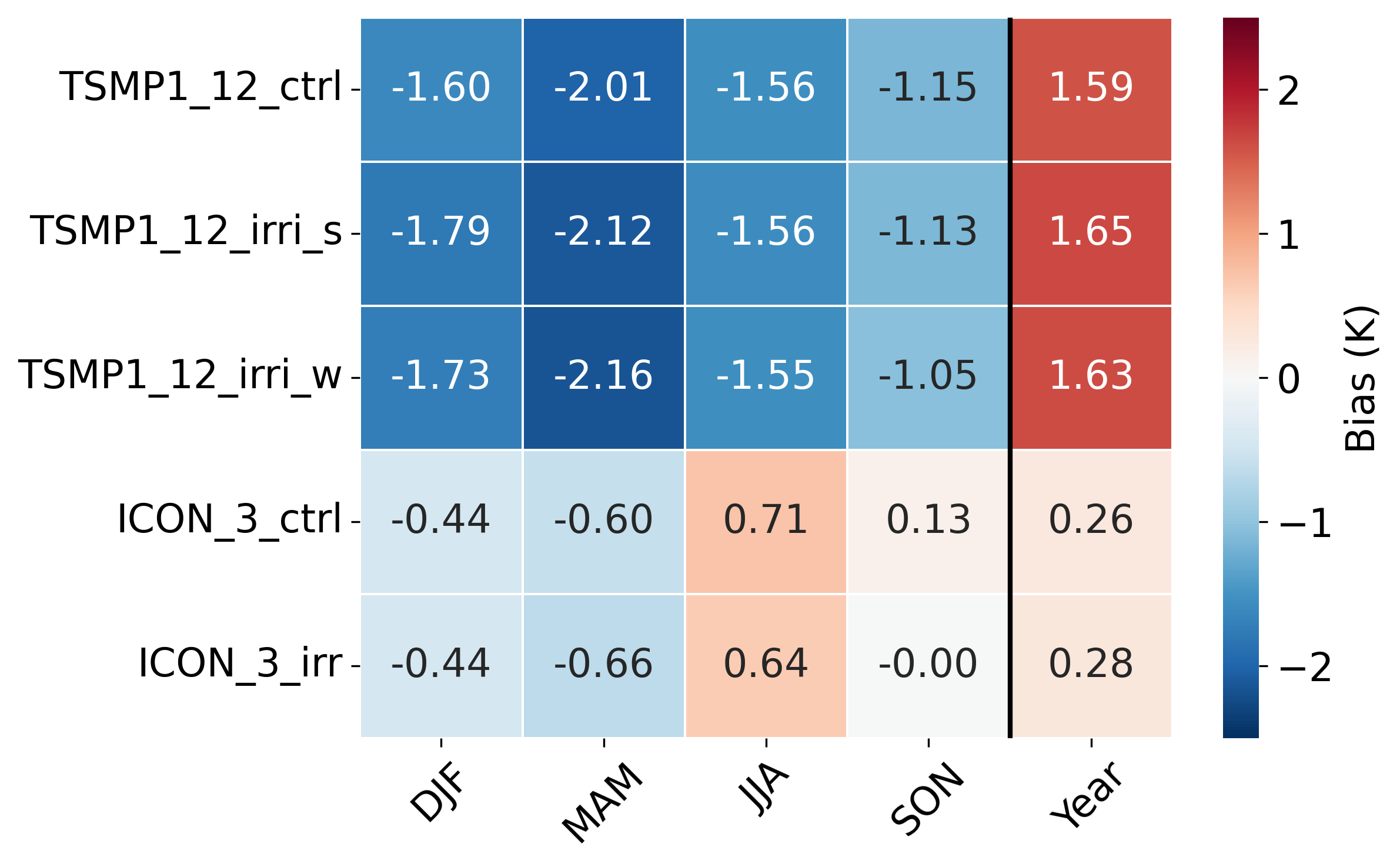}
  \label{tm2_Rhein_11}
\end{subfigure}\hspace{1mm} 
\begin{subfigure}{0.45\textwidth}
  \vspace{-5mm}
  \caption{Tisa}
  \includegraphics[width=\linewidth]{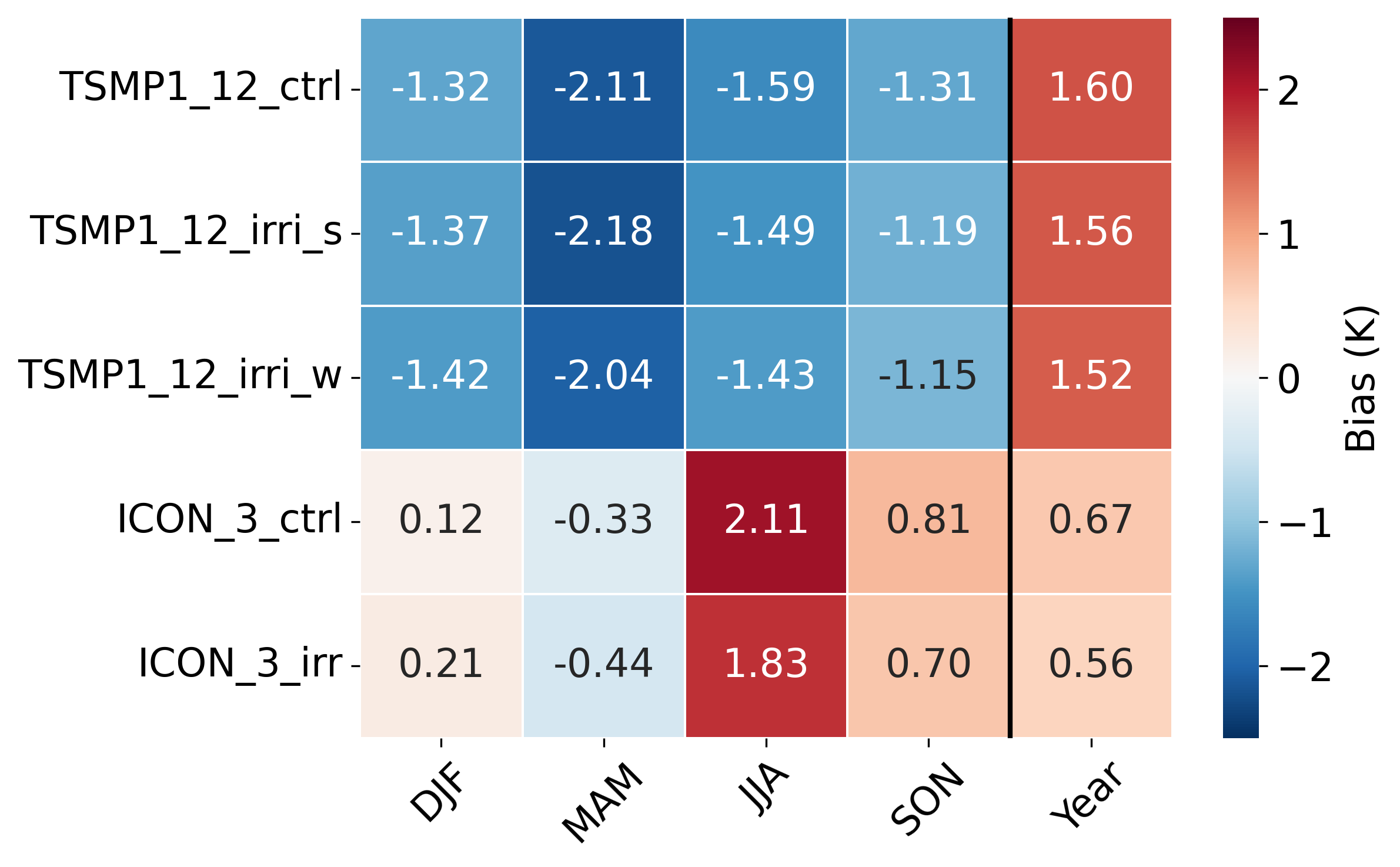}
  \label{tm2_Tisa_11}
\end{subfigure}\hspace{1mm} 
  \caption{Mean seasonal T2m for different watersheds for the period 2011-2020. The column "Year", shows the yearly MAE for the whole period.}
    \label{T2m_SEASDIFF_2011_mean}
\end{figure}


\begin{figure}[H]
\centering
\begin{subfigure}{0.92\textwidth}
    \caption{Ebro}
    \includegraphics[width=0.9\linewidth]{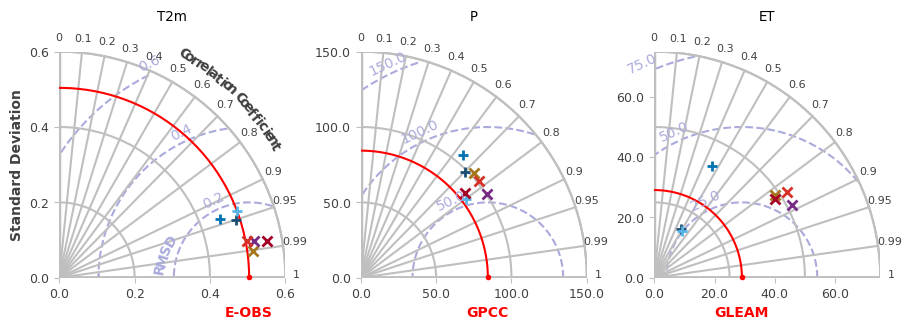}
    \label{Taylor_Ebro}
\end{subfigure}
\vfill
\begin{subfigure}{0.92\textwidth}
    \caption{Po}
    \includegraphics[width=0.9\linewidth]{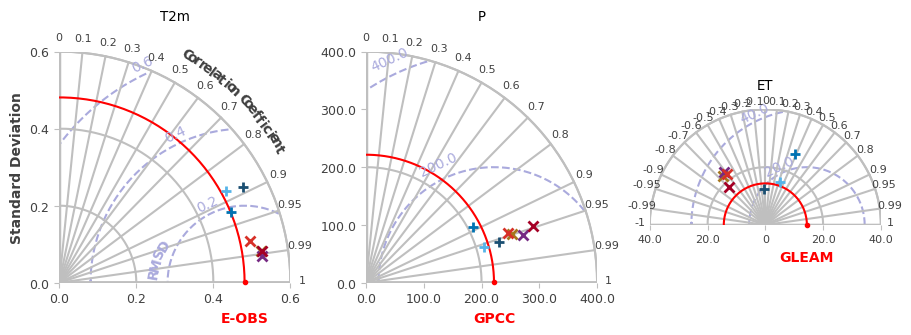}
    \label{Taylor_Po}
\end{subfigure}
\vfill
\begin{subfigure}{0.92\textwidth}
    \caption{Rhine}
    \includegraphics[width=0.9\linewidth]{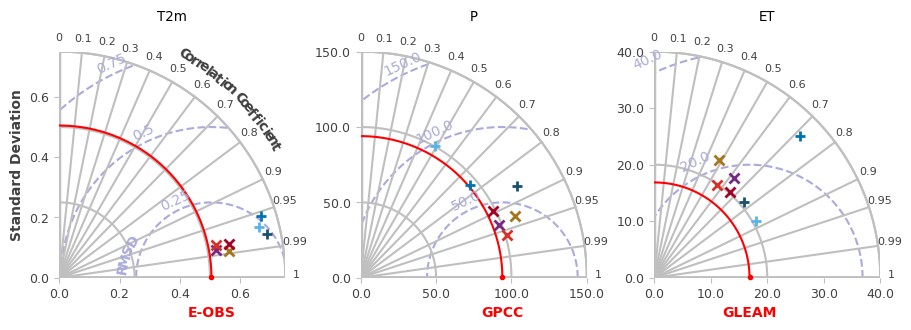}
    \label{Taylor_Rhein}
\end{subfigure}
\vfill
\begin{subfigure}{0.92\textwidth}
    \caption{Tisa}
    \includegraphics[width=0.9\linewidth]{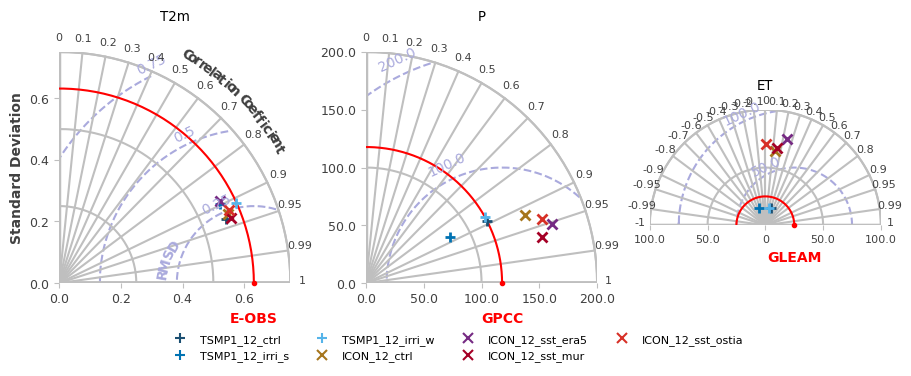}
    \label{Taylor_Tisa}
\end{subfigure}
    \caption{Taylor diagrams comparing simulation results to reference datasets for 2 m temperature (T2m), precipitation (P) and evapotranspiration (ET), over the period 2003–2010, for each basin. The simulations are compared against E-OBS (for T2m), GPCC (for P) and GLEAM (for ET)}
    \label{Taylor_stat_all_2003}
\end{figure}

\begin{figure}[H]
\centering
\begin{subfigure}{0.92\textwidth}
    \caption{Ebro}
    \includegraphics[width=0.9\linewidth]{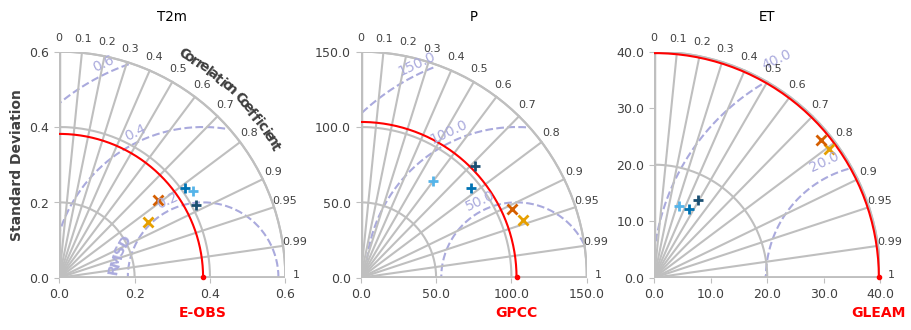}
    \label{Taylor_Ebro_2011}
\end{subfigure}
\vfill
\begin{subfigure}{0.92\textwidth}
    \caption{Po}
    \includegraphics[width=0.9\linewidth]{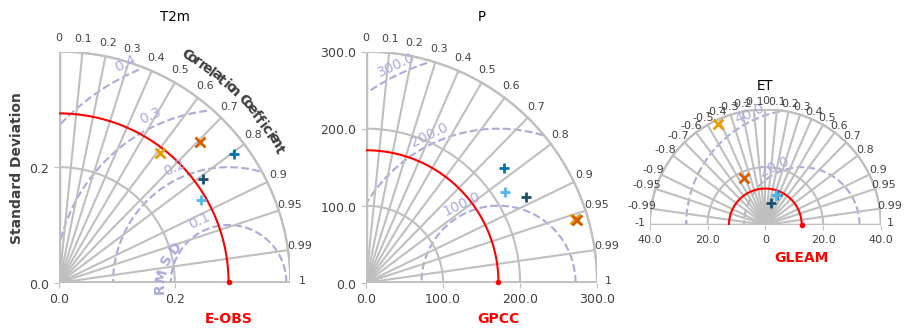}
    \label{Taylor_Po_2011}
\end{subfigure}
\vfill
\begin{subfigure}{0.92\textwidth}
    \caption{Rhine}
    \includegraphics[width=0.9\linewidth]{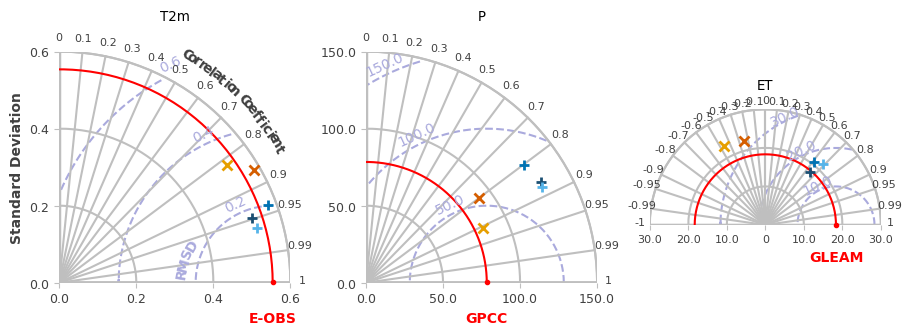}
    \label{Taylor_Rhein_2011}
\end{subfigure}
\vfill
\begin{subfigure}{0.92\textwidth}
    \caption{Tisa}
    \includegraphics[width=0.9\linewidth]{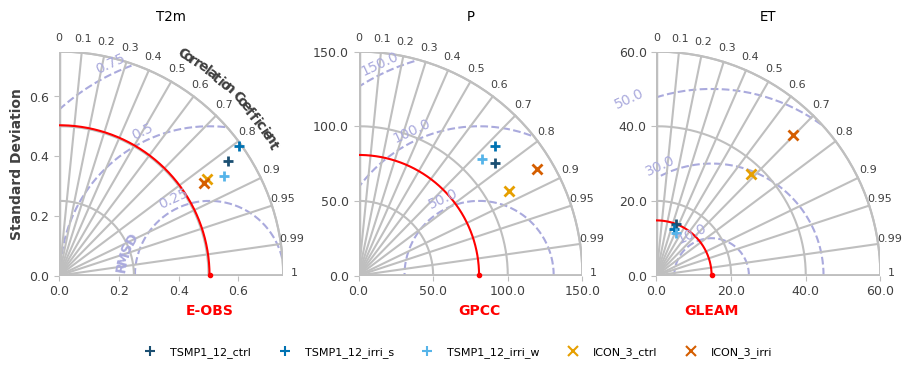}
    \label{Taylor_Tisa_2011}
\end{subfigure}
    \caption{Taylor diagrams comparing simulation results to reference datasets for 2 m temperature (T2m), precipitation (P) and evapotranspiration (ET), over the period 2011–2020, for each basin. The simulations are compared against E-OBS (for T2m), GPCC (for P) and GLEAM (for ET)}
    \label{Taylor_stat_all_2011}
\end{figure}

\begin{figure}[H]
\centering
\begin{subfigure}{0.9\textwidth}
    \caption{Ebro}
    \includegraphics[width=0.9\linewidth]{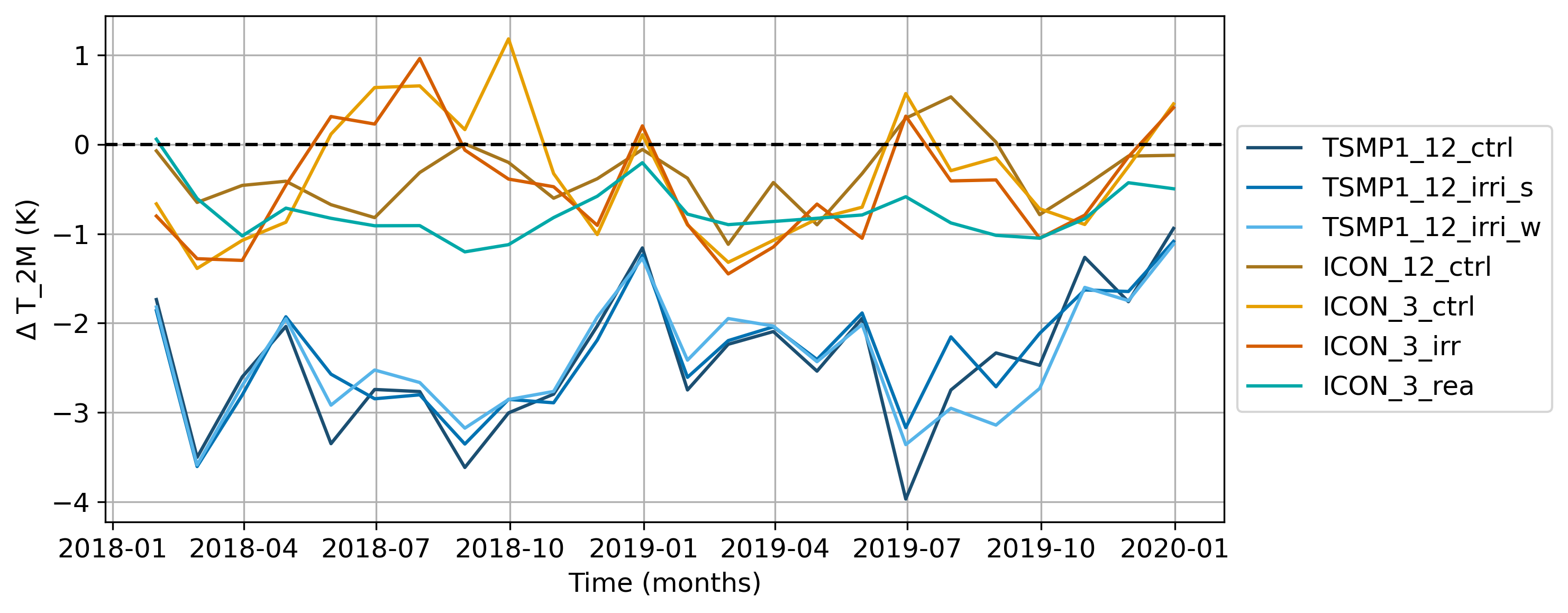}
    \label{18_T_2M_bias_Ebro}
\end{subfigure}
\vfill
\begin{subfigure}{0.9\textwidth}
    \caption{Po}
    \includegraphics[width=0.9\linewidth]{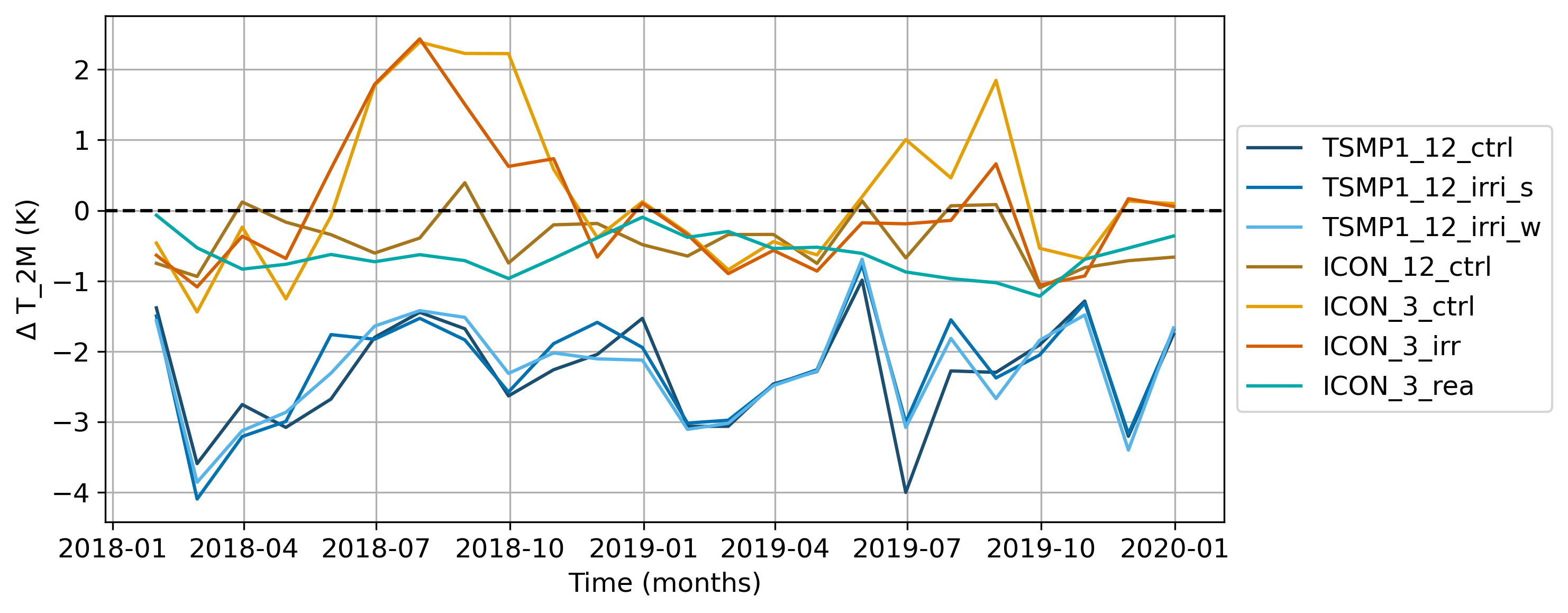}
    \label{18_T_2M_bias_Po}
\end{subfigure}
\vfill
\begin{subfigure}{0.9\textwidth}
    \caption{Rhine}
    \includegraphics[width=0.9\linewidth]{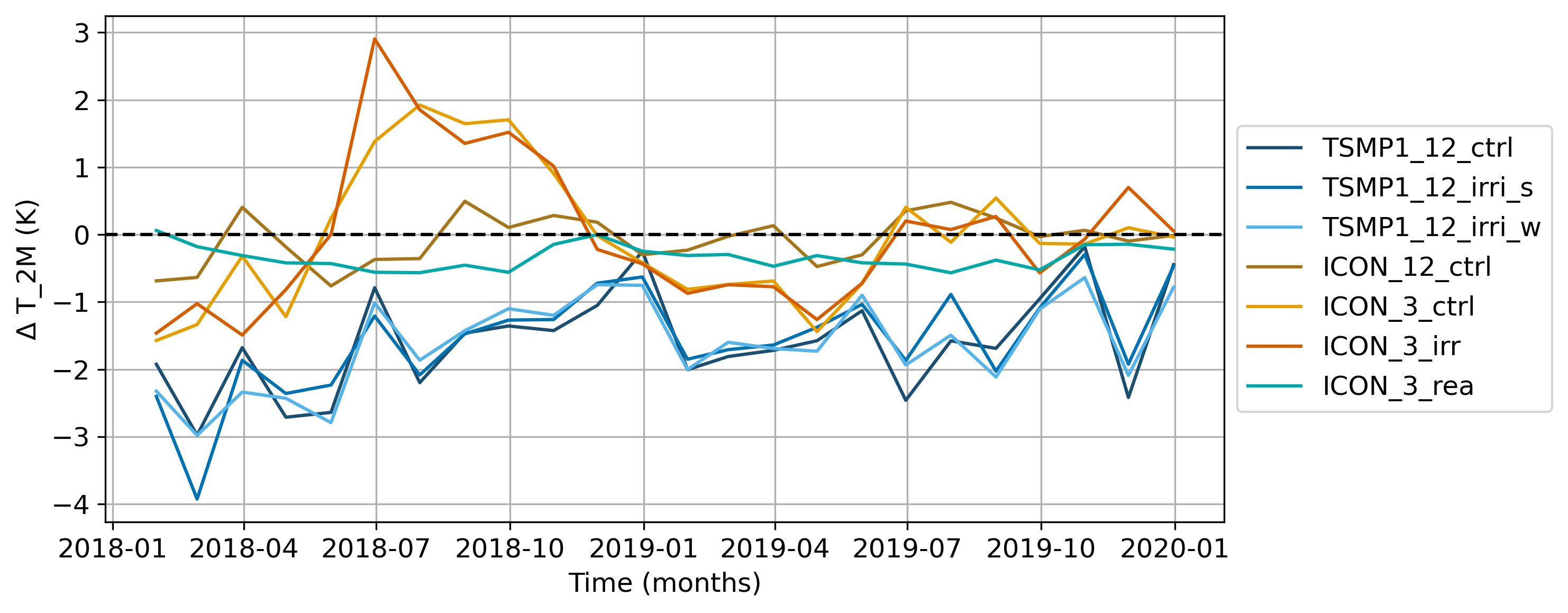}
    \label{18_T_2M_bias_Rhine}
\end{subfigure}
\begin{subfigure}{0.9\textwidth}
    \caption{Tisa}
    \includegraphics[width=0.9\linewidth]{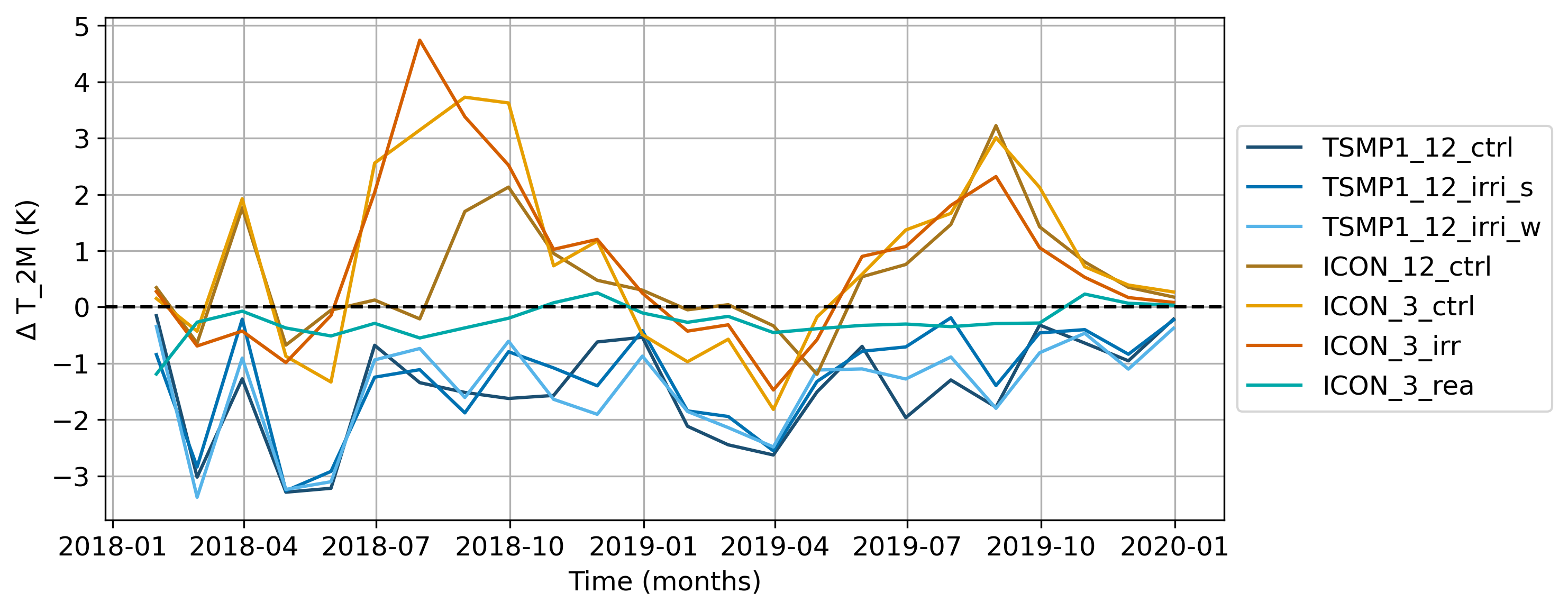}
    \label{18_T_2M_bias_Tisa}
\end{subfigure}
    \caption{Timeseries of monthly T2m biases (Sim. - E-OBS) for representative watersheds.}
    \label{Time_T_2M_bias_2018}
\end{figure}

\begin{figure}[H]
\begin{subfigure}{0.9\textwidth}
    \caption{Ebro}
    \includegraphics[width=0.9\linewidth]{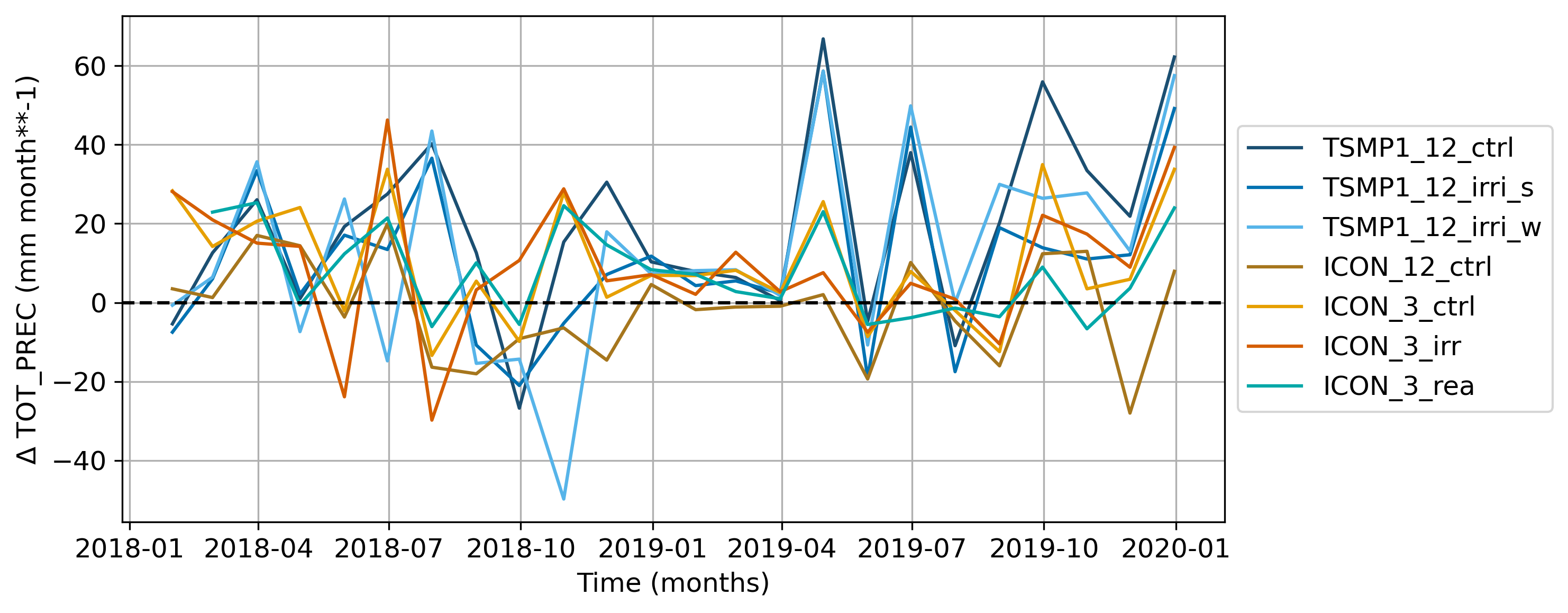}
    \label{18_PP_bias_Ebro}
\end{subfigure}
\vfill
\begin{subfigure}{0.9\textwidth}
    \caption{Po}
    \includegraphics[width=0.9\linewidth]{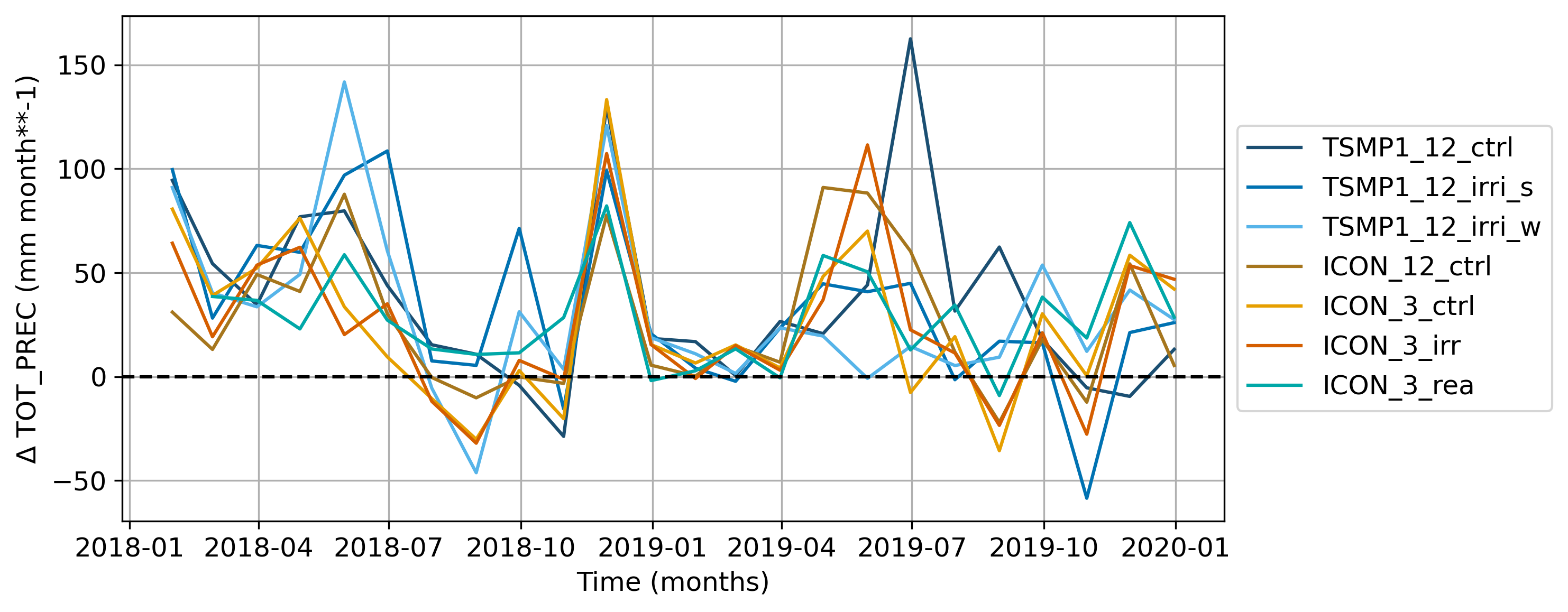}
    \label{18_PP_bias_Po}
\end{subfigure}
\vfill
\begin{subfigure}{0.9\textwidth}
    \caption{Rhine}
    \includegraphics[width=0.9\linewidth]{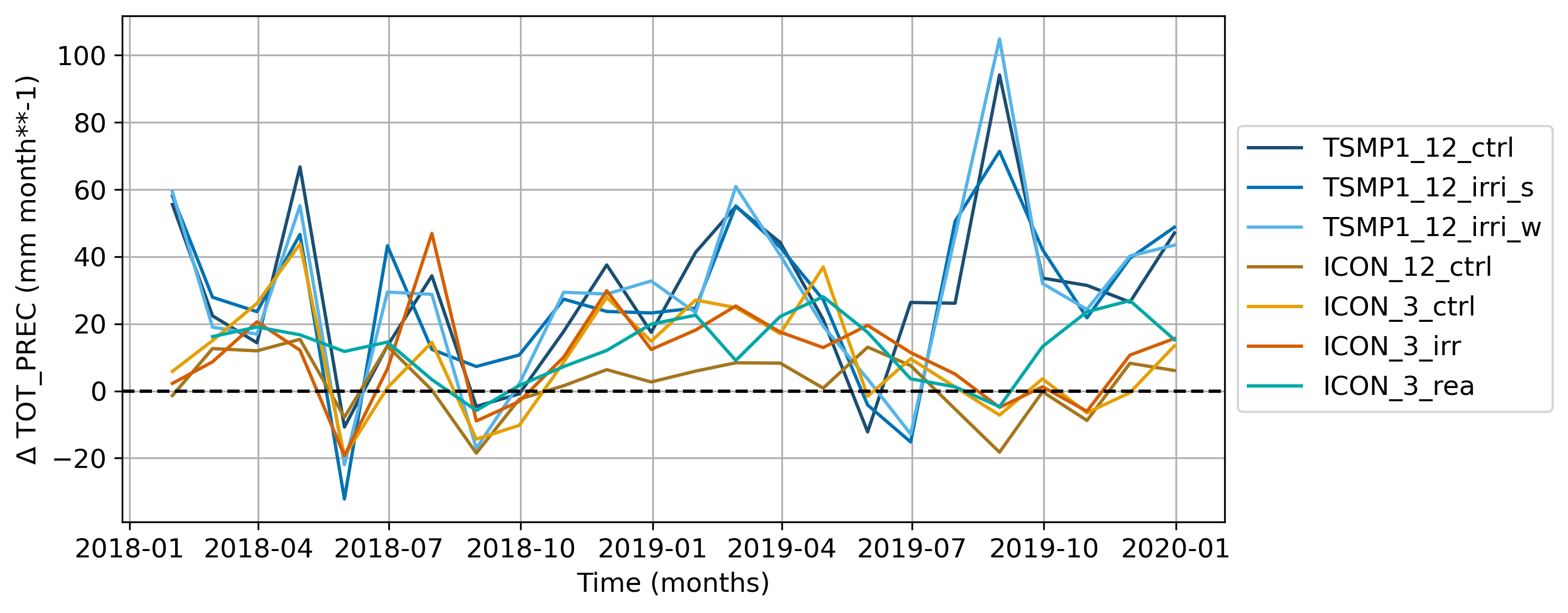}
    \label{18_PP_bias_Rhine}
\end{subfigure}
\vfill
\begin{subfigure}{0.9\textwidth}
    \caption{Tisa}
    \includegraphics[width=0.9\linewidth]{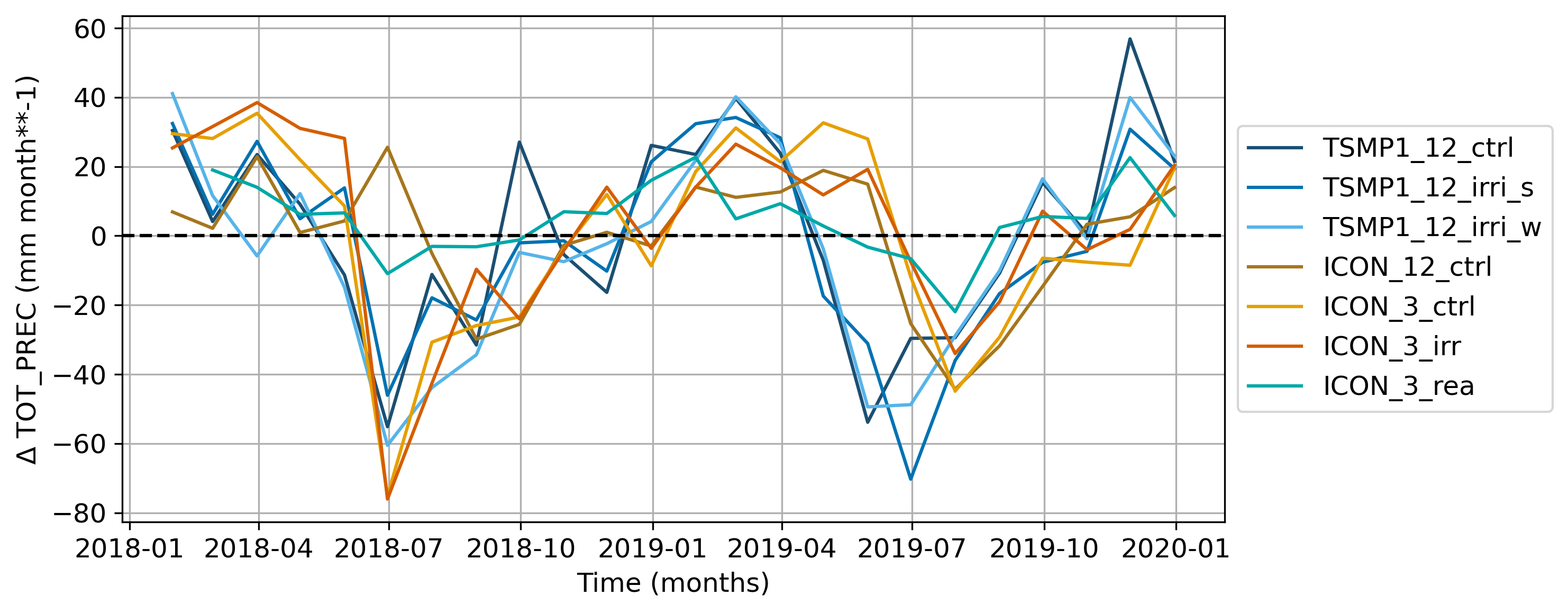}
    \label{18_PP_bias_Tisa}
\end{subfigure}
    \caption{Timeseries of monthly precipitation biases (Sim. - GPCC) for the representative watersheds.}
    \label{Time_PP_bias_2018}
\end{figure}

\begin{figure}[H]
\begin{subfigure}{0.9\textwidth}
    \caption{Ebro}
    \includegraphics[width=0.9\linewidth]{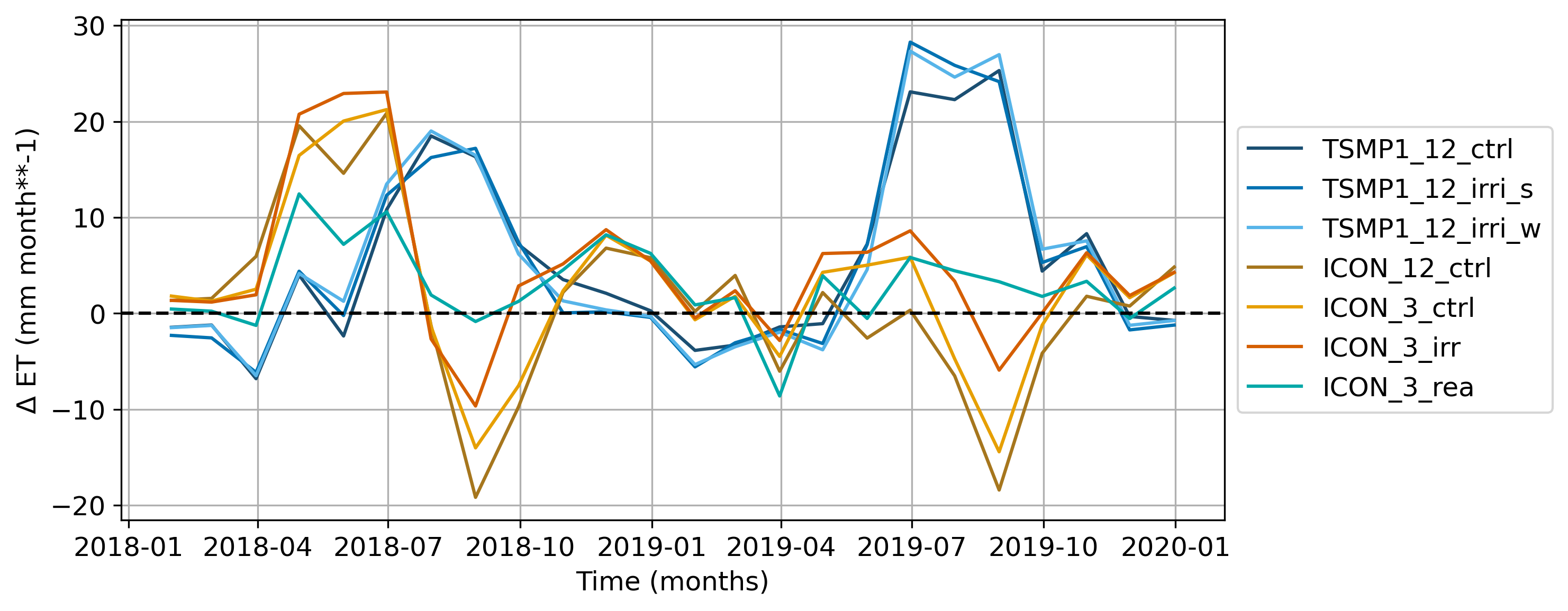}
    \label{18_ET_bias_Ebro}
\end{subfigure}
\vfill
\begin{subfigure}{0.9\textwidth}
    \caption{Po}
    \includegraphics[width=0.9\linewidth]{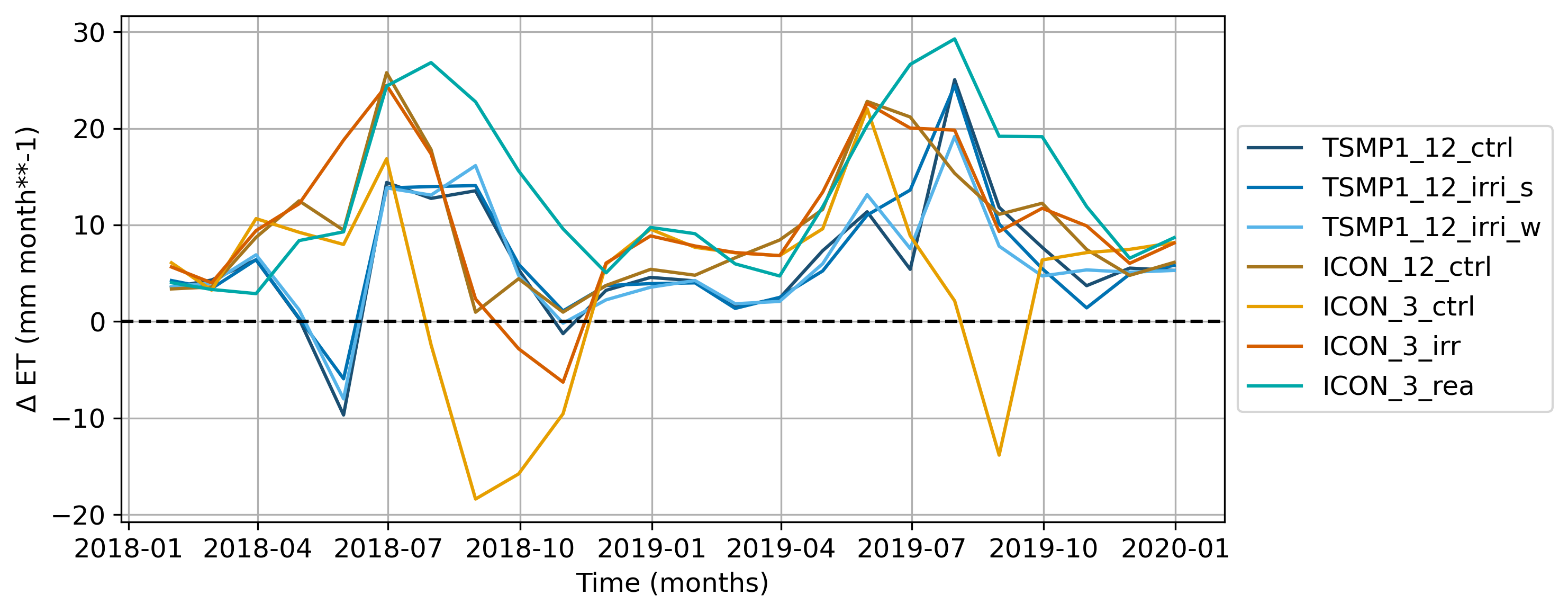}
    \label{18_ET_bias_Po}
\end{subfigure}
\vfill
\begin{subfigure}{0.9\textwidth}
    \caption{Rhine}
    \includegraphics[width=0.9\linewidth]{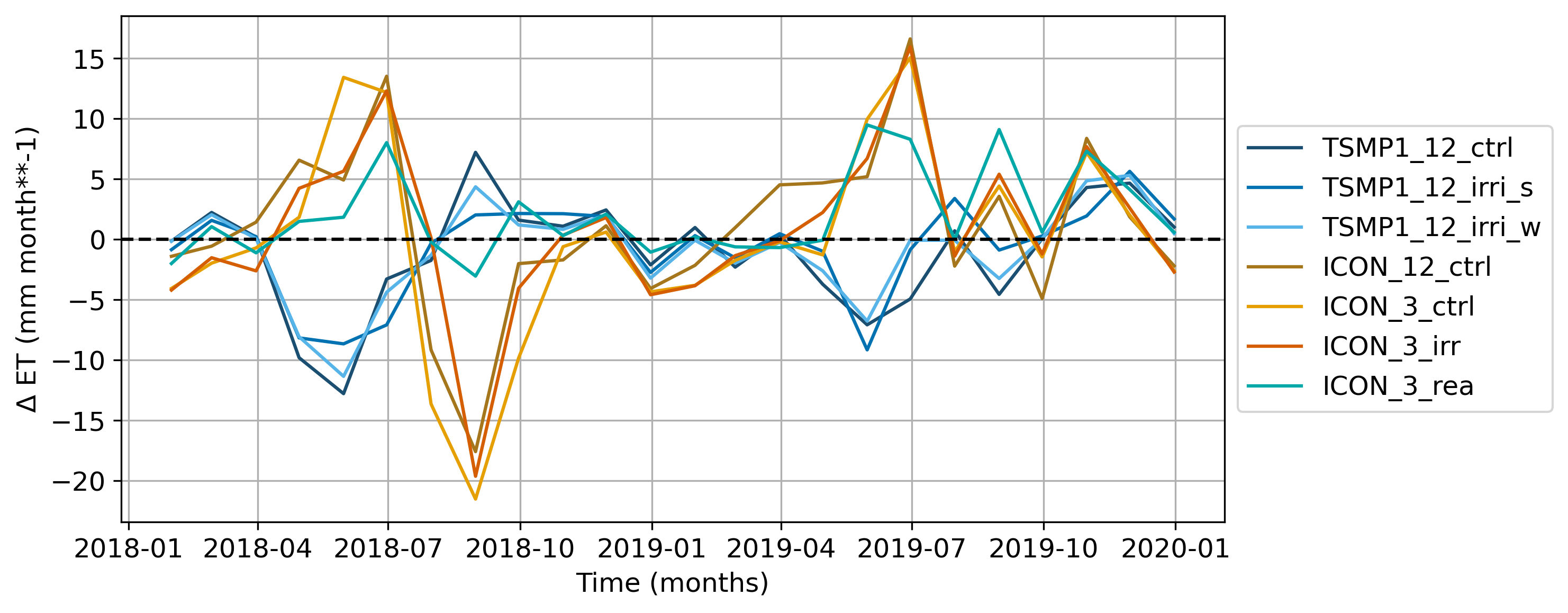}
    \label{18_ET_bias_Rhine}
\end{subfigure}
\vfill
\begin{subfigure}{0.9\textwidth}
    \caption{Tisa}
    \includegraphics[width=0.9\linewidth]{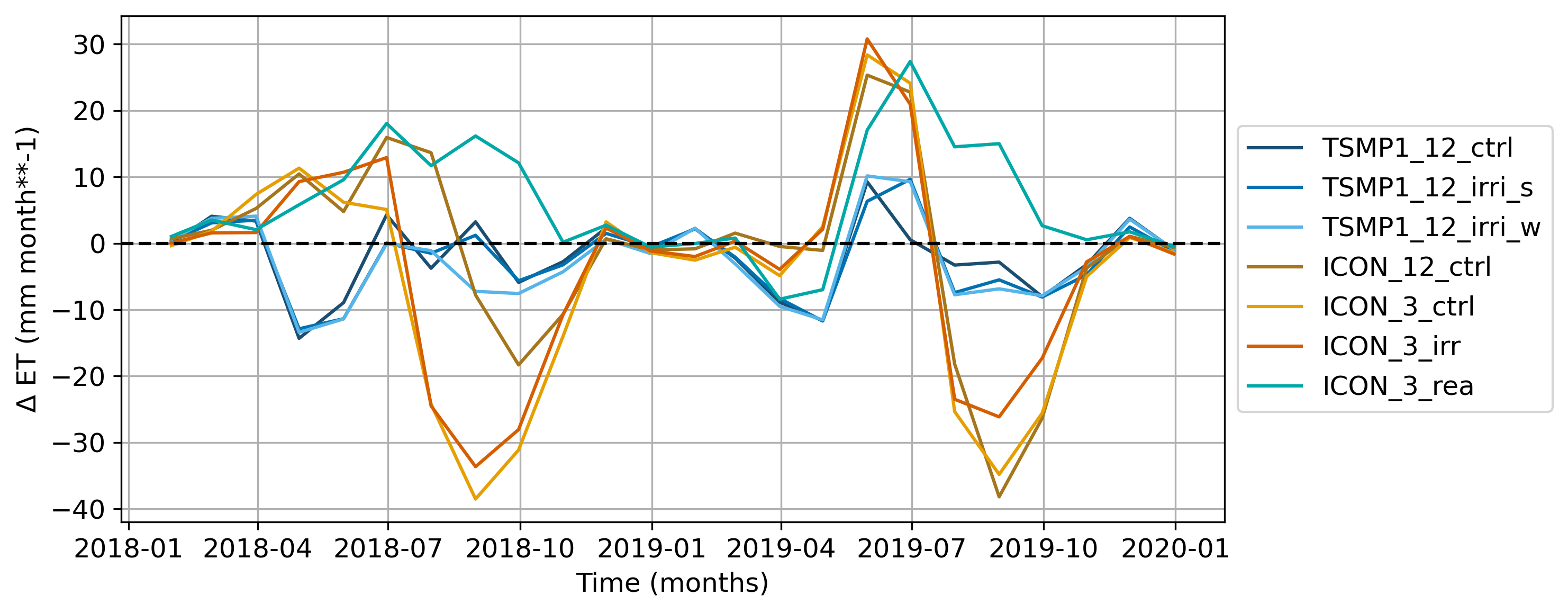}
    \label{18_ET_bias_Tisa}
\end{subfigure}
    \caption{Timeseries of monthly evapotranspiration biases (Sim. - GLEAM) for the representative watersheds.}
    \label{Time_ET_bias_2018}
\end{figure}


\begin{sidewaystable}
    \caption{Evaluation metrics for T2m (2018--2020).}
    \begin{threeparttable}
        \begin{tabular}{clllcccllllll}
        \headrow
        \thead{Simulations}&    \multicolumn{3}{c}{Ebro}&\multicolumn{3}{c}{Po}& \multicolumn{3}{c}{Rhein}& \multicolumn{3}{c}{Tisa}\\
        \headrow
        &    \thead{Bias} &\thead{STD} &\thead{RMSE} &\thead{Bias} & \thead{STD} & \thead{RMSE}  & \thead{Bias} & \thead{STD} &\thead{RMSE}   & \thead{Bias} & \thead{STD} &\thead{RMSE}   \\
        TSMP1\_12\_ctrl    &    -2.43&0.75&2.55&-2.31& 0.76& 2.43 & -1.60& 0.74&1.77 & -1.48& 0.92&1.74\\
        TSMP1\_12\_irri\_s &    -2.36&0.63&2.44&-2.23& 0.76& 2.35 & -1.57& 0.76&1.74 & -1.28& 0.88&1.55\\
        TSMP1\_12\_irri\_w &    -2.40&0.64&2.49&-2.27& 0.74& 2.39 & -1.63& 0.66&1.76 & -1.45& 0.87&1.69\\
        ICON\_12\_ctrl     &    -0.35&0.37&0.51&-0.42& 0.38& 0.56 & -0.07& 0.36&0.37 & 0.56& 0.98&1.13\\
        ICON\_3\_ctrl      &    -0.36&0.69&0.78&0.24& 1.09& 1.12 & -0.04& 0.98&0.98 & 0.85& 1.56&1.78\\
        ICON\_3\_irr       &    -0.47&0.62&0.78&0.003& 0.91& 0.91 & -0.02& 1.09&1.09 & 0.76& 1.44&1.63\\
        ICON\_3\_rea       &    -0.76&0.28&0.81&-0.63& 0.27& 0.68 & -0.34& 0.18&0.38 & -0.25& 0.29&0.38\\
        \hline  
    \end{tabular}
    \end{threeparttable}
    \label{table_T2m_metrics_2018}
\end{sidewaystable}

\begin{sidewaystable}
    \caption{Evaluation metrics for P (2018--2020).}
    \begin{threeparttable}
        \begin{tabular}{clllcccllllll}
        \headrow
        \thead{Simulations}&    \multicolumn{3}{c}{Ebro}&\multicolumn{3}{c}{Po}& \multicolumn{3}{c}{Rhein}& \multicolumn{3}{c}{Tisa}\\
        \headrow
        &    \thead{Bias} &\thead{STD} &\thead{RMSE} &\thead{Bias} & \thead{STD} & \thead{RMSE}  & \thead{Bias} & \thead{STD} &\thead{RMSE}   & \thead{Bias} & \thead{STD} &\thead{RMSE}   \\
        TSMP1\_12\_ctrl    &    19.19&22.43&29.52&37.82& 44.12& 58.11& 29.09& 24.17&37.82& 1.67& 28.10&28.15\\
        TSMP1\_12\_irri\_s &    11.12&20.63&23.43&34.24& 39.99& 52.65& 28.18& 23.45&36.67& -1.44& 27.24&27.28\\
        TSMP1\_12\_irri\_w &    12.72&25.49&28.48&31.61& 39.94& 50.93& 28.68& 26.93&39.34& -3.28& 28.72&28.91\\
        ICON\_12\_ctrl     &    -1.42&12.52&12.60&26.61& 33.61& 42.87& 2.49& 9.08&9.42& -1.01& 18.19&18.22\\
        ICON\_3\_ctrl      &    10.14&14.68&17.84&26.46& 37.85& 46.19& 9.68& 15.69&18.43& 0.44& 28.03&28.03\\
        ICON\_3\_irr       &    9.45&17.06&19.50&25.44& 36.43& 44.43& 10.19& 13.56&16.96& 2.69& 26.65&26.79\\
        ICON\_3\_rea       &    7.69&11.11&13.51&28.31& 23.61& 36.86& 12.84& 9.35&15.88& 4.62& 10.24&11.23\\
        \hline 
    \end{tabular}
    \end{threeparttable}
    \label{table_P_metrics_2018}
\end{sidewaystable}

\begin{sidewaystable}
    \caption{Evaluation metrics for ET (2018--2020).}
    \begin{threeparttable}
        \begin{tabular}{clllcccllllll}
        \headrow
        \thead{Simulations}&    \multicolumn{3}{c}{Ebro}&\multicolumn{3}{c}{Po}& \multicolumn{3}{c}{Rhein}& \multicolumn{3}{c}{Tisa}\\
        \headrow
        &    \thead{Bias} &\thead{STD} &\thead{RMSE} &\thead{Bias} & \thead{STD} & \thead{RMSE}  & \thead{Bias} & \thead{STD} &\thead{RMSE}   & \thead{Bias} & \thead{STD} &\thead{RMSE}   \\
        TSMP1\_12\_ctrl    &    5.45&9.11&10.62&6.20& 6.44& 8.94& -1.07& 4.45&4.57& -1.86& 5.55&5.86\\
        TSMP1\_12\_irri\_s &    5.31&9.99&11.31&6.46& 6.14& 8.91& -0.73& 3.80&3.86& -2.25& 5.67&6.10\\
        TSMP1\_12\_irri\_w &    5.58&10.08&11.52&5.98& 5.68& 8.24& -0.93& 3.83&3.94& -2.64& 6.22&6.76\\
        ICON\_12\_ctrl     &    1.04&9.31&9.37&9.55& 6.67& 11.65& 1.06& 6.82&6.90& -0.96& 14.08&14.11\\
        ICON\_3\_ctrl      &    2.49&8.50&8.86&4.29& 9.54& 10.47& -0.09& 8.03&8.03& -4.95& 16.92&17.63\\
        ICON\_3\_irr       &    4.64&7.89&9.16&10.11& 7.41& 12.54& 0.74& 6.67&6.71& -3.40& 15.40&15.77\\
        ICON\_3\_rea       &    2.88&4.25&5.14&13.13& 8.15& 15.46& 2.04& 3.62&4.16& 6.09& 8.51&10.47\\
        \hline 
    \end{tabular}
    \end{threeparttable}
    \label{table_ET_metrics_2018}
\end{sidewaystable}
\begin{figure}[H]
\centering
\begin{subfigure}{0.9\textwidth}
    \caption{Ebro}
    \includegraphics[width=0.9\linewidth]{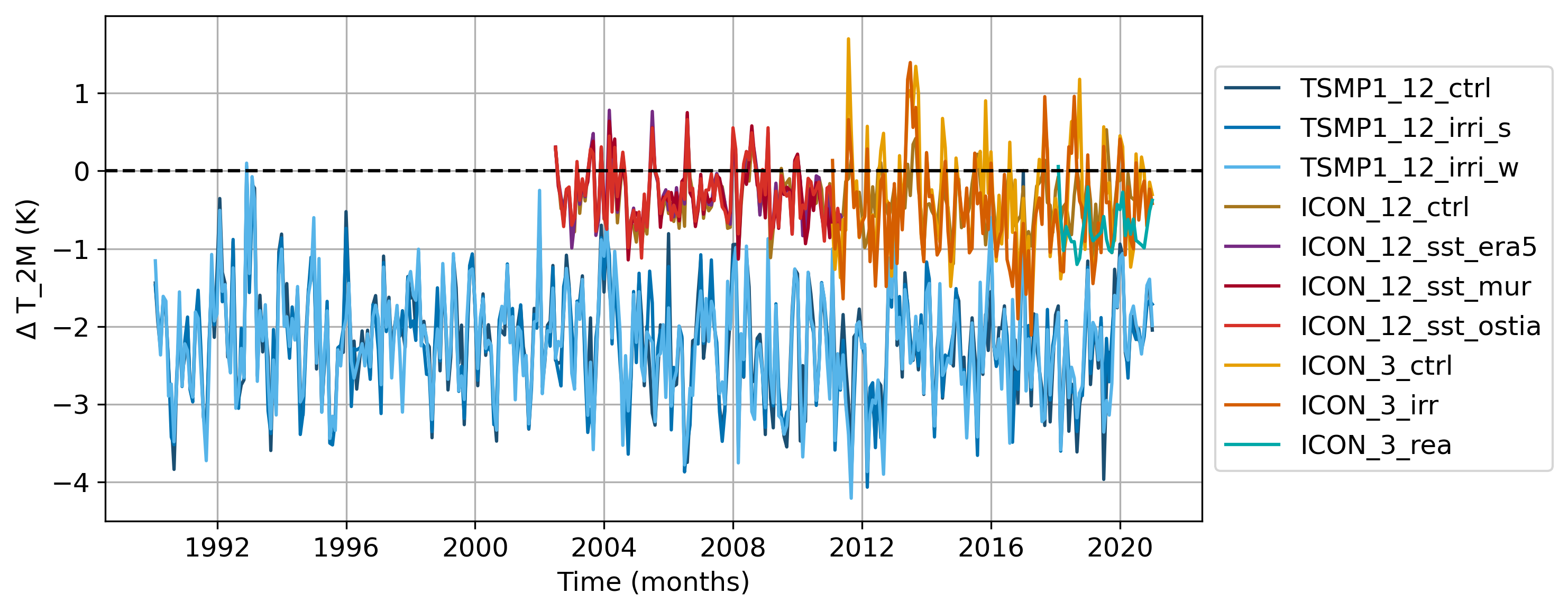}
    \label{Time_T_2M_bias_Ebro}
\end{subfigure}
\vfill
\begin{subfigure}{0.9\textwidth}
    \caption{Po}
    \includegraphics[width=0.9\linewidth]{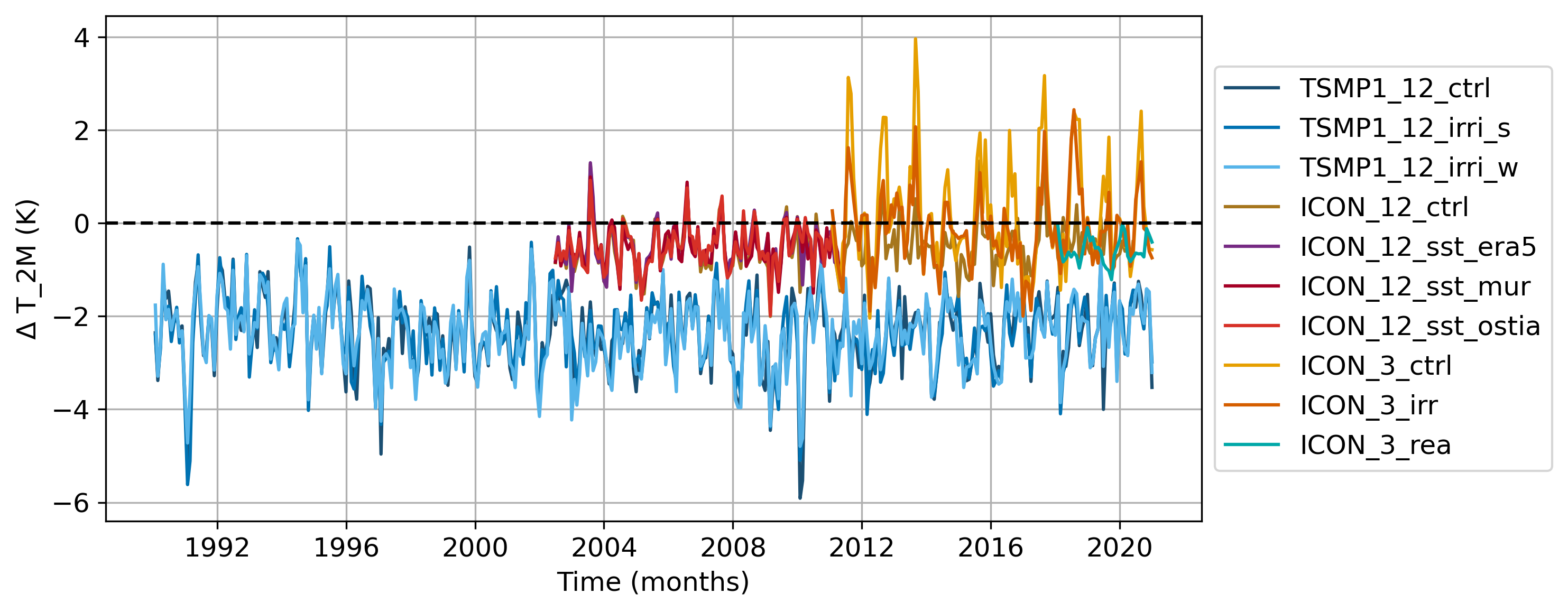}
    \label{Time_T_2M_bias_Po}
\end{subfigure}
\vfill
\begin{subfigure}{0.9\textwidth}
    \caption{Rhine}
    \includegraphics[width=0.9\linewidth]{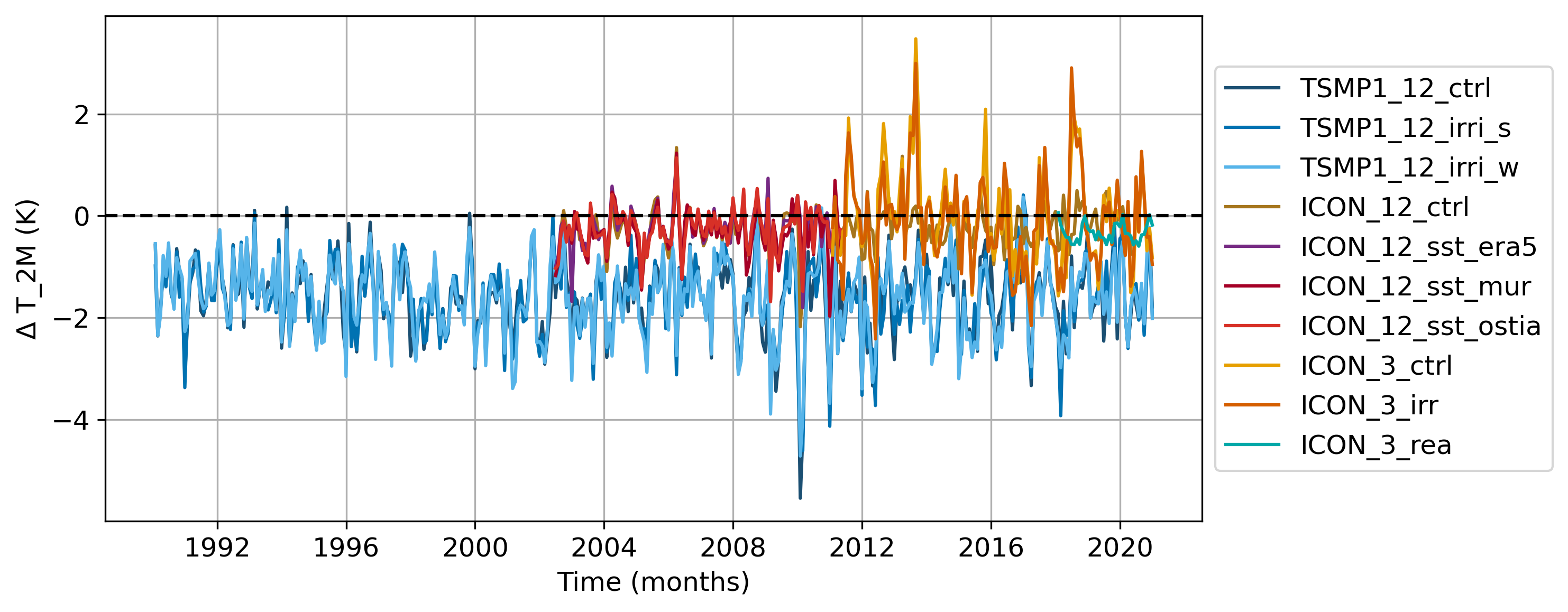}
    \label{Time_T_2M_bias_Rhine}
\end{subfigure}
\vfill
\begin{subfigure}{0.9\textwidth}
    \caption{Tisa}
    \includegraphics[width=0.9\linewidth]{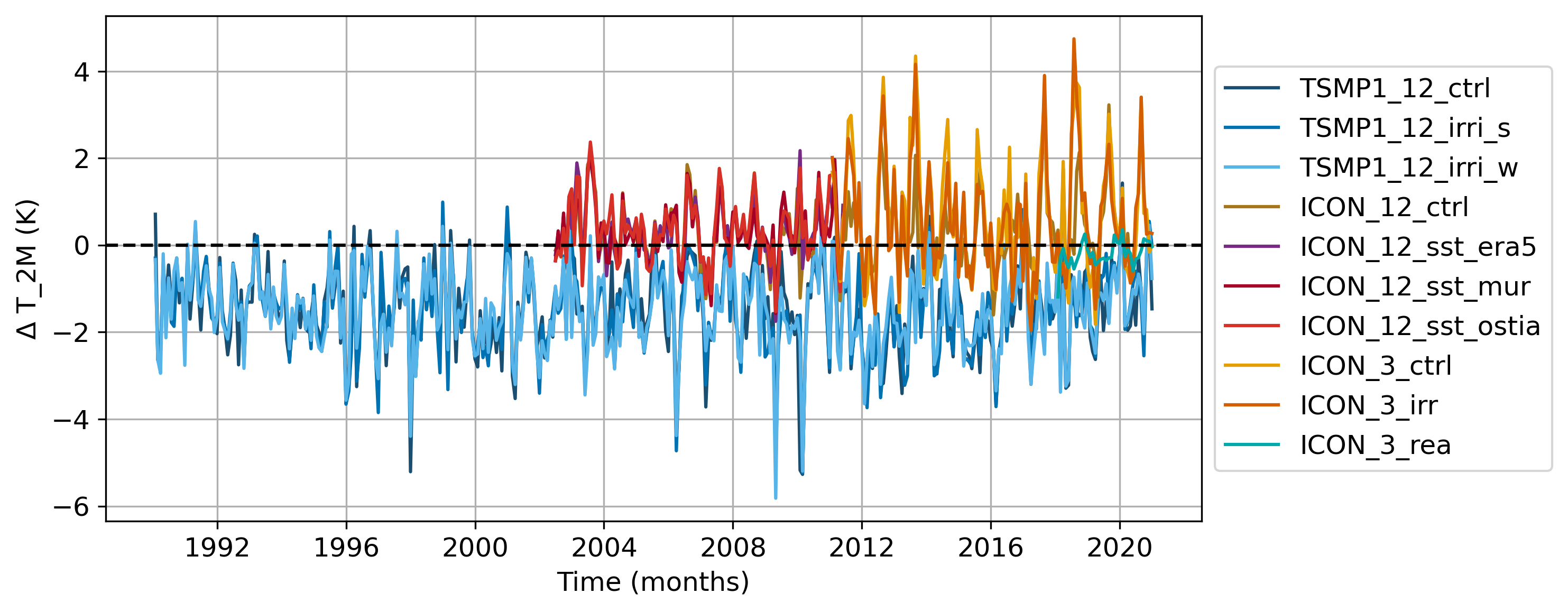}
    \label{Time_T_2M_bias_Tisa}
\end{subfigure}
    \caption{Timeseries of monthly T2m biases (Sim. - E-OBS) for the representative watersheds.}
    \label{Time_T_2M_bias_all}
\end{figure}

\begin{figure}[H]
\centering
\begin{subfigure}{0.9\textwidth}
    \caption{Ebro}
    \includegraphics[width=0.9\linewidth]{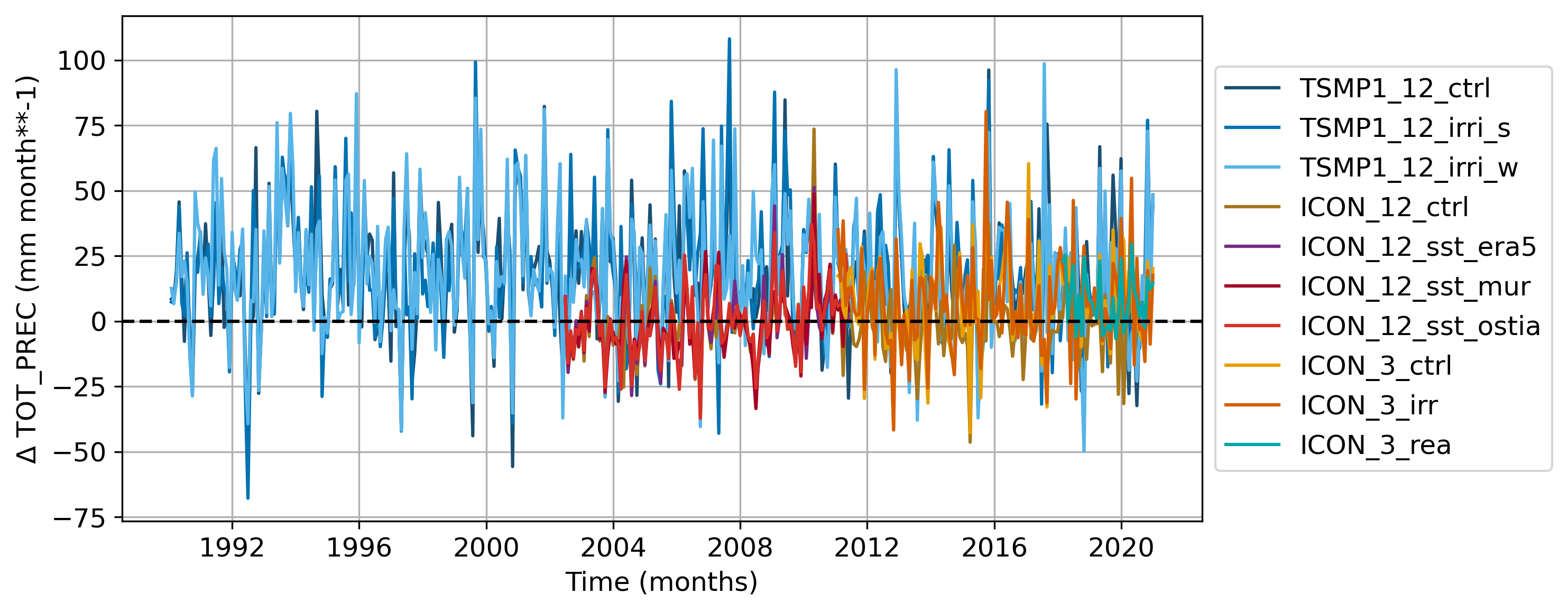}
    \label{Time_PP_bias_Ebro}
\end{subfigure}
\vfill
\begin{subfigure}{0.9\textwidth}
    \caption{Po}
    \includegraphics[width=0.9\linewidth]{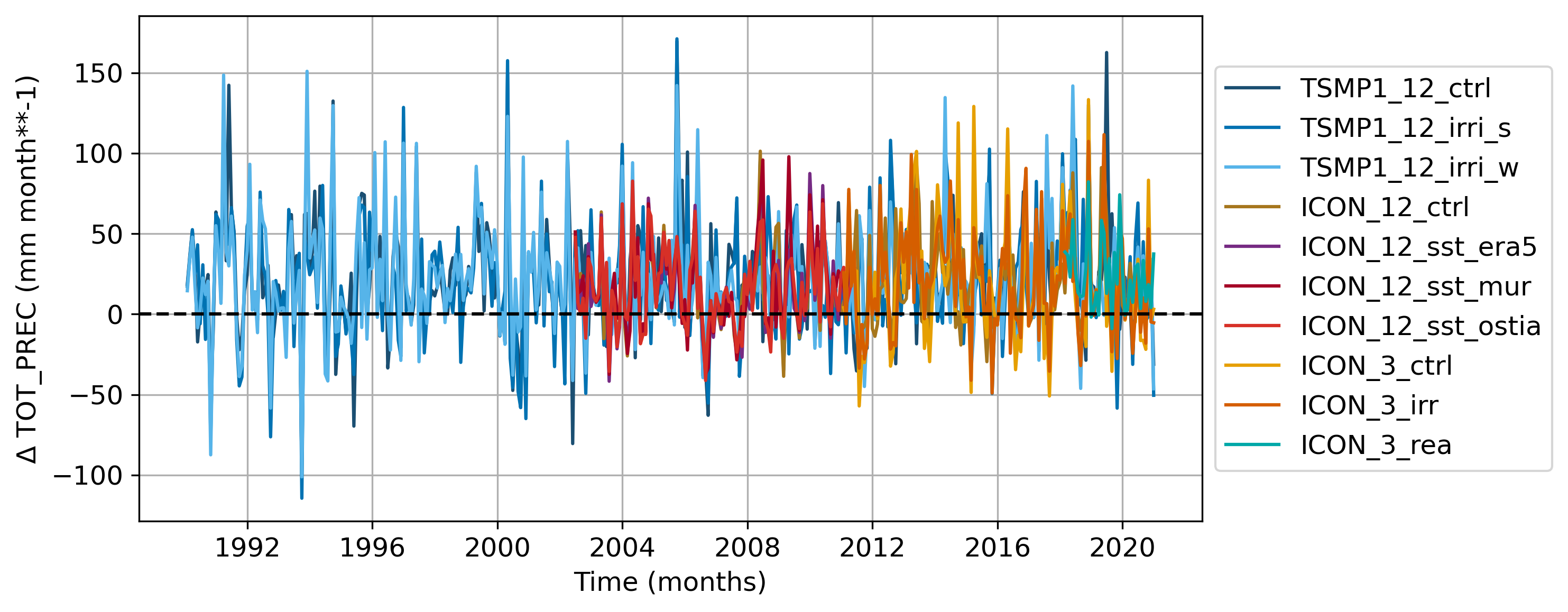}
    \label{Time_PP_bias_Po}
\end{subfigure}
\vfill
\begin{subfigure}{0.9\textwidth}
    \caption{Rhine}
    \includegraphics[width=0.9\linewidth]{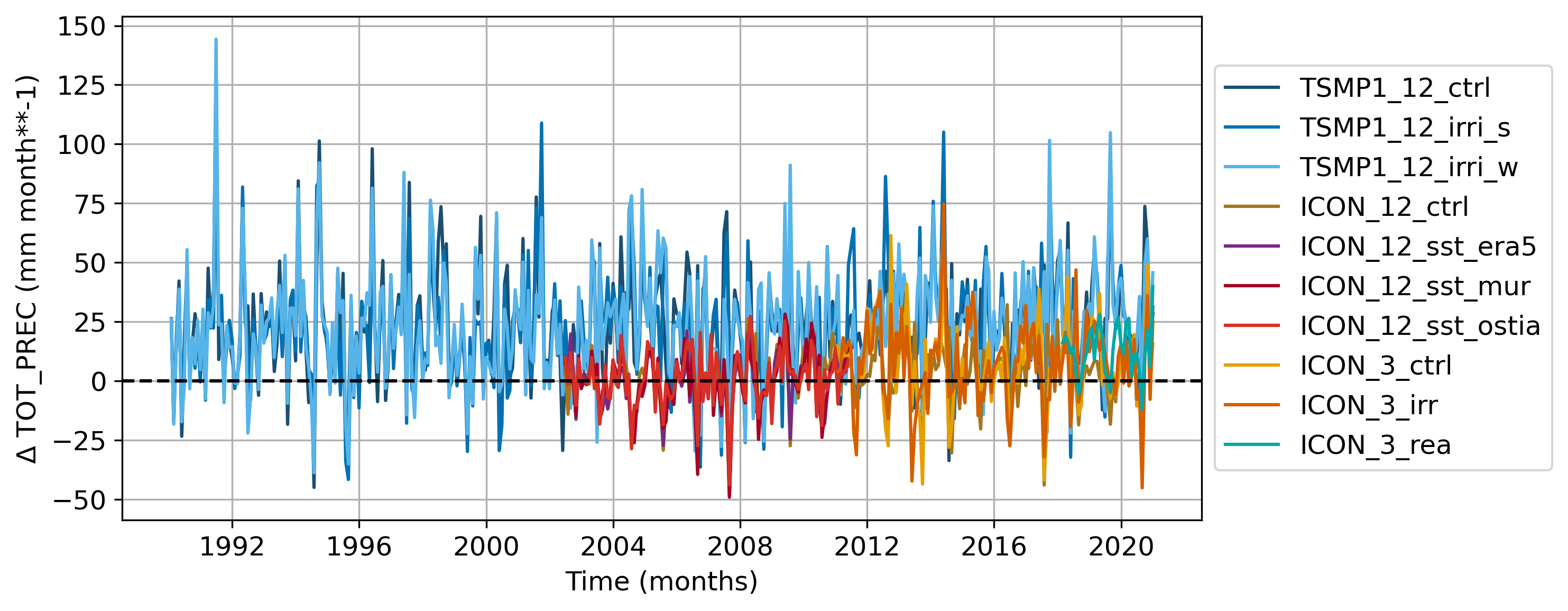}
    \label{Time_PP_bias_Rhine}
\end{subfigure}
\vfill
\begin{subfigure}{0.9\textwidth}
    \caption{Tisa}
    \includegraphics[width=0.9\linewidth]{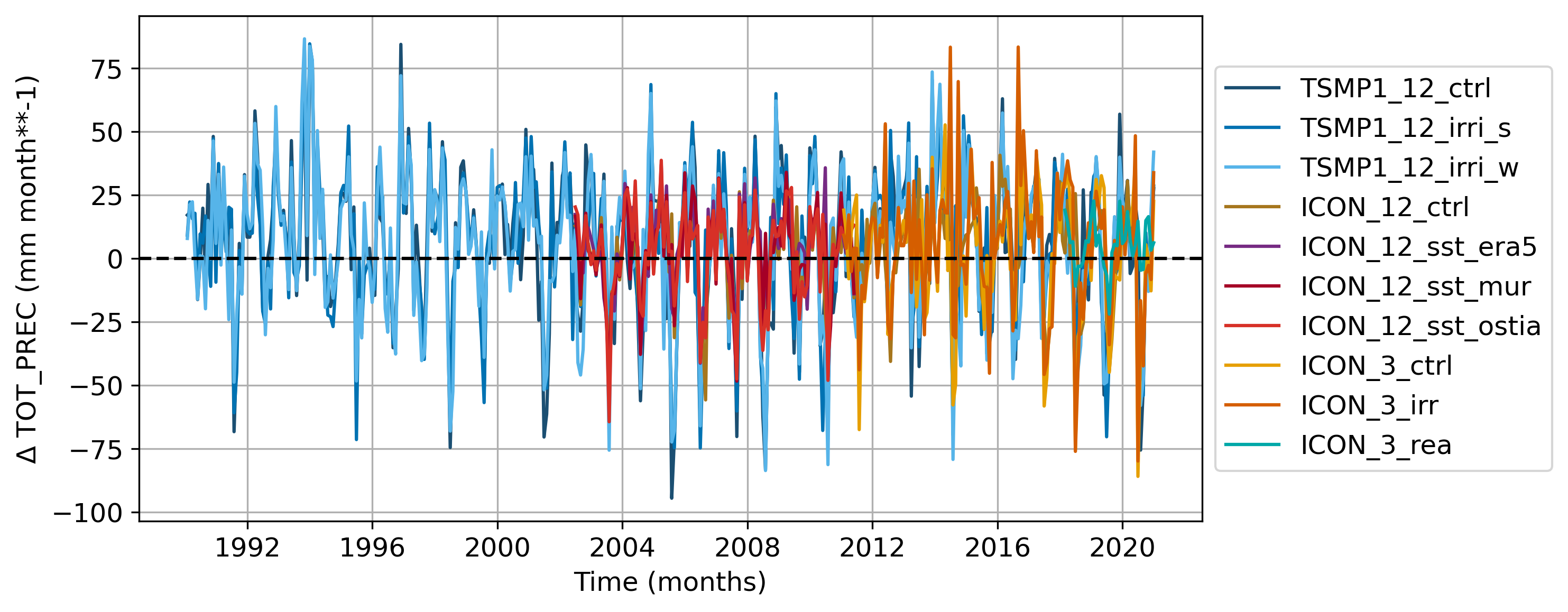}
    \label{Time_PP_bias_Tisa}
\end{subfigure}
    \caption{Timeseries of monthly precipitation biases (Sim. - GPCC) for the representative watersheds.}
    \label{Time_PP_bias_all}
\end{figure}

\begin{figure}[H]
\centering
\begin{subfigure}{0.9\textwidth}
    \caption{Ebro}
    \includegraphics[width=0.9\linewidth]{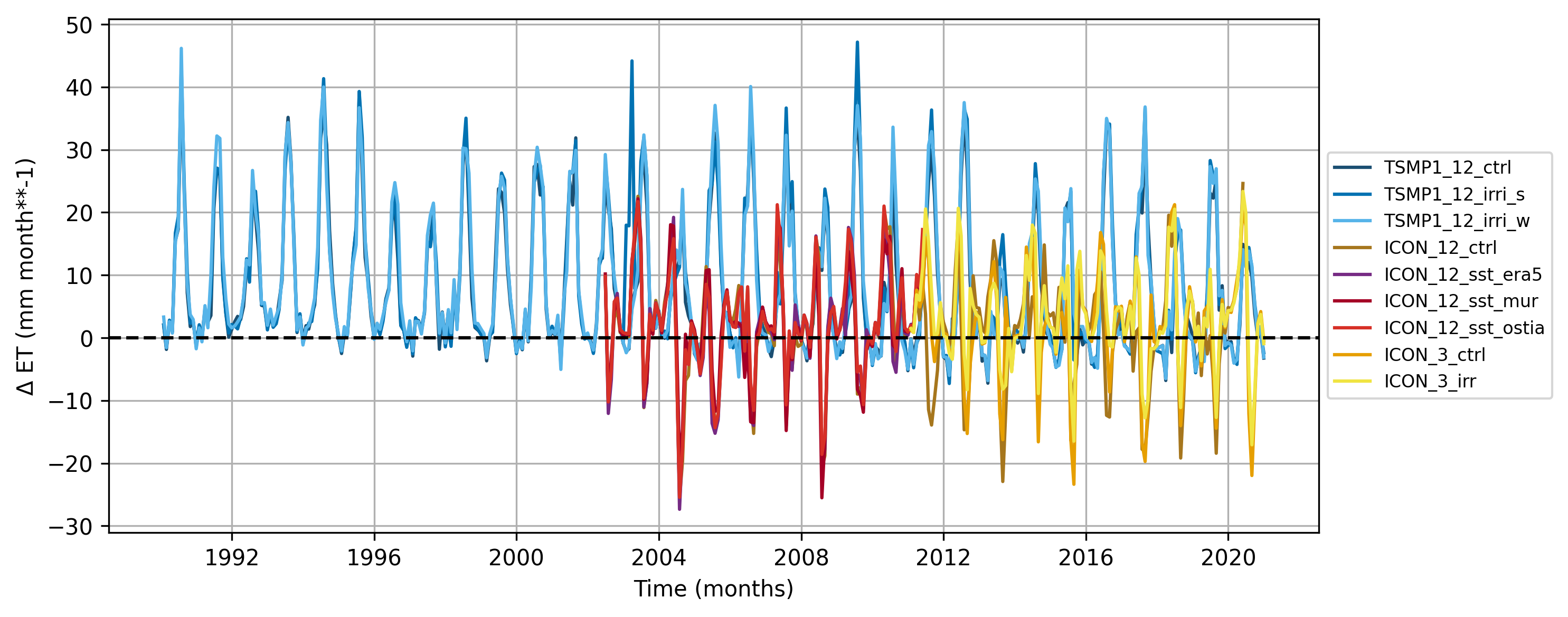}
    \label{Time_ET_bias_Ebro}
\end{subfigure}
\vfill
\begin{subfigure}{0.9\textwidth}
    \caption{Po}
    \includegraphics[width=0.9\linewidth]{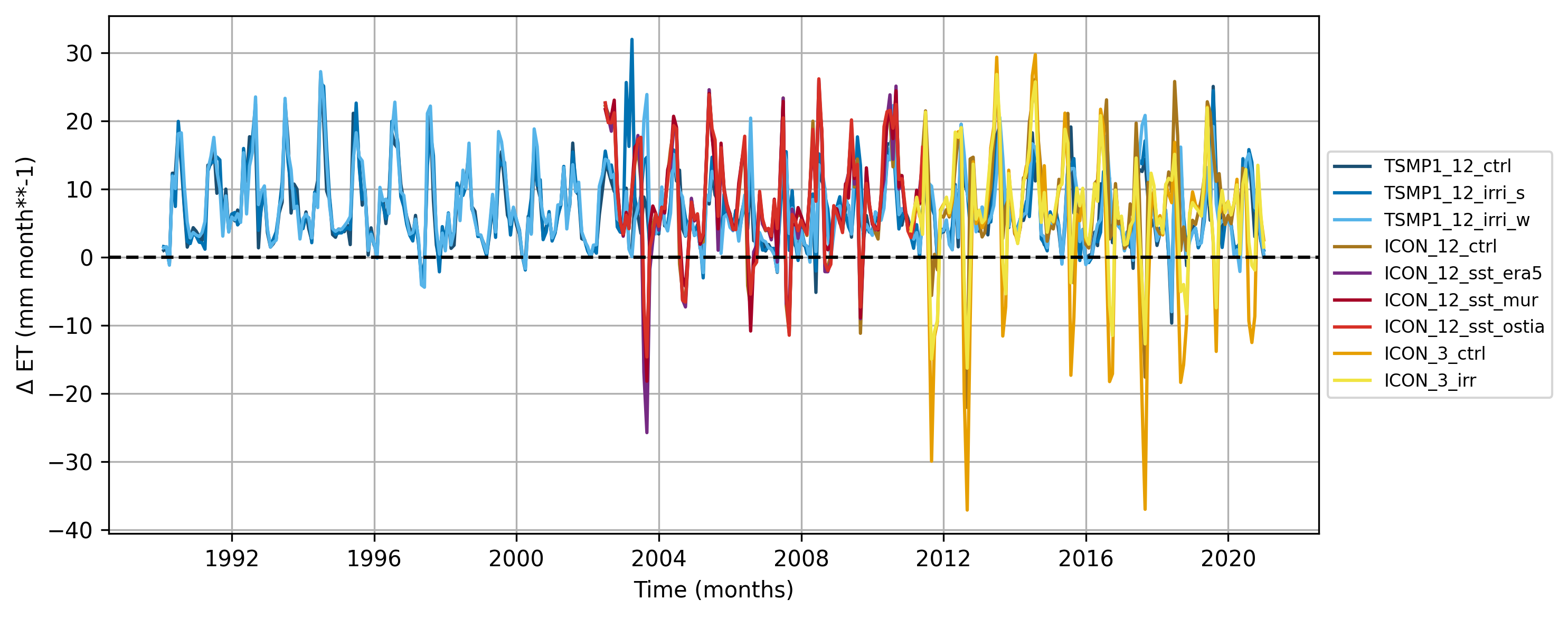}
    \label{Time_ET_bias_Po}
\end{subfigure}
\vfill
\begin{subfigure}{0.9\textwidth}
    \caption{Rhine}
    \includegraphics[width=0.9\linewidth]{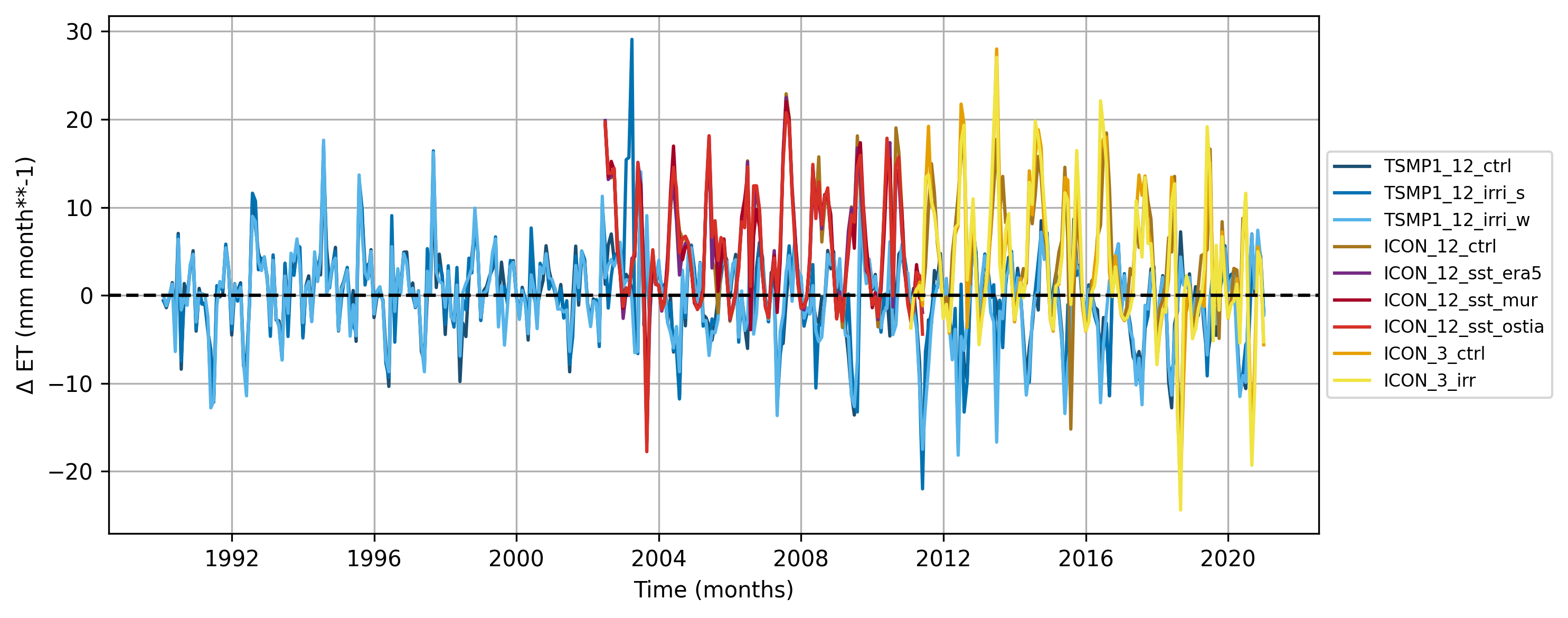}
    \label{Time_ET_bias_Rhein}
\end{subfigure}
\vfill
\begin{subfigure}{0.9\textwidth}
    \caption{Tisa}
    \includegraphics[width=0.9\linewidth]{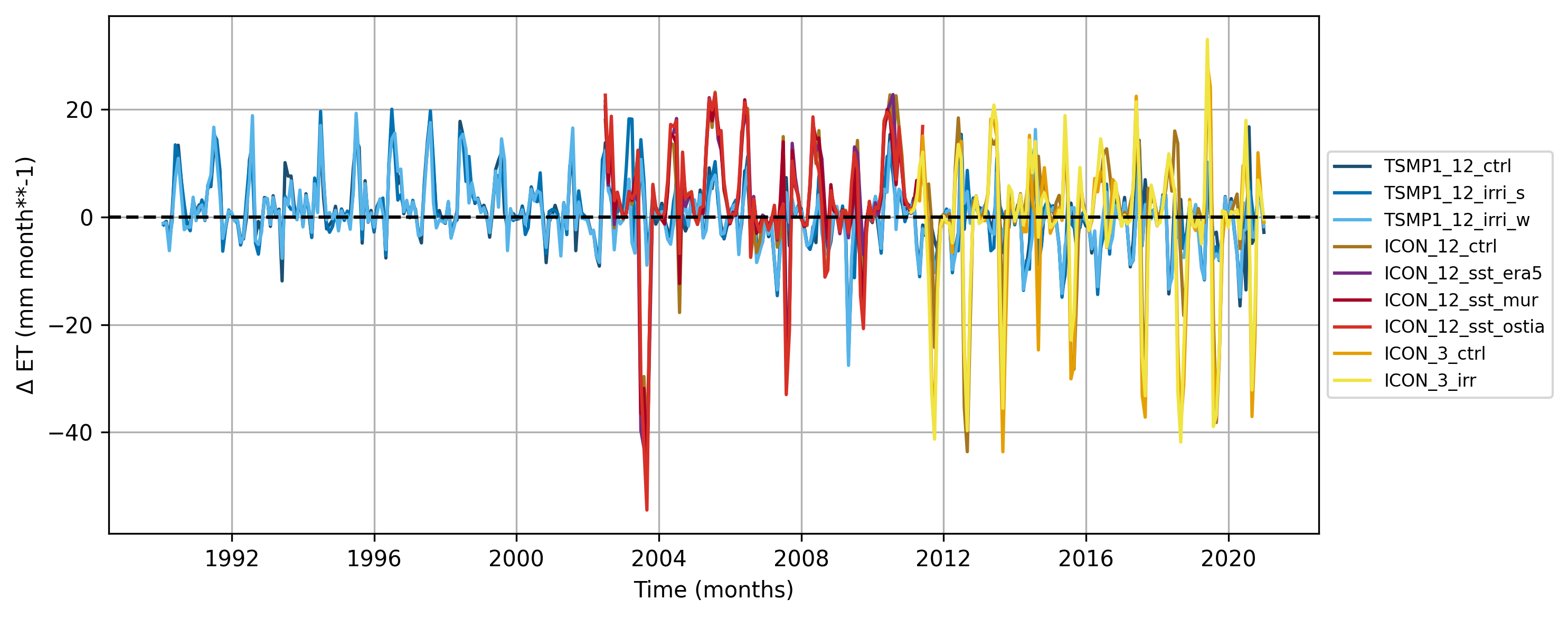}
    \label{Time_ET_bias_Tisa}
\end{subfigure}
    \caption{Timeseries of monthly ET biases (Sim. - GLEAM) for the representative watersheds.}
    \label{Time_ET_bias_all}
\end{figure}

\end{document}